\documentclass[oneside,12pt]{IIScthesisPSnPDF}

\usepackage{amssymb}
\usepackage{amsmath}
\usepackage{latexsym}
\usepackage{multicol}
\usepackage{booktabs}
\usepackage{multirow}
\usepackage{fancyhdr}
\usepackage[numbers]{natbib}
\usepackage{cleveref}
\usepackage{wrapfig}
\usepackage{algorithm}
\usepackage{algorithmic}
\usepackage{enumerate}
\usepackage{cancel}
\usepackage{mathrsfs}
\usepackage{color}
\usepackage[compact]{titlesec}
\usepackage{mdwlist}
\usepackage{tikz}
\usepackage{varwidth}
\usepackage[vertfit]{breakurl}
\usepackage{datetime}
\usepackage{pdfpages}
\usepackage{subcaption}
\usepackage{dsfont}
\usepackage{graphicx}
\usepackage{setspace}
\usepackage[skins]{tcolorbox}

\usetikzlibrary{shapes,arrows, trees}

\newcommand{\x}{{\boldsymbol{x}}}

\newdateformat{monthyeardate}{ \monthname[\THEMONTH], \THEYEAR}

\newcommand{\blankpage}{
	\newpage
	\thispagestyle{empty}
	\mbox{}
	\newpage
}

\crefname{observation}{observation}{observations}
\crefname{algorithm}{algorithm}{algorithms}
\crefname{align}{equation}{equations}
\crefname{eqnarray}{equation}{equations}

\newcommand{\remove}[1]{}

\newcommand{\notes}[1]{}

\newcommand{\first}[1]{$1^{\mathrm{st}}$}
\newcommand{\second}[1]{$2^{\mathrm{nd}}$}

\newcommand{\squishlisttwo}{
\begin{list}{$\blacktriangleright$}
{ \setlength{\itemsep}{0.5pt}
\setlength{\parsep}{0pt}
\setlength{\topsep}{0pt}
\setlength{\partopsep}{0.5pt}
\setlength{\leftmargin}{1em}
\setlength{\labelwidth}{1em}
\setlength{\labelsep}{0.5em} } }

\newcommand{\squishend}{
\end{list} }
\allowdisplaybreaks[1]

\begin{document}
\title{Acoustic and Semantic Modeling of Emotion in Spoken Language} 

\submitdate{\monthyeardate\today} 
\phd
\dept{Department of Electrical Engineering}
\faculty{Faculty of Engineering}
\author{Soumya Dutta}

\maketitle

\newpage\null\thispagestyle{empty}\newpage
\begin{center}
\LARGE{\underline{\textbf{Declaration of Originality}}}
\end{center}
\noindent I, \textbf{Soumya Dutta}, with SR No. \textbf{04-03-00-10-12-20-2-19108} hereby declare that
the material presented in the thesis titled

\begin{center}
\textbf{Acoustic and Semantic Modeling of Emotion in Spoken Language}
\end{center}

\noindent represents original work carried out by me in the \textbf{Department of Electrical Engineering} at \textbf{Indian Institute of Science} during the years \textbf{2021-2026}.

\noindent With my signature, I certify that:
\begin{itemize}
	\item I have not manipulated any of the data or results.
	\item I have not committed any plagiarism of intellectual
	property.
	I have clearly indicated and referenced the contributions of
	others.
	\item I have explicitly acknowledged all collaborative research
	and discussions.
	\item I have understood that any false claim will result in severe
	disciplinary action.
	\item I have understood that the work may be screened for any form
	of academic misconduct.
\end{itemize}

\vspace{15mm}

\noindent {\footnotesize{Date:	\hfill	Student Signature}} \qquad

\vspace{15mm}

\noindent In my capacity as supervisor of the above-mentioned work, I certify that the above statements are true to the best of my knowledge, and I have carried out due diligence to ensure the originality of the
thesis.

\vspace{15mm}

\noindent  {\footnotesize{Advisor Name: \hfill Advisor Signature}} \qquad

\blankpage

\vspace*{\fill}
\begin{center}
\large\bf \textcopyright \ Soumya Dutta\\
\large\bf \monthyeardate\today\\
\large\bf All rights reserved
\end{center}
\vspace*{\fill}
\thispagestyle{empty}

\blankpage

\vspace*{\fill}
\begin{center}
\Large DEDICATED TO \\[2em]
\Large\it My Parents \& Sister \\[1em]
for their unconditional support \\ [1em]
\Large\it My Wife \\[1em]
for her companionship \\ [1em]
\end{center}
\vspace*{\fill}
\thispagestyle{empty}




\setcounter{secnumdepth}{3}
\setcounter{tocdepth}{2}

\frontmatter 
\pagenumbering{roman}

\newpage\null\thispagestyle{empty}\newpage
\prefacesection{Acknowledgements}

The completion of this thesis would not have been possible without the guidance, support, and encouragement of many individuals.\\
First and foremost, I would like to express my deepest gratitude to my advisor, Dr. Sriram Ganapathy, for giving me the opportunity to work at the Learning and Extraction of Acoustic Patterns (LEAP) Lab. He entrusted me with the freedom to pursue a research direction that was new to both of us, allowing me to grow intellectually and independently. There was a prolonged period during the last five years when publications did not materialize despite sustained effort. I am especially grateful to Sriram for his patience, encouragement, and unwavering support during that phase. I will always cherish our discussions - both technical and non-technical - which have profoundly shaped my thinking and perspective.\\
I would also like to thank the Ministry of Education (MoE), Government of India, for the financial support provided through the Prime Minister’s Research Fellowship (PMRF). I am equally grateful to the Qualcomm Innovation Fellowship (QIF) for supporting my research. I would also like to thank my mentors during my internship stints at Samsung Research and Adobe Research.\\
I would like to thank my colleagues at the LEAP Lab - Samrat, Ameen, Prachi, Shreyas, Anurenjan, and others - for the time we spent together in the lab. The small words of encouragement at crucial moments often went a long way in keeping me motivated throughout this journey. I am also thankful to my co-authors - Avni, Smruthi, Viveka, Varada - and my current collaborators, Payal and Paras, for the stimulating discussions and collaborative efforts that shaped this work.\\
Life at IISc would have been incomplete without the wonderful group of friends at B Mess - Debarpan, Aniket, Arnab, Sambit, Ujjwal Da, and Akash. Except for Debarpan (who is also my lab colleague) and Sambit, the others left IISc in 2022/2023, but the long conversations we shared - often with me enthusiastically dominating the discussion - in the mess and hostels remain some of my fondest memories of IISc. A big thank you to my friends from my life before IISc - Arup, Ayan, Suman, Kaustav D., and Kaustav C. With time, our interactions have reduced significantly, but it is always good to know that some of my well-wishers are still just a call away.\\
Finally, I would like to thank my family. I am deeply grateful to my parents and my elder sister for their constant support and encouragement over the last five years. Knowing that they have always had my back is something I may have grown accustomed to, but it is a privilege I will never take for granted.\\
And to my wife, Abhipsa - we entered IISc as friends and became life partners during the course of this journey. We have witnessed each other’s frequent lows and occasional highs over these five years, and have hopefully emerged stronger together. I look forward to building a beautiful partnership with her in the years to come.

\prefacesection{Abstract}

Emotions are fundamental psychological states that lend humanness to everyday interactions, shaping trust, engagement, and social bonding. As artificial intelligence systems powered by large language models become increasingly integrated into daily life, enabling them to accurately recognize, characterize, and generate fine-grained human emotions remains a critical challenge. Although emotional expression is inherently multimodal, this thesis focuses on emotions as conveyed through spoken language. The thesis investigates how acoustic and semantic information can be jointly modeled to advance both emotion understanding and emotion synthesis from speech.

The first part of this thesis focuses on emotion-aware representation learning through pre-training. While existing self-supervised learning objectives for speech and text primarily emphasize linguistic reconstruction, they often under-represent affective cues critical for emotion recognition. This thesis proposes pre-training strategies that incorporate both acoustic and semantic supervision, yielding representations that are better aligned with emotional characteristics in spoken language. In addition, a speech-driven supervised pre-training pipeline is introduced for text-based emotion recognition, enabling the construction of large-scale emotion-aware text models even in the absence of manually annotated text corpora.

The second part addresses emotion recognition from spoken language in realistic and conversational settings. Emotion Recognition in Conversations (ERC) introduces challenges such as contextual dependency, speaker turns, and modality imbalance, which require principled integration of acoustic and semantic cues across conversational turns. Hierarchical modeling frameworks based on cross-modal attention and mixture-of-experts fusion are proposed to capture conversational structure and modality-specific dynamics, resulting in robust performance across multiple benchmark datasets.

The final part of the thesis investigates emotion style transfer in speech synthesis. A novel formulation of emotion style transfer is introduced in a textless and non-parallel speech-to-speech setting, where the goal is to transfer emotional speaking style from a reference utterance to a source utterance while preserving the source speaker’s identity and linguistic content. By framing emotion transfer as a disentangled representation learning problem, the proposed approach enables controllable emotional transformations under realistic conditions. Extensive objective and subjective evaluations demonstrate effective emotion transfer, speaker preservation, and content invariance. The thesis further shows that emotion style transfer can be leveraged as a data augmentation strategy to improve emotion recognition performance in low-resource settings.

Overall, this thesis highlights key limitations in current speech and language technologies, including recent large language models, in their ability to robustly model emotions from spoken language. By adopting an acoustic–semantic modeling perspective, the proposed methods contribute toward endowing future AI systems with more reliable emotional understanding and expressive capabilities. The thesis concludes by outlining extensions of this work to clinically relevant applications such as automatic depression detection, underscoring the broader impact of emotion-aware modeling of spoken language.

\prefacesection{Publications based on this Thesis}
\noindent \textbf{Peer-reviewed Journal Papers}

\begin{enumerate}
    \item S. Dutta, S. Balaji, S. Ganapathy, ``A Mixture-of-Experts Model for Multimodal Emotion Recognition in Conversations ,'' \textsl{Computer Speech \& Language} (accepted for publication).
      \item S. Dutta, S. Ganapathy, ``Leveraging Content and Acoustic Representations for Speech Emotion Recognition,'' \textsl{IEEE Transactions on Audio, Speech and Language Processing}, vol. 33, pp. 3678-3689, 2025.
      \item S. Dutta, A. Jain, S. Ganapathy, ``Textless and Non-Parallel Speech-to-Speech Emotion Style Transfer,'' \textsl{IEEE Transactions on Audio, Speech and Language Processing} (under review).
    \end{enumerate}

\noindent \textbf{Peer-reviewed Conference Papers}

\begin{enumerate}
\item  S. Dutta, S. Balaji, Varada R, V. Salinamakki, S. Ganapathy ``ABHINAYA - A System for Speech Emotion Recognition In Naturalistic Conditions Challenge,'' \textsl{Proc. Interspeech 2025}: pp. 4663-4667.
\item S. Dutta, S. Ganapathy, ``LLM supervised Pre-training for Multimodal Emotion Recognition in Conversations,'' \textsl{Proc. ICASSP 2025}: pp. 1-5.
\item S. Dutta, S. Ganapathy, ``Zero Shot Audio To Audio Emotion Transfer With Speaker Disentanglement,'' \textsl{Proc. ICASSP 2024}: pp. 10371-10375.
  \item S. Dutta, S. Ganapathy, ``Multimodal Transformer with Learnable Frontend and Self Attention for Emotion Recognition,'' \textsl{Proc. ICASSP 2022}: pp. 6917-6921.

    \end{enumerate}

\noindent \textbf{Preprints}
\begin{enumerate}
    \item S. Dutta, S. Ganapathy, ``HCAM--hierarchical cross attention model for multi-modal emotion recognition,'' \textsl{arXiv preprint arXiv:2304.06910 (2023).}
\end{enumerate}
\newpage\null\thispagestyle{empty}\newpage
\tableofcontents
\listoffigures
\listoftables

\mainmatter 
\setcounter{page}{1}
\chapter{Introduction}
\label{chap:introduction}
Emotions play a central role in human communication, shaping how messages are conveyed, perceived, and responded to in social interactions. Humans express emotions in a number of ways, such as facial expressions~\cite{tarnowski2017emotion}, speech~\cite{scherer2003vocal}, gestures~\cite{navarretta2012individuality}, physiological signals~\cite{knapp2010physiological}, or using a combination of these. Among these modalities, spoken language occupies a particularly important role in conveying emotions during interactive communication. As conversational artificial intelligence systems increasingly adopt voice-based interfaces, recent studies~\cite{phang2025investigating} indicate a strong user preference for spoken interaction when expressing or querying emotional states, compared to text-based communication. Supporting such interactions requires not only the ability to understand emotions from speech, but also the ability to generate speech with emotional expressiveness. 

\section{Emotion Understanding and Synthesis}
\begin{figure}[t]
    \centering
    \includegraphics[width=0.9\textwidth,trim={7cm 6cm 7cm 4cm},clip]{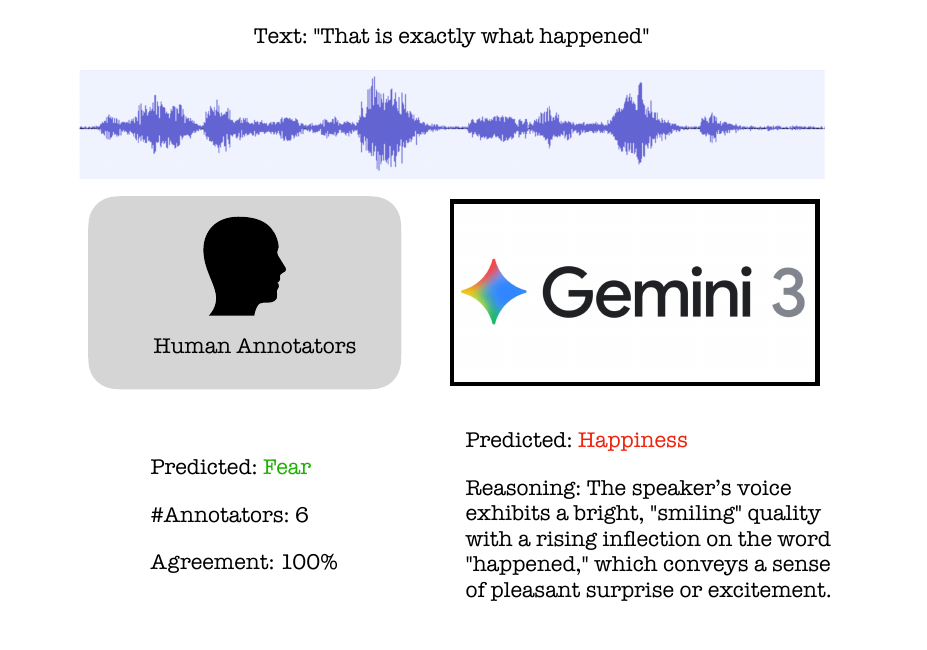}
    \caption{Illustration of a limitation in emotion understanding from spoken language by a multimodal conversational model. Although the utterance is perceived as conveying fear by six human annotators with $100\%$ agreement, Gemini $3$ (with thinking enabled) predicts a happy emotion and provides an incorrect explanation.}
    \label{fig:intro_example}
\end{figure}

Despite substantial progress in speech and language technologies, robust understanding of emotions from spoken language remains a difficult task. Large language model (LLM)-based conversational agents have demonstrated strong capabilities in text-based interaction and are increasingly used for emotionally sensitive applications, including emotional support and mental health-related queries~\cite{vaidyam2019chatbots, zheng2025customizing}. Increasingly, these models are trained in a multimodal manner, thereby enabling them to process both speech and textual inputs~\cite{team2023gemini}.\\
However, access to speech input alone does not guarantee reliable emotion understanding. Figure~\ref{fig:intro_example} illustrates a representative failure case in which a spoken utterance conveying fear is incorrectly inferred as expressing happiness by a large language model (LLM), despite the availability of acoustic cues. Notably, the same utterance is annotated as fear by six independent human annotators with $100\%$ agreement, indicating that the emotional intent is unambiguous to human listeners. The model’s incorrect prediction, accompanied by flawed explanatory reasoning, highlights a limitation of current systems: emotion inference is often driven by shallow semantic heuristics or incomplete acoustic reasoning rather than principled modeling of emotion-relevant acoustic and semantic factors in speech. Beyond understanding, this limitation also has implications for speech generation: producing emotionally appropriate spoken responses requires explicit control over how acoustics in speech interact with its linguistic content.\\
This points to a central challenge in both understanding and synthesis of emotions in speech: emotional intent is not determined by linguistic content or acoustic realization in isolation, but emerges from their interaction. Addressing this challenge requires a principled understanding of how acoustic and semantic information are jointly encoded and perceived in speech.

\section{Acoustics and Semantics in Spoken Language}
Throughout this thesis, we view emotion modeling in speech through this acoustic-semantic lens, and investigate how these two sources of information can be effectively learned, combined, and disentangled for emotion understanding and synthesis.\\ 
While acoustic and semantic cues jointly contribute to emotion understanding, learning representations that generalize across datasets remains a central challenge in speech emotion recognition. Self-supervised speech models such as wav2vec~\cite{schneider2019wav2vec}, HuBERT~\cite{hsu2021hubert}, and WavLM~\cite{chen2022wavlm} have demonstrated strong generalization across a wide range of speech tasks. However, their pre-training objectives are primarily optimized for linguistic reconstruction while ignoring the fine-grained acoustic variations crucial for emotional expression. This gap motivates the exploration of emotion-aware pre-training strategies that jointly model acoustic and semantic factors aligned with affective cues.\\
Beyond utterance-level emotion recognition, many real-world applications require understanding emotions as they evolve over the course of a conversation. This setting, commonly referred to as Emotion Recognition in Conversations (ERC)~\cite{poria2019emotion}, involves modeling contextual dependencies across multiple utterances and speakers. Although combining acoustic and semantic information has proven effective for tasks such as multimodal emotion recognition~\cite{poria2017context, poria2017review, yoon2018multimodal} and depression detection~\cite{shen2022automatic, wu2023self}, accurately modeling emotions in conversational settings remains challenging due to the strong contextual dependence. Further, most existing ERC datasets are collected under controlled or acted conditions, limiting their ability to reflect the variability encountered in naturalistic speech.\\
Beyond emotion understanding, generating emotionally appropriate speech requires control over how acoustic and semantic factors are realized in the output signal. This motivates the study of disentangled representation learning~\cite{wang2024disentangled}, which aims to separate components such as linguistic content, speaker identity, and emotional speaking style. While such factorization has been explored for tasks including speaker recognition~\cite{kwon2020intra} and automatic speech recognition~\cite{chan2022content}, it has been most prominently applied in voice conversion (VC), where the objective is to modify speaker identity while preserving linguistic content~\cite{sisman2020overview}. Building on this idea, a new paradigm of controllable speech synthesis can be envisaged in which only the emotional speaking style is transferred from a reference signal to a target utterance, while keeping both the speaker identity and linguistic content intact. This task, referred to as \textit{emotion style transfer}~\cite{dutta2024zero}, requires a principled factorization of speech into its acoustic and semantic components.\\
In this thesis, we build upon these developments and investigate three interrelated directions for advancing the modeling of emotion in spoken language. First, we study emotion-aware representation learning techniques that capture emotion-relevant variations in speech by jointly leveraging acoustic and semantic information. Second, we examine how speech and text representations can be effectively used for emotion recognition from spoken language, with a particular focus on conversational and naturalistic settings. Third, we explore the disentanglement of acoustic and semantic factors to enable controllable emotional transformations in speech, exemplified by the task of emotion style transfer. Together, these directions provide a unified framework for understanding and synthesizing emotional information encoded in spoken language.
\section{Problem Statements}\label{sec:intro_probstat}
The different problem statements considered in this thesis are described in detail.
\subsection{Emotion Aware Pre-training in Spoken Language}\label{sec:intro_pretrain}
Most pre-training algorithms for speech and text involve masking portions of the raw signal or text and training models to predict the masked regions. This paradigm, popularized as masked language modeling (MLM)~\cite{devlin2019bert}, has led to substantial advances in self-supervised representation learning. However, existing pre-training objectives are primarily optimized for semantic reconstruction, which biases models toward content-related information while downplaying expressive acoustic variations. Since emotional intent in speech is often conveyed through prosody, energy, and temporal dynamics, this limits the effectiveness of such representations for emotion recognition. In this thesis, we consider two problems related to designing pre-training objectives that are better aligned with emotion recognition:
\begin{itemize}
    \item Both semantic and acoustic cues contribute critically to emotion understanding; yet current self-supervised learning frameworks seldom model these factors jointly. While some recent approaches explore emotion-aware pre-training for speech~\cite{chen2024vesper, ma2024emotion2vec}, they still overlook the complementary roles of semantics and acoustics. In this thesis, we investigate pre-training strategies that leverage both factors, and propose a unified recipe for emotion-aware self-supervised learning in speech.
    \item Recent work has shown that task-specific knowledge can be transferred from large pre-trained models to smaller or task-specific models through supervision or distillation~\cite{hinton2015distilling}. We formulate a complementary problem setting in which spoken utterances are first transcribed into text and annotated with emotion labels inferred using a large language model (LLM). The resulting emotion-labeled text corpus is then used to pre-train semantic representations for emotion recognition, enabling emotion-aware text pre-training even when manually annotated emotion data is unavailable.
\end{itemize}
\subsection{Multimodal Emotion Recognition}\label{sec:intro_emorec}
The problems considered under the purview of emotion recognition from spoken language are:
\begin{itemize}
    \item Emotion Recognition in Conversations (ERC) is a sequence-to-sequence classification problem in which a conversation, consisting of multiple utterances from two or more speakers, is mapped to a sequence of emotion labels. The task requires modeling long-range contextual dependencies as well as interactions between speakers. Most existing approaches address ERC using end-to-end architectures that jointly perform contextual modeling and multimodal fusion. In this thesis, we study ERC through frameworks that explicitly separate contextual modeling from multimodal fusion via hierarchical modeling or modular network architectures.

    \item Traditional Speech Emotion Recognition (SER) systems have primarily relied on acoustic features such as mel-spectrograms or hand-crafted descriptors extracted using toolkits like openSMILE~\cite{eyben2010opensmile}. While several prior works incorporate linguistic information, these approaches are typically evaluated in controlled or acted settings with access to ground-truth transcripts. In this thesis, we focus on SER in naturalistic conditions and explore the use of ASR-generated transcripts to integrate acoustic and semantic cues derived directly from speech.

\end{itemize}
\subsection{Emotion Style Transfer in Speech-to-Speech Settings}\label{sec:intro_est}
With the growing adoption of speech synthesis in applications such as personal assistants~\cite{lopez2017alexa}, entertainment~\cite{wang2019comic}, and robotics~\cite{marge2022spoken}, the demand for controllable and expressive speech generation has increased significantly. Consequently, recent research in text-to-speech (TTS) has focused on modeling expressive attributes such as pitch, energy, speaking style, and emotion~\cite{xie2025towards}. While these approaches enable fine-grained control over synthesized speech, they typically rely on textual input.\\
In contrast, this thesis explores a different paradigm: \emph{emotion style transfer} (EST) in speech-to-speech settings. Given a source speech signal and a reference speech signal conveying a target emotion, the goal is to transfer the emotional speaking style from the reference to the source while preserving the speaker identity and linguistic content of the source. We investigate this problem in a \emph{textless} and \emph{non-parallel} setting, where transcriptions are unavailable and paired recordings of the same speaker uttering identical content in different emotions do not exist. This setting presents unique challenges and necessitates effective disentanglement of emotional, semantic, and speaker-related factors.
\section{Outline of Contributions}\label{sec:intro_contr}
\begin{figure}[t]
    \centering
    \includegraphics[width=0.9\textwidth,trim={0cm 9cm 0cm 3cm},clip]{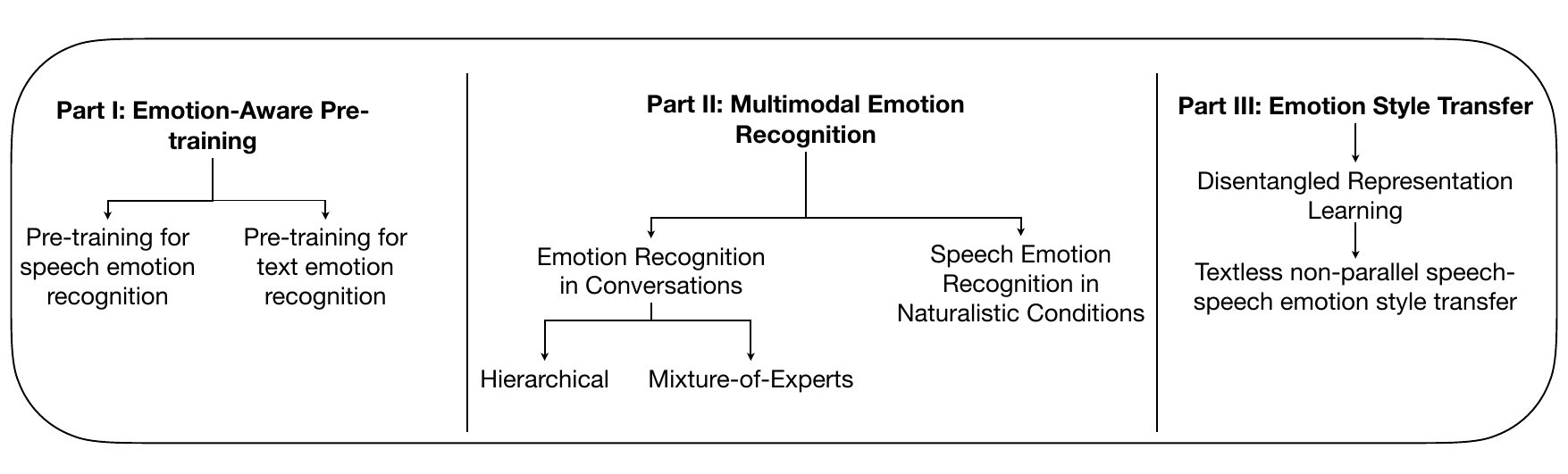}
    \caption{Summary of the thesis contributions}
    \label{fig:intro_contr}
\end{figure}

This thesis makes contributions along three broad and interconnected research directions: 
(i) emotion-aware pre-training of speech and text models, 
(ii) emotion recognition from spoken language, and 
(iii) emotion style transfer in speech-to-speech settings. 
An overview of these contributions is illustrated in Fig.~\ref{fig:intro_contr}.\par
\begin{itemize}
    \item \textbf{Emotion-Aware Pre-training:}
    We study the design of pre-training objectives that are better aligned with emotion understanding from spoken language. Moving beyond conventional masked prediction objectives, we investigate pre-training strategies that explicitly account for both acoustic and semantic factors of speech. Our proposal is shown to achieve better generalization across eight benchmark datasets on speech emotion recognition, while maintaining computational efficiency. 
    Additionally, we propose a framework that leverages automatic speech transcription and large language models to generate emotion-labeled text data, enabling emotion-aware pre-training of text-based models even when only spoken language is available. This establishes a practical strategy for emotion-aware text representation learning in scenarios where manually annotated text resources are limited.
    
    \item \textbf{Multimodal Emotion Recognition} 
    We introduce hierarchical modeling approaches for handling the complexity of Emotion Recognition in Conversations (ERC). We study modular fusion strategies, including mixture-of-experts architectures, to effectively integrate acoustic and semantic cues across conversational turns. Our proposed architectures establish new state-of-the-art results on popular ERC datasets.
    Further, we investigate speech emotion recognition (SER) under realistic and naturalistic conditions, focusing on robustness to data imbalance. As part of the \textit{Speech Emotion Recognition in Naturalistic Conditions Challenge} at \textit{Interspeech 2025}, we explore the use of speech large language models (SLLMs), large language models (LLMs), and conventional self-supervised speech models such as WavLM for emotion classification. We explore different loss functions to mitigate class imbalance. Our performance in the challenge establish the robustness of the proposed modeling strategies in realistic scenarios.

    \item \textbf{Emotion Style Transfer in Speech:}
    We formulate emotion style transfer as a disentangled representation learning problem in a textless, non-parallel speech-to-speech setting. To the best of our knowledge, this work is among the first to address zero-shot emotion style transfer without requiring parallel emotional recordings of identical content. The proposed framework enables emotional style transfer while preserving speaker identity and linguistic content, and generalizes to unseen speakers and unseen emotion combinations. 
    
\end{itemize}
\section{Road Map for the Rest of the Thesis}\label{sec:intro_roadmap}
\begin{figure}[t]
    \centering
    \includegraphics[width=0.9\textwidth,trim={3cm 7cm 4cm 5cm},clip]{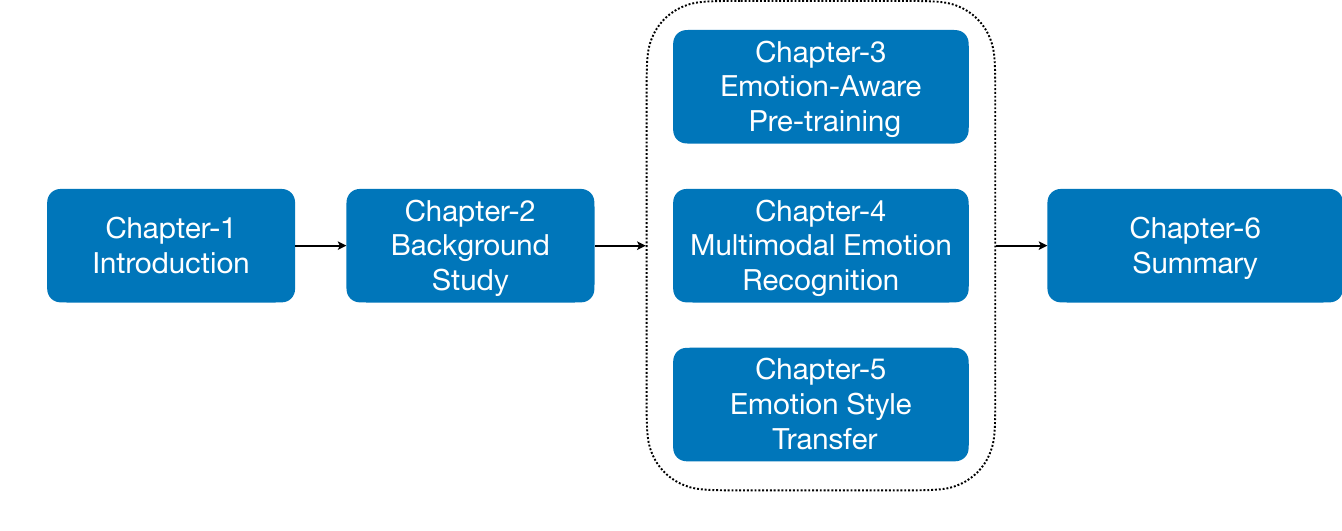}
    \caption{Road map for the thesis chapters}
    \label{fig:intro_roadmap}
\end{figure}
The organization of the chapters in this thesis is illustrated schematically in Fig.~\ref{fig:intro_roadmap}. The remainder of the thesis is structured as follows.\par
Chapter~\ref{chap:setting} provides the background for the experimental studies by introducing the datasets used and some baseline systems employed across the different problems considered.\par
Chapter~\ref{chap:pretraining} introduces an emotion-aware pre-training framework that incorporates both acoustic and semantic supervision for speech representations. The proposed pre-trained model is shown to outperform significantly larger models across multiple speech emotion recognition benchmarks. The chapter further presents an LLM-based supervised pre-training strategy for text-based emotion recognition. 
\par
Chapter~\ref{chap:recognition} focuses on the problem of emotion recognition from spoken language in conversational and naturalistic conditions. 
A hierarchical modeling paradigm is explored for the task of Emotion Recognition in Conversations (ERC). Further, a modular mixture-of-experts framework is proposed to fuse acoustic and semantic representations derived from speech large language models and text-based large language models for ERC. We then discuss the system proposed for the \textit{Interspeech 2025} Speech Emotion Recognition in Naturalistic Conditions Challenge. \par
Chapter~\ref{chap:styletransfer} addresses the problem of emotion style transfer in a speech-only, non-parallel setting. The chapter first investigates the disentanglement of semantic, speaker, and emotional factors in speech, and subsequently proposes a framework for emotion style transfer. Extensive objective and subjective evaluations demonstrate the effectiveness of the proposed approach. Additionally, the method is applied as a data augmentation strategy for improving emotion recognition performance.\par
Finally, Chapter~\ref{chap:summary} summarizes the key contributions of this thesis and discusses its limitations. The chapter concludes by outlining potential future research directions, including extensions of emotion recognition and pre-training methods toward clinically relevant tasks such as automatic depression detection.

\chapter{Background Study}
\label{chap:setting}
This chapter presents the background and prior work relevant to the experimental studies conducted in this thesis. We begin by reviewing the speech and text representations commonly used for emotion recognition, ranging from hand-crafted descriptors to task-driven and self-supervised feature extractors. We then survey prior work on emotion-aware pre-training for speech and text models, highlighting limitations in capturing affective cues. Next, we review approaches to emotion recognition in conversations (ERC), with particular emphasis on methods that integrate acoustic and semantic information. While existing studies focus on controlled or acted datasets, we discuss representative works that move toward emotion recognition using multimodal cues. 
In the context of emotion synthesis, we first review related research on expressive speech generation through text-to-speech (TTS) and voice conversion (VC), which serve as important precursors to the problem addressed in this thesis. We then focus on prior work in emotion style transfer (EST), the task of transferring emotional expressiveness while preserving linguistic content and speaker identity. 
Finally, we describe the datasets used across different experimental settings considered in this thesis.

\section{Speech and Text Features for Emotion modeling}\label{sec:setting_rep}
 Here, we discuss the speech and textual representations used for emotion recognition.
\subsection{Features for Speech Emotion Recognition}
\label{sec:setting_speechfeats}

Over the years, a wide variety of features have been explored for emotion recognition from speech, ranging from hand-crafted descriptors to large-scale pre-trained models. These features can be broadly categorized as follows:

\begin{itemize}
    \item \textbf{Hand-crafted features}: 
    Various acoustic characteristics of the speech signal are known to correlate with the expression of emotion~\cite{scherer2003vocal}. Building on this observation, time-domain features such as speech rate, frequency-domain features such as the fundamental frequency ($F_0$), and amplitude-related features such as intensity have been extensively used for Speech Emotion Recognition (SER)~\cite{lieberman1962some, petrushin1999emotion}. Over time, several feature sets were standardized by the community, most notably through the Interspeech Paralinguistic Challenges~\cite{schuller09_interspeech, schuller2013interspeech}. However, these feature sets were often high-dimensional, motivating the development of more compact representations such as the eGeMAPS feature set, which consists of 88 carefully selected parameters~\cite{eyben2015geneva}. These hand-crafted feature sets are made available through the openSMILE toolkit~\cite{eyben2010opensmile} and have been widely used for SER across diverse settings~\cite{syed2018computational, triantafyllopoulos2019towards, haider2021emotion, leem2021separation}.
    
    \item \textbf{Learnable feature extractors}: 
    As different speech tasks often require task-specific representations, several works have explored learning feature extractors directly from raw speech waveforms. Early efforts replaced fixed mel-filterbank representations with learnable components~\cite{sainath2013learning}. This idea was further extended by directly learning convolutional filters from raw audio, as proposed in multiple subsequent studies~\cite{sainath2015learning, hoshen2015speech, ravanelli2018speaker, zeghidour2021leaf}. Such learnable representations have been shown to outperform hand-crafted features across a range of speech processing tasks, including SER.
    
    \item \textbf{Self-supervised models}: 
    While hand-crafted and learnable features improved performance under controlled conditions, they were often insufficient for SER in diverse and real-world settings. This motivated the development of general-purpose speech representations through self-supervised learning. Representative approaches include Contrastive Predictive Coding (CPC)~\cite{oord2018representation}, PASE~\cite{pascual2019learning, ravanelli2020multi}, wav2vec~\cite{schneider2019wav2vec, baevski2020wav2vec}, HuBERT~\cite{hsu2021hubert}, and WavLM~\cite{chen2022wavlm}. These models learn rich representations that capture both acoustic and linguistic properties of speech. Pre-trained and fine-tuned variants of such models have consistently demonstrated strong performance improvements for SER and have been widely adopted in recent works~\cite{pepino2021emotion, wang2021fine, morais2022speech, diatlova2024adapting}.
\end{itemize}

\subsection{Features for Text Emotion Recognition}
\label{sec:setting_textfeats}

Similar to Speech Emotion Recognition (SER), the task of emotion recognition from text has attracted significant attention from the research community. The features used for this task can be broadly categorized as follows:

\begin{itemize}
    \item \textbf{Lexicon-based features}: 
    A wide range of affective lexicons has been developed for English, including WordNet~\cite{fellbaum1998wordnet}, WordNet-Affect~\cite{strapparava2004wordnet}, and SenticNet~\cite{biagioni2016senticnet}, among others. These resources associate words or concepts with predefined emotional categories. Based on such associations, text is commonly represented using term frequency-inverse document frequency (TF-IDF) vectors, which are then used for emotion classification. Lexicon-based approaches have demonstrated satisfactory performance in several prior studies~\cite{alm2005emotions, chaffar2011using, plaza2018sinai}.
    
    \item \textbf{Self-supervised models}: 
    One of the earliest approaches for learning distributed text representations was proposed by Mikolov et al.~\cite{mikolov2013efficient} in the form of the word2vec model. Along with similar embedding methods such as GloVe~\cite{pennington2014glove}, these models learn fixed-dimensional word representations from large text corpora and have been widely used for text emotion recognition~\cite{polignano2019comparison, gupta2021decoding, shaday2024application, shukla2024word}. While these embeddings outperform lexicon-based approaches, they suffer from an important limitation: each word is represented by a single static vector, independent of its contextual usage.
    
    To address this limitation, contextualized word representations were introduced through models such as ELMo~\cite{peters-etal-2018-deep}, GPT~\cite{radford2018improving}, and BERT~\cite{devlin2019bert}, all of which are based on the transformer architecture~\cite{vaswani2017attention}. Trained on large-scale text corpora, these models provide context-dependent representations that generalize well across a wide range of downstream tasks, including emotion recognition from text. Several studies have explored both feature extraction and fine-tuning strategies using these models for improved performance. For a comprehensive discussion of transformer-based approaches to text emotion recognition, we refer the reader to recent survey works~\cite{al2024challenges, acheampong2021transformer, kusal2023systematic}.
\end{itemize}

\section{Prior Work on Emotion-Aware Pre-training}\label{sec:setting_priorpre}
This section reviews prior work on pre-training speech and text models with objectives tailored for improved emotion understanding. Unlike generic self-supervised learning paradigms, these approaches incorporate emotion-related inductive biases either through supervision, distillation, or task-specific loss functions.

\subsection{Pre-training Speech Models for Emotion}\label{sec:setting_priorprespeech}
\begin{itemize}
    \item \textbf{Chen et al.~\cite{chen2024vesper}} proposed an emotion-aware speech representation learning framework based on knowledge distillation from a pre-trained WavLM-large~\cite{chen2022wavlm} model. The distillation loss was selectively applied to frames exhibiting high or low pitch and energy, under the assumption that such regions are more informative of emotional content. This targeted distillation strategy resulted in a compact student model that achieved strong performance across three SER benchmarks.
    
    \item \textbf{Ma et al.~\cite{ma2024emotion2vec}} similarly adopted a knowledge distillation paradigm, with data2vec~\cite{baevski2022data2vec} as the teacher model. Their approach incorporates both an utterance-level loss, which aligns pooled student and teacher representations, and a frame-level loss that minimizes the mean-squared error (MSE) between student and teacher representations over a masked subset of frames. The resulting model demonstrates competitive performance across multiple SER benchmarks, including multilingual settings.
\end{itemize}
Despite their effectiveness, these approaches primarily rely on implicitly transferring affective cues from large teacher models, without explicitly modeling the complementary roles of acoustic and semantic factors in emotion understanding.
\subsection{Pre-training Text Models for Emotion}\label{sec:setting_priorpretext}
Pre-training of text models is typically carried out using a masked language modeling (MLM) objective~\cite{devlin2019bert}, where the model is trained to predict masked-out words in a sentence. While such objectives yield general-purpose textual representations, relatively few works have explored pre-training strategies explicitly aimed at improving the emotion understanding capabilities of text models. This subsection reviews prior efforts in this direction.

\begin{itemize}
    \item \textbf{Tian et al.~\cite{tian2020skep}} proposed a sentiment-aware pre-training framework that assigns higher masking probabilities to a small set of seed words known to carry sentiment. In addition to the MLM loss, auxiliary objectives such as polarity prediction and aspect-sentiment pair prediction were incorporated during pre-training. Although the work focused primarily on sentiment analysis rather than fine-grained emotion recognition, the proposed strategy was shown to improve the performance of pre-trained models such as RoBERTa~\cite{liu2019roberta}.
    
    \item \textbf{Sosea et al.~\cite{sosea2021emlm}} introduced a similar approach in which words from emotion and sentiment lexicons are masked with higher probability during pre-training. This strategy, referred to as emotional MLM (eMLM), was demonstrated to yield improvements on downstream emotion classification tasks.
\end{itemize}
In contrast to these lexicon-driven masking strategies, this thesis explores emotion-aware text pre-training using supervision derived from spoken language via large language models, enabling scalable emotion annotation beyond curated text corpora.
\section{Prior Work on Multimodal Emotion Recognition}\label{sec:setting_prioremorec}
We now discuss the prior work related to multimodal emotion recognition from spoken language.
\subsection{Emotion Recognition in Conversations}
\label{sec:setting_priorerc}

In this subsection, we review prior work on Emotion Recognition in Conversations (ERC), which extends utterance-level emotion recognition by incorporating conversational context and speaker interactions.

\begin{itemize}
    \item \textbf{Poria et al.~\cite{poria2017context}} introduced one of the earliest solutions to the ERC problem by proposing a bidirectional LSTM~\cite{hochreiter1997long} framework for contextual modeling and multimodal fusion. Although the unimodal feature extractors were not jointly trained with the contextual model, the study demonstrated that decoupling contextual modeling from multimodal fusion can be more effective than training a single unified network.
    
    
    \item \textbf{Yun et al.~\cite{yun2024telme}} introduced a knowledge distillation framework to align non-text modalities with text representations. The proposed approach leverages speech and visual modalities and explicitly incorporates speaker information to address the varying number of speakers and interaction patterns present in conversational data.
    
    \item \textbf{Li et al.~\cite{li2024cfn}} modelled the dynamics of emotional transitions across consecutive utterances by jointly predicting emotion categories and emotion shifts. Explicitly modeling emotion evolution over conversational turns resulted in improved performance across multiple ERC benchmark datasets.
\end{itemize}
While these works have significantly advanced contextual and multimodal modeling for ERC, most existing approaches tightly couple representation learning, contextual modeling, and multimodal fusion within end-to-end architectures. As a result, the individual contributions of acoustic and semantic cues to emotion recognition are often difficult to interpret or control. In contrast, this thesis adopts a modular and hierarchical perspective on ERC~\cite{dutta2022multimodal}, explicitly separating utterance-level representation learning from conversational context modeling and multimodal fusion. By doing so, we enable a more systematic investigation of how acoustic and semantic information contribute to emotion understanding in conversations, and how these cues can be effectively combined through flexible fusion mechanisms such as mixture-of-experts models.
\subsection{Speech Emotion Recognition: A Multimodal Perspective}\label{sec:setting_priorser}
Speech emotion recognition has been extensively studied using a wide range of acoustic features and modeling techniques. While early work predominantly focused on acoustic cues alone, recent advances have highlighted the importance of incorporating semantic information derived from spoken language. In this subsection, we review prior work on speech emotion recognition with a focus on approaches that integrate acoustic and semantic cues, as this perspective is most relevant to the contributions presented in this thesis. Most of these approaches rely on the availability of manual transcriptions that are commonly provided with SER datasets. In scenarios where such transcriptions are unavailable, automatic speech recognition (ASR) systems can be employed to convert speech signals into text, enabling the application of these multimodal techniques.
\begin{itemize}
    
    \item \textbf{Lee et al.~\cite{lee2005toward}} incorporated semantic information by introducing the notion of \textit{emotional salience} of words within a corpus. Mutual information between individual words and emotion categories was used to quantify the relevance of words to specific emotions. Similar to earlier work, the task was limited to classifying emotions into positive and negative categories.
    
    \item \textbf{Jin et al.~\cite{jin2015speech}} extended bag-of-words representations by introducing emotion-specific word co-occurrence statistics, resulting in an \textit{emotion vector} representation. These semantic features, when combined with acoustic information, led to improved performance on speech emotion recognition benchmarks.
    
    \item \textbf{Sun et al.~\cite{sun2024fine}} proposed a recent transformer-based multimodal framework that processes text and speech using BERT~\cite{devlin2019bert} and wav2vec~2.0~\cite{baevski2020wav2vec}, respectively. To address differences in information content across modalities, the authors introduced a disentangled representation learning strategy to separate shared and modality-specific features. The proposed method demonstrated improved SER performance on a benchmark dataset.
\end{itemize}
Despite demonstrating the benefits of integrating acoustic and semantic information, the aforementioned approaches are predominantly evaluated on controlled or acted speech datasets, where recording conditions, speaking style, and emotional expression are carefully curated. As a result, their performance and robustness under naturalistic conditions—characterized by spontaneous speech, background noise, speaker variability, and imbalanced emotion distributions—remain largely unexplored. In contrast, this thesis focuses explicitly on speech emotion recognition in naturalistic settings, and investigates how acoustic and semantic cues can be effectively leveraged when such assumptions no longer hold.

\section{Prior Work on Expressive Speech Synthesis}
The control of emotion and expressiveness in synthesized speech has received significant attention in recent years, driven by applications such as conversational agents, entertainment, and human-robot interaction. This section reviews prior work related to expressive speech synthesis, focusing on text-to-speech, voice conversion, and emotion style transfer.
\subsection{Text-to-Speech}\label{sec:setting_priortts}
A large body of work exists in the domain of expressive text-to-speech (TTS), where the goal is to synthesize speech from text while controlling attributes such as speaker identity, prosody, and emotional expressiveness. In this subsection, we review representative works that derive speaking style from a reference speech signal.

\begin{itemize}
    \item \textbf{Min et al.~\cite{min2021meta}} proposed an early framework for reference-based expressive TTS, where the speaking style is inferred from a reference speech signal. The authors introduced Style-Adaptive Layer Normalization (SALN), in which the text embeddings are modulated using gain and bias parameters predicted from a learned style vector extracted from the reference speech.
    
    \item \textbf{Huang et al.~\cite{huang2022generspeech}} argued that representing expressive style using a single global vector is overly restrictive. To address this, they proposed a multi-level style modeling approach that incorporates phoneme-level, frame-level, and utterance-level style representations. The resulting system was shown to generalize effectively to unseen reference speakers.
    
    \item \textbf{Li et al.~\cite{li2023styletts}} employed latent diffusion models~\cite{rombach2022high} to capture expressive style information from reference speech. The learned latent style representations enabled high-quality expressive speech synthesis, achieving performance comparable to human speech in perceptual evaluations.
\end{itemize}

\subsection{Voice Conversion}\label{sec:setting_priorvc}
While expressive TTS systems rely on text transcripts for synthesis, a growing body of work focuses on modifying speech signals directly in an speech-to-speech setting. The task of voice conversion (VC) aims to transform the speaker identity of a source speech signal to that of a reference speaker while preserving the linguistic content, without requiring textual supervision. An extension of this problem, referred to as expressive voice conversion (EVC), additionally seeks to transfer the emotional speaking style of the reference speech. This subsection reviews representative works in this domain.

\begin{itemize}
    \item \textbf{Rizos et al.~\cite{rizos2020stargan}} proposed one of the earliest approaches to joint speaker and emotion conversion using a StarGAN~\cite{choi2018stargan} framework. The model enables transformation of both speaker identity and emotional style in speech. The authors further demonstrated the utility of the converted samples as a data augmentation strategy for speech emotion recognition.
    
    \item \textbf{Maimon et al.~\cite{maimon2023speaking}} employed a pre-trained self-supervised speech model to decompose speech into content and style representations. A learnable embedding was used to capture speaker-specific style information, which was then leveraged for emotional voice conversion. The study additionally highlighted speaking rate as a critical factor for effective emotional style transfer in VC.
    
    \item \textbf{Yao et al.~\cite{yao2024promptvc}} incorporated a reference-based style encoder, similar to that proposed by Min et al.~\cite{min2021meta}, to model speaking style in a voice conversion framework. Style-Adaptive Layer Normalization (SALN) was used to condition the conversion process, resulting in improved naturalness and style similarity compared to prior approaches.
\end{itemize}

\subsection{Style Transfer}\label{sec:setting_priorest}
Emotion style transfer (EST) refers to the task of transferring the emotional speaking style from a reference speech signal to a source speech signal while preserving the linguistic content and speaker identity of the source. Unlike expressive voice conversion (EVC), EST requires an explicit disentanglement of emotional style from speaker identity, making it a more challenging problem. While a few attempts have addressed this task in a parallel setting\footnote{When speech from the same speaker with identical content but different emotional expressions is available during training.}, this thesis presents one of the first solutions to emotion style transfer in a non-parallel, speech-only setting~\cite{dutta2024zero}. In this subsection, we review subsequent works that have explored related problem formulations.

\begin{itemize}
    \item \textbf{Zhang et al.~\cite{zhangvevo}} proposed VEVO, a unified framework that modifies content representations using reference speech while conditioning acoustic representations on reference speaker characteristics. The resulting model supports multiple expressive speech tasks, including emotion style transfer. Although VEVO is trained using transcribed speech, it does not require textual input during inference.
    
    \item \textbf{Yao et al.~\cite{yao2025stablevc}} utilized the style encoder from NaturalSpeech~3~\cite{ju2024naturalspeech} to extract style representations from reference speech. Speaker embeddings were derived using multiple utterances from the same speaker and disentangled from the style representation, enabling transfer of emotional speaking style across speakers. However, the proposed method was not evaluated under conditions where either the source speaker or linguistic content is unseen during training.
    
    \item \textbf{Yoon et al.~\cite{yoon2025maestro}} proposed a framework that enables fine-grained control over content, pitch, energy, rhythm, and speaker identity in converted speech. Emotion style transfer emerges as a natural application of this model, which employs a prosody extractor to transfer emotional characteristics from reference to source speech. While a content classifier is introduced to enhance disentanglement, the evaluation remains limited to scenarios where the source speech content is observed during training, a limitation shared with~\cite{yao2025stablevc}.
\end{itemize}
Taken together, prior work on expressive TTS, voice conversion, and emotion style transfer highlights steady progress toward controllable speech synthesis. However, most existing approaches either rely on textual supervision, assume parallel emotional data, or entangle emotional style with speaker identity or content. In contrast, the focus of this thesis is on emotion style transfer in a speech-only, non-parallel setting, where neither aligned emotional utterances nor text transcriptions are available, and where explicit disentanglement of emotion, content, and speaker factors is essential.

\section{Datasets}\label{sec:setting_datasets}
The datasets used in this thesis span multiple tasks, including emotion-aware pre-training, emotion recognition, and emotion style transfer. Each dataset is described below along with its primary role in the thesis, while secondary usages are noted where applicable. A summary of the different datasets is also provided in Table~\ref{tab:dataset_summary}.

\begin{itemize}
    \item \textbf{MSP-PODCAST}~\cite{busso2025msp} is a large-scale emotional speech corpus consisting of approximately $409$ hours of podcast recordings. The data are annotated into $10$ emotion categories—angry, happy, sad, neutral, disgust, fear, contempt, surprise, other, and none—and are additionally labeled with continuous valence, arousal, and dominance scores. Owing to its scale, naturalistic recording conditions, and rich emotional annotations, this dataset is used both for emotion-aware pre-training (sections~\ref{sec:care_pre},~\ref{sec:merits_pre}) and for evaluating speech emotion recognition under naturalistic conditions (Sec.~\ref{sec:abhinaya_data}).

    \item \textbf{IEMOCAP}~\cite{busso2008iemocap} is a multimodal dataset consisting of $151$ recorded conversations organized into five sessions, where each session involves a dyadic interaction between one male and one female speaker. The recordings are segmented into utterances, resulting in a total of $10{,}039$ utterances annotated by human evaluators into one of $10$ emotion categories: angry, happy, sad, neutral, frustrated, excited, fearful, surprised, disgusted, and ``other''. Following standard evaluation protocols, we consider two classification settings. In the first, a four-class setup is used with angry, happy, sad, and neutral categories, where the happy and excited classes are merged, resulting in $5{,}531$ utterances. In the second setting, a six-class classification task is adopted as in prior work~\cite{lian2022smin}, using the first six emotion categories and yielding $7{,}433$ utterances. In both settings, Session~5 is held out for testing. In this thesis, IEMOCAP is used as a benchmark dataset for evaluating emotion-aware pre-trained representations (sections~\ref{sec:care_down},~\ref{sec:merits_down}) and for experiments on Emotion Recognition in Conversations (ERC) (sections~\ref{sec:hcam_down},~\ref{sec:misterd_down}).

    \item \textbf{MELD}~\cite{poria2019meld} is a multi-party conversational emotion recognition dataset derived from the TV show \textit{Friends}. It consists of $1{,}433$ conversations comprising a total of $13{,}708$ utterances. Each utterance is annotated into one of seven emotion categories: angry, joy, sadness, fear, disgust, surprise, and neutral. The dataset is partitioned into $1{,}153$ conversations ($11{,}098$ utterances) for training and validation, and $280$ conversations ($2{,}610$ utterances) for testing. In this thesis, MELD is used as a benchmark dataset for evaluating emotion-aware pre-trained representations (sections~\ref{sec:care_down},~\ref{sec:merits_down}) and for experiments on Emotion Recognition in Conversations (ERC), complementing the dyadic conversational setting provided by IEMOCAP with a more complex multi-party interaction scenario (sections~\ref{sec:hcam_down},~\ref{sec:misterd_down}).

    \item \textbf{CMU-MOSI}~\cite{zadeh2016mosi} is a sentiment analysis dataset consisting of $93$ spoken monologues segmented into $2{,}199$ utterances, each annotated with a continuous sentiment score in the range $[-3, 3]$. Following prior work, the task is formulated as a binary classification problem, where utterances with sentiment scores in the range $[-3, 0)$ are labeled as negative and those in the range $[0, 3]$ as positive. We adopt the same dataset partitioning protocol as Lian et al.~\cite{lian2022smin}, using $49$ monologues for training, $13$ for validation, and the remaining $31$ for testing, resulting in $1{,}188$ training utterances, $325$ validation utterances, and $686$ test utterances. In this thesis, CMU-MOSI is used to evaluate the generalization of emotion-aware pre-trained representations (sections~\ref{sec:care_down},~\ref{sec:merits_down}) and multimodal emotion recognition models in a monologue-based setting, complementing conversational benchmarks such as IEMOCAP and MELD (sections~\ref{sec:hcam_down},~\ref{sec:misterd_down}).

    \item \textbf{DAIC-WOZ}~\cite{gratch2014distress} is a benchmark dataset for automatic depression detection, consisting of $189$ clinical interviews conducted between a participant and an interviewer. Of these, $107$ interviews are designated for training and $35$ for development. The dataset exhibits significant class imbalance, with only $30$ training interviews labeled as ``depressed''. Following prior work~\cite{wu2023self}, utterances are randomly sampled from each interview across five splits, and performance is reported on the development set. In this thesis, DAIC-WOZ is used to evaluate the effectiveness of emotion-aware pre-trained speech representations on a clinically relevant downstream task, assessing their generalization beyond standard emotion recognition benchmarks (Sec.~\ref{sec:care_down}).

    \item \textbf{RAVDESS-Song}~\cite{livingstone2018ryerson} is an emotional singing dataset comprising $1{,}012$ recordings performed by $23$ different singers. Each recording is annotated with one of six emotion categories: neutral, calm, happy, sad, angry, and fear. We conduct a speaker-independent evaluation by creating $5$ different splits. For each split, recordings from $16$ singers are used for training, recordings from $3$ singers are used for validation, and recordings from the remaining $4$ singers are reserved for evaluation. In this thesis, RAVDESS-Song is used to evaluate the generalization of emotion-aware pre-trained speech representations in a singing voice setting, providing a complementary affective evaluation scenario beyond spoken language (Sec.~\ref{sec:care_down}).

    \item \textbf{CaFE}~\cite{gournay2018canadian} is a Canadian French emotional speech dataset comprising $936$ utterances spoken by $12$ speakers. Each utterance is annotated with one of seven emotion categories: neutral, angry, disgust, sad, surprise, fear, and happy. Similar to RAVDESS-Song, we conduct a speaker-independent evaluation using $5$ different splits. For each split, utterances from $8$ speakers are used for training, while the remaining $4$ speakers are evenly divided between validation and testing. The speaker assignments are randomized across splits. In this thesis, CaFE is used to evaluate the generalization of emotion-aware pre-trained speech representations across languages, providing a non-English benchmark complementary to English emotion recognition datasets (Sec.~\ref{sec:care_down}).

    \item \textbf{Emo-DB}~\cite{burkhardt2005database} is a German emotional speech dataset consisting of $535$ utterances spoken by $10$ speakers. Each utterance is annotated with one of seven emotion categories: neutral, angry, disgust, sad, boredom, fear, and joy. Similar to CaFE and RAVDESS-Song, we conduct a speaker-independent evaluation using $5$ different splits. For each split, utterances from $6$ speakers are used for training, while the remaining $4$ speakers are evenly divided between validation and testing. Speaker assignments are randomized across splits. In this thesis, Emo-DB is used to evaluate the cross-lingual generalization of emotion-aware pre-trained speech representations, complementing English and French emotion recognition benchmarks (Sec.~\ref{sec:care_down}).

   \item \textbf{MSP-IMPROV}~\cite{busso2016msp} is an audio-visual emotional dataset comprising $8{,}438$ utterances spoken by $12$ actors, where each utterance is annotated with continuous valence, arousal, and dominance (VAD) scores in the range $[1,5]$. To ensure speaker-independent evaluation, we adopt a leave-two-speakers-out protocol by splitting the dataset into $12$ folds. In each fold, utterances from $10$ speakers are used for training, while utterances from the remaining two speakers are used for validation and testing. Final performance is reported as the average across all folds. In this thesis, MSP-IMPROV is used to evaluate the effectiveness of emotion-aware pre-trained speech representations for continuous affect prediction, specifically valence, arousal, and dominance estimation (Sec.~\ref{sec:care_improv}).

    \item \textbf{ESD}~\cite{zhou2021seen} is an emotional speech dataset comprising utterances spoken by $10$ native English and $10$ native Chinese speakers, covering five emotion categories: neutral, happy, angry, sad, and surprise. In this thesis, we use only the English subset of the dataset and follow the predefined train-validation-test splits, using $300$ utterances per speaker per emotion for training. This results in a total of $15{,}000$ training utterances, with $2{,}500$ unseen utterances ($50$ per speaker per emotion) used for validation. In this thesis, ESD is used for training the emotion style transfer system; however, the parallel nature of the dataset (i.e., identical content spoken in different emotions) is not exploited during training (Sec.~\ref{est_dataset}). Parallel utterances are used only during evaluation to enable controlled assessment of emotion transfer quality while keeping speaker identity and linguistic content fixed (Sec.~\ref{sec:est_evaluation}).

    \item \textbf{CREMA-D}~\cite{cao2014crema} is an acted emotional speech dataset consisting of $7{,}442$ audio-visual recordings of actors uttering $12$ fixed sentences across six emotion categories: angry, sad, happy, fear, disgust, and neutral. In this thesis, CREMA-D is used exclusively for the evaluation of emotion style transfer systems (Sec.~\ref{sec:est_evaluation}). 

    \item \textbf{TIMIT}~\cite{garofolo1993darpa} is a phonetically balanced speech corpus containing approximately $5$ hours of speech from $630$ speakers, where each speaker utters a fixed set of $10$ sentences. The dataset was originally designed for automatic speech recognition and does not contain expressive or emotional speech. In this thesis, TIMIT is used for the evaluation of emotion style transfer systems as a source of neutral, non-expressive speech (Sec.~\ref{sec:est_evaluation}).

\end{itemize}

\begin{table*}[t]
\centering
\caption{Summary of datasets used in this thesis and their roles across different chapters.}
\label{tab:dataset_summary}
\resizebox{\textwidth}{!}{
\begin{tabular}{l|l|c|l}
\toprule
\textbf{Dataset} & \textbf{Task} & \textbf{\# Samples} & \textbf{Section} \\
\midrule

\multirow{2}{*}{MSP-PODCAST~\cite{busso2025msp}}
& Emotion-aware speech pre-training
& \multirow{2}{*}{$\sim$409 hours}
& ~\ref{sec:care_pre},~\ref{sec:merits_pre} \\ \cline{2-2} \cline{4-4}
& Naturalistic speech emotion recognition
& 
& ~\ref{sec:abhinaya_data}\\
\midrule

\multirow{2}{*}{IEMOCAP~\cite{busso2008iemocap}}
& Evaluation of pre-trained representations
& \multirow{2}{*}{10{,}039 utt.}
& ~\ref{sec:care_down},~\ref{sec:merits_down} \\ \cline{2-2} \cline{4-4}
& Emotion Recognition in Conversations (ERC)
& 
& ~\ref{sec:hcam_down},~\ref{sec:misterd_down} \\
\midrule

\multirow{2}{*}{MELD~\cite{poria2019meld}}
& Evaluation of pre-trained representations
& \multirow{2}{*}{13{,}708 utt.}
& ~\ref{sec:care_down},~\ref{sec:merits_down} \\ \cline{2-2} \cline{4-4}
& Emotion Recognition in Conversations (ERC)
& 
& ~\ref{sec:hcam_down},~\ref{sec:misterd_down} \\
\midrule

\multirow{2}{*}{CMU-MOSI~\cite{zadeh2016mosi}}
& Evaluation of pre-trained representations
& \multirow{2}{*}{2{,}199 utt.}
& ~\ref{sec:care_down},~\ref{sec:merits_down} \\ \cline{2-2} \cline{4-4}
& Emotion Recognition in Conversations (ERC)
& 
& ~\ref{sec:hcam_down},~\ref{sec:misterd_down} \\
\midrule

DAIC-WOZ~\cite{gratch2014distress}
& Depression detection (representation evaluation)
& 189 interviews
& ~\ref{sec:care_down}\\
\midrule

RAVDESS-Song~\cite{livingstone2018ryerson}
& Evaluation of pre-trained speech representations
& 1{,}012 recordings
& ~\ref{sec:care_down} \\
\midrule

CaFE~\cite{gournay2018canadian}
& Evaluation of pre-trained speech representations
& 936 utt.
& ~\ref{sec:care_down} \\
\midrule

Emo-DB~\cite{burkhardt2005database}
& Evaluation of pre-trained speech representations
& 535 utt.
& ~\ref{sec:care_down} \\
\midrule

MSP-IMPROV~\cite{busso2016msp}
& Valence--Arousal--Dominance prediction
& 8{,}438 utt.
& ~\ref{sec:care_improv} \\
\midrule

\multirow{2}{*}{ESD~\cite{zhou2021seen}}
& Emotion style transfer (training; non-parallel use)
& \multirow{2}{*}{15{,}000 utt.}
& ~\ref{est_dataset} \\ \cline{2-2} \cline{4-4}
& Evaluation of emotion style transfer
& 
& ~\ref{sec:est_evaluation} \\
\midrule

CREMA-D~\cite{cao2014crema}
& Evaluation of emotion style transfer
& 7{,}442 recordings
& ~\ref{sec:est_evaluation} \\
\midrule

TIMIT~\cite{garofolo1993darpa}
& Evaluation of emotion style transfer
& 5 hours
& ~\ref{sec:est_evaluation} \\
\bottomrule
\end{tabular}}
\end{table*}

\section{Chapter Summary}

This chapter presented a comprehensive background of prior work relevant to the three core research directions addressed in this thesis: emotion-aware representation learning, multimodal emotion recognition from spoken language, and expressive speech synthesis through emotion style transfer.

We first reviewed the evolution of speech and text features for emotion recognition, ranging from hand-crafted acoustic descriptors and lexicon-based textual features to large-scale self-supervised and transformer-based models. While these representations have led to steady improvements in performance, they largely rely on generic pre-training objectives that are not explicitly aligned with affective cues in spoken language.

Next, we surveyed existing approaches to emotion-aware pre-training for speech and text models. Prior work primarily incorporates emotional inductive biases through knowledge distillation or modified masking strategies. However, these methods often focus on either acoustic or semantic aspects in isolation, motivating the need for unified pre-training strategies that jointly model both factors for emotion understanding.

We then reviewed research on multimodal emotion recognition, with particular emphasis on Emotion Recognition in Conversations (ERC). Existing methods demonstrate the importance of contextual modeling and multimodal fusion but often struggle to balance modality dominance, scalability, and robustness. Furthermore, most multimodal emotion recognition benchmarks are based on acted or controlled speech, limiting their applicability to real-world, naturalistic scenarios.

Finally, we discussed prior work on expressive speech synthesis, covering text-to-speech, voice conversion, and emotion style transfer. While significant progress has been made in reference-based expressive synthesis, explicit disentanglement of emotional style from speaker identity and linguistic content remains challenging—especially in non-parallel, speech-only settings.

Together, this background highlights several open challenges:  
(i) the lack of emotion-aware pre-training objectives that jointly leverage acoustic and semantic cues,  
(ii) the need for robust multimodal emotion recognition frameworks capable of handling conversational structure and naturalistic speech, and  
(iii) the difficulty of achieving controllable emotion transfer in speech without parallel supervision.  

These gaps directly motivate the methodologies and contributions presented in the subsequent chapters of this thesis.

\chapter{Emotion-Aware Pre-training for Spoken Language}
\label{chap:pretraining}

Self-supervised pre-training has become a cornerstone of modern speech and language processing, enabling models to learn general-purpose representations from large amounts of unlabeled data. In speech, models such as wav2vec2.0~\cite{baevski2020wav2vec}, HuBERT~\cite{hsu2021hubert}, and WavLM~\cite{chen2022wavlm} have demonstrated remarkable transferability across a wide range of downstream tasks, including automatic speech recognition, speaker recognition, and speech emotion recognition. Despite these advances, the objectives used during pre-training are primarily designed to capture linguistic structure or short-term acoustic regularities, rather than affective characteristics of speech.

Emotion perception in spoken language, however, relies on a subtle interplay between acoustic realization and semantic context. Prosodic variations such as pitch, energy, and speaking rate interact with linguistic content to convey emotional intent, and these cues often manifest differently across speakers, recording conditions, and datasets. As a result, models pre-trained with objectives optimized for semantic reconstruction may fail to adequately encode emotion-relevant variations, leading to limited generalization when transferred to emotion recognition tasks.


In this chapter, we investigate emotion-aware pre-training strategies for spoken language that aim to bridge this gap. We study how acoustic and semantic factors can be jointly leveraged during pre-training to produce representations that are better aligned with downstream emotion recognition tasks. Specifically, we explore pre-training objectives that emphasize emotion-relevant regions of the speech signal, and analyze their impact on generalization across multiple speech emotion recognition benchmarks.

Beyond speech representations, this chapter also examines emotion-aware pre-training for text models derived from spoken language. Rather than relying solely on large text corpora with implicit emotional content, we consider a speech-driven supervision paradigm in which spoken utterances are transcribed and annotated with emotion labels inferred using a large language model. These emotion-labeled text corpora are then used to pre-train text-based representations, enabling emotion-aware semantic modeling even in scenarios where labeled text data is scarce.


\section{Pre-training for Speech Emotion Recognition}\label{sec:pretraining_ser}
We introduce a self-supervised model for speech emotion recognition (SER) called \textbf{C}ontent and \textbf{A}coustic \textbf{R}epresentations of \textbf{E}motions (CARE)~\cite{dutta2025leveraging}. To the best of our knowledge, our approach  is the first effort to pre-train a self-supervised  model that integrates both semantic and acoustic components of speech. CARE leverages a dual encoding framework for processing speech signals: a semantic encoder, which aligns speech representations with sentence-level transcripts, and a non-semantic encoder, which aligns speech representations with low-level acoustic features from the PASE+ model~\cite{ravanelli2020multi}. The outputs of both encoders are combined, and a lightweight classification head is then trained to perform emotion recognition.
\subsection{Proposed Approach}\label{sec:pretraining_method_care}
\subsubsection{Background}
\textbf{RoBERTa}: One of the significant contributions in creating a text representation model was proposed by Devlin et al.~\cite{devlin2019bert}.
Liu et. al \cite{liu2019roberta} trained this architecture on a larger corpus of textual data without the next sentence prediction task. This pre-trained model, known as robust optimized BERT approach (RoBERTa), was shown to outperform BERT in a number of downstream tasks.

\begin{figure}
    \centering
    \includegraphics[width=0.9\textwidth,trim={1.5cm 5cm 0cm 2.5cm},clip]{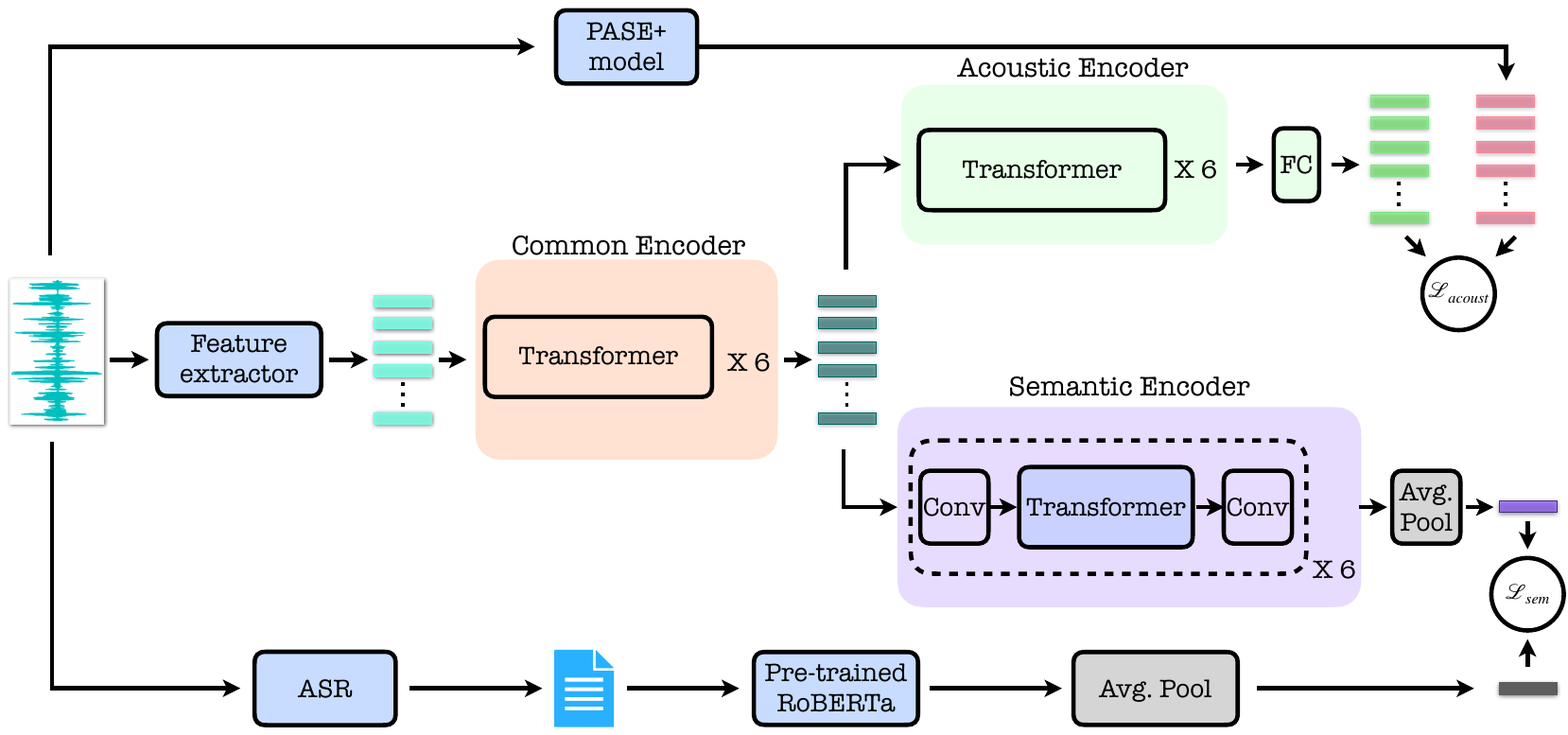}
    \caption{Block diagram of the proposed CARE model.   The acoustic encoder of the model is trained with  PASE+ features as targets. Blocks in blue indicate either frozen components or those with no learnable parameters. For the semantic encoder the transformer layers are frozen while the convolutional adapters are trained. As the dimension of the output from the acoustic encoder is $768$, a FC layer is attached to match the PASE+ feature dimension of $256$. This FC layer and the average pool block after the semantic encoder are not used during inference. }
    \label{fig:entire model}
    \vspace{-0.1in}
\end{figure}

\subsubsection{CARE Model}\label{sec:caremodel}
 We propose a dual encoding scheme (semantic and acoustic encoders) to  process the speech signal through distinct supervisory signals suited to their respective objectives. The chosen supervision for each encoder is detailed as follows:\\
\textbf{Semantic supervision:} We do not assume the availability of   ground-truth text transcripts for the pre-training data.  In such a scenario, pre-trained automatic speech recognition (ASR) systems (Whisper-large-v3~\cite{radford2023robust}) offer an alternative for generating these transcripts. Typically, ASR systems have been shown  to exhibit higher word error rates (WER) on emotional speech compared to neutral speech datasets~\cite{li2023asr}. Podcast recordings, on the other hand,  provide sufficiently long context and offer a broad content variety suitable for pre-training the semantic encoder. Specifically, we observe a WER of $12.53\%$, which may be reasonable for SER tasks. Since the semantic encoder’s purpose is to align the speech signal with its content to facilitate emotion recognition, an ASR-style alignment loss could be applied. However, a sentence-level representation for text is more appropriate for the task of emotion recognition as established by Fan et al.~\cite{fan2022sentiment}. Therefore, we extract contextual word-level embeddings from the transcripts using a pre-trained RoBERTa model~\cite{liu2019roberta} and mean-pool these embeddings to obtain a single feature vector representing the entire transcript. These utterance-level embeddings serve as the supervisory signal, or ``teacher'', for the semantic encoder in our CARE model. We denote these utterance-level embeddings by $\boldsymbol{y}_{text}$.\\
\noindent \textbf{Acoustic Supervision:} In prior works, mean-pooled   representations have shown to encode characteristics like speaker identity, accent, and language~\cite{krishna2024towards}. However,  we speculate that emotion in speech is often contained in fine-grained acoustic attributes  such as pitch, rhythm, and their modulations~\cite{petrushin1999emotion}. Thus, a frame-level target is chosen for the acoustic encoder.

A direct approach for the frame level acoustic targets would involve masking parts of the speech signal and reconstructing them.
However, prior works show that random masking is less effective for emotion recognition than selectively masking high-energy or high-pitch regions, as demonstrated by Chen et al.~\cite{chen2024vesper}. 
Based on these observations, we choose to predict PASE+ features, which encompass filter-bank energies, pitch, and other low-level descriptors essential for capturing emotion. Specifically, we use frame-level PASE+ features with $256$ dimensions as targets for the acoustic encoder in our CARE model. These features are down-sampled by a factor of $2$, producing target descriptors at a frequency of $50$ Hz. We denote the acoustic targets from the PASE+ model by $\boldsymbol{y}_{pase}$.\\

\noindent \textbf{Model Architecture:} The speech signal is first processed through a series of convolutional layers designed to produce frame representations every $20$ ms. These are followed by a stack of six transformer layers, forming the common encoder that serves both the acoustic and semantic encoder pathways in the proposed model.

The semantic encoder is designed to align the speech representations with its corresponding generated transcript. This encoder consists of six transformer layers which are initialized with the weights from a pre-trained text representation model. Being trained with textual data, the transformer layers in the semantic encoder do not generalize to speech representations. To address this, we propose a novel adaptation strategy by introducing two 1D-convolutional blocks—one placed before and one after each transformer layer. 

The first block adjusts the speech representations from the common encoder to align them with the internal representations expected by the text-based model. The second block refines these representations post-transformer processing. Additionally, following established practice for processing speech in text models~\cite{yu2024connecting, fathullah2024prompting, tang2023salmonn}, the time resolution of the speech sequence is reduced before processing by the transformer layers in the semantic encoder. Specifically, the convolutional block preceding each transformer layer down-samples the sequence length by a factor of three, while the block following it up-samples it by the same factor. Each convolutional block consists of a single convolutional layer, with a kernel size of $5$, and input and output channels set to $768$ in order to match the dimension of the pre-trained transformer layers. While adaptation of speech SSL models with convolution layers has been explored in prior works~\cite{li2023evaluating, kim2024convolution}, adapting pre-trained text models for speech tasks, using convolutional adapters, is explored for the first time in this work. Importantly, the transformer layers themselves are not updated during training. Finally, the semantic encoder’s output representations are average-pooled to produce an utterance-level representation. \par

The acoustic encoder also consists of six transformer layers, with its output subsequently mapped to $256$ dimensions, using a fully-connected layer, to match the PASE+ feature targets. Figure \ref{fig:entire model} provides a block diagram of the CARE model.\\

\noindent \textbf{Loss:} A semantic loss, $L_{sem}$ and a frame-level acoustic loss, $L_{acoust}$, are employed for training the semantic and acoustic encoders, respectively. 
 Denoting the semantic supervision by $\boldsymbol{y}_{text}$ and the output from the semantic encoder as $\hat{\boldsymbol{y}}_{sem}$, the semantic loss is the mean square error (MSE) loss:
\begin{equation}\label{eq:utt}
    L_{sem.} = \frac{1}{N}\sum_{i=1}^{N}||\boldsymbol{y}_{text}^{i} - \hat{\boldsymbol{y}}_{sem}^{i}||_{2}^{2}
\end{equation}
where $N$ denotes the batch size.\par
For the frame level loss, let $\hat{\boldsymbol{y}}_{acoust} \in \mathbb{R}^{N\times T \times D}$ denote the output of the acoustic encoder, where $N$, $T$ and $D$ denote the batch size, number of frames per utterance and the dimension of the representation,  respectively. The loss is defined as:
\begin{equation}\label{eq:frame}
    L_{acoust.} =  \frac{1}{NT}\sum_{i=1}^{N}\sum_{j=1}^{T}||\boldsymbol{y}_{pase}^{ij} - \hat{\boldsymbol{y}}_{acoust}^{ij}||_{2}^{2}
\end{equation}
where $\boldsymbol{y}_{pase}$ denotes the acoustic target.
\textcolor{black}{The total loss during pre-training is given by}
\begin{equation}\label{eq:total}
    \textcolor{black}{L_{tot.} = L_{sem.} + \lambda L_{acoust.}}
\end{equation}
\textcolor{black}{where $\lambda$ is decided based on the validation performance.}\\

\noindent \textbf{Inference:} For evaluating the model across various downstream tasks, we adopt the paradigm proposed in the SUPERB benchmark~\cite{yang2021superb}. 
The outputs from each transformer layer of the acoustic encoder are concatenated with the outputs from the convolution block following each transformer layer in the semantic encoder. These are then combined with layer-wise outputs from the common encoder and the convolutional feature extractor. This process yields a total of $13$ layer representations—one from the convolutional feature extractor, six from the common encoder, and six from the concatenated semantic and acoustic encoders. A convex combination of these layer representations is then fed into a classification head. It is to be noted that, during inference, the fully-connected layer in the acoustic encoder and the average pooling block in the semantic encoder are not used.

In this setup, the only learnable parameters for the downstream tasks are the weights for the convex combination and those of the lightweight classification head.

\subsection{Experiments and Results}
\subsubsection{Pre-training}\label{sec:care_pre}
The MSP-PODCAST corpus~\cite{busso2025msp} is used for the task of pre-training. A total of $149,307$ samples amounting to $230$ hours of emotional speech data are used. Out of these, $80\%$ of the data is randomly chosen as the training set while the remaining $20\%$ serves as the validation set. The Whisper-large-v3 model is used for generating the transcripts (the WER observed is $12.53\%$), while the pre-trained RoBERTa model is used for encoding the transcripts. The common encoder is initialized with first $6$ layers of the WavLM-base model, while the acoustic encoder is initialized with the last $6$ layers of the same. The convolutional feature extractor is also initialized from the WavLM-base model. The $6$ transformer layers of the semantic encoder are initialized with the weights of the last $6$ layers of a pre-trained RoBERTa base model, while the convolutional adapters are randomly initialized.

\begin{table}[t]
\centering
\caption{Summary of the evaluation datasets. The last column indicates if the training/test data have common speakers.}
\label{tab:datasets}
\resizebox{\linewidth}{!}{
\begin{tabular}{l|c|c|c|c|c|c|c}
\toprule
Dataset & \# Utt. & \# Train & \# Val & \# Test &\# Classes & Evaluation & Spkr. Ind. \\
\midrule
IEMOCAP-4~\cite{busso2008iemocap} 
& $5531$ & $3860$ & $430$ & $1241$ & $4$ & Predefined & \Checkmark \\

IEMOCAP-6~\cite{busso2008iemocap} 
& $7433$ & $5230$ & $580$ & $1623$ & $6$ & Predefined & \Checkmark \\

MELD~\cite{poria2019meld} 
& $13706$ & $9988$ & $1108$ & $2610$ & $7$ & Predefined & \XSolid \\

CMU-MOSI~\cite{zadeh2016mosi} 
& $2199$ & $1188$ & $325$ & $686$ & $2$ & Predefined & \Checkmark \\

DAIC-WOZ \cite{gratch2014distress} & $10197$ &$6003$ & $2097$ & $2097$ & $2$ & Predefined & \Checkmark \\

RAVDESS-Song~\cite{livingstone2018ryerson} 
& $1012$ & -- & -- & -- & $6$ & $5$-fold CV & \Checkmark \\
CaFE \cite{gournay2018canadian}  & $936$ & -- & -- & -- & $7$ & $5$-fold CV & \Checkmark  \\
EmoDB \cite{burkhardt2005database} & $535$ & -- & -- & -- & $7$ & $5$-fold CV & \Checkmark \\
\bottomrule
\end{tabular}}
\end{table}


  


\subsubsection{Downstream Tasks}\label{sec:care_down}
A summary of the different datasets used for evaluation is mentioned in Table~\ref{tab:datasets}. For a detailed description of the different datasets we refer the reader to Sec.~\ref{sec:setting_datasets}.

\subsubsection{Loss and Evaluation Metrics}
The cross-entropy loss is used for training the downstream model weights (the convex combination weights and the lightweight classification head parameters). 

For testing, we use the weighted F1-score as the evaluation metric as many of the datasets are class-imbalanced  (Table~\ref{tab:datasets}). Denoting the F1 score of class $c$ with $N_c$ samples, by $F1_c$, the weighted F1-score is
\begin{equation}
    WF1 = \frac{1}{\sum_{c=1}^{C}N_c}\sum_{c=1}^{C}N_c\times F1_c
\end{equation}
\textcolor{black}{We also report the unweighted average recall (UAR) for all cases, which is the mean of the class-wise recall scores.}

\subsubsection{Implementation Details}
\noindent \textbf{Pre-training:} During pre-training, all the speech utterances from MSP-PODCAST are padded or randomly cropped to a duration of $5$ seconds. The model is trained with a learning rate of $1e$-$5$ and a batch size of $128$ with AdamW~\cite{loshchilov2017decoupled} as the optimizer. The model is trained for a total of $200,000$ steps and the best model parameters based on validation set performance are  chosen for evaluation of downstream datasets. \textcolor{black}{We experiment with different values of $\lambda$ (Eq.~\ref{eq:total}) to balance the two losses during pre-training. Setting $\lambda=0.1$ results in degraded performance, while increasing it to $\lambda=10$ does not yield any significant improvement over $\lambda=1$. Therefore, we fix $\lambda=1$  for all the subsequent  experiments.}\\

\noindent \textbf{Fine-tuning and evaluation:} For the downstream task training,  the speech signals are cropped to a maximum duration of $30$ seconds or padded to a minimum duration of $1$ second. For the depression detection dataset, DAIC-WOZ, each speech segment has a duration of $10$ seconds~\cite{wu2023self}.

Each layer output in the common, semantic, and acoustic encoders has a dimensionality of $T\times768$, where $T$ denotes the number of frames in the speech signal, sampled at $50$Hz. For the CARE model, as outputs from the $6$ semantic and acoustic encoder layers are concatenated, the combined output dimension is $6\times T \times1536$. To align with this dimensionality, the output from the convolutional feature extractor and the common encoder's $6$ layers are duplicated to yield features of dimension $7 \times T \times 1536$. Representations from these $13$ layers are combined through a convex combination approach with learnable weights producing features of dimension $T \times 1536$. Following this, features are mean-pooled along the temporal dimension, producing a single $1536$-dimensional vector per audio file. This is input into a classification head consisting of a two-layer feed-forward neural network that employs ReLU activation~\cite{nair2010rectified}. Only the weights for the convex combination of layer representations and those in the two-layer fully connected classification head are trained on each downstream dataset, consistent with the SUPERB framework~\cite{yang2021superb}.

We use a batch size of $32$ with a learning rate of $1e$-$4$ and train the model for   $50$ epochs. The hidden dimension of the two-layer classification head is set to be $256$. The AdamW optimizer is used here as well. All the models, including the CARE and the baseline systems, utilize the same classification backend. Thus, the  design allows fair comparison of the different representations\footnote{Code available at \url{https://github.com/iiscleap/CARE}.}.
\begin{table}[t!]
\centering
\caption{Comparison with other works for downstream datasets. $^{\#}$  models which include downstream dataset in pre-training. Results in \textbf{bold}, \underline{underlined} indicate the best  and the second-best model, respectively. All numbers are weighted F1-scores computed over $5$ random initializations (mean and standard deviation shown). \textcolor{black}{The Unweighted Average Recall is also shown in brackets.}\label{tab:results-cat-oth}  }
    \resizebox{0.99\textwidth}{!}{%
\begin{tabular}{@{}l|c|c|c|c|c|c|c||c@{}}
\toprule
\multicolumn{1}{l|}{\multirow{2}{*}{Datasets}} &WavLM \cite{chen2022wavlm} &HuBERT \cite{hsu2021hubert} & data2vec \cite{baevski2022data2vec} & emotion2vec \cite{ma2024emotion2vec} & SONAR \cite{duquenne2023sonar} & \multicolumn{2}{c||}{SALMONN \cite{tang2023salmonn}}  & CARE \\ 
& \textbf{Params}:$94$M & \textbf{Params}:$94$M & \textbf{Params}:$94$M 
& \textbf{Params}:$94$M & \textbf{Params}:$600$M & \textbf{Params}:$7$B & \textbf{Params}:$13$B & \textbf{Params}:$160$M \\ \midrule
IEMOCAP-4 & $65.9^{\pm0.5}\textcolor{black}{(67.2)}$ & $65.0^{\pm0.2}\textcolor{black}{(68.0)}$ & $62.7^{\pm0.7}\textcolor{black}{(64.0)}$ & $67.5^{\pm0.6\#}\textcolor{black}{(69.0)}$ & $59.4^{\pm0.4}\textcolor{black}{(61.0)}$ & $\mathbf{75.8}^{\pm0.6\#}\textcolor{black}{(76.9)}$ & $\underline{72.9}^{\pm2.3\#}\textcolor{black}{(74.9)}$ & $69.4^{\pm0.5}\textcolor{black}{(70.1)}$ \\ \midrule
IEMOCAP-6 & $51.7^{\pm0.5}\textcolor{black}{(48.4)}$ & $50.7^{\pm0.9}\textcolor{black}{(46.5)}$ & $46.0^{\pm0.4}\textcolor{black}{(42.3)}$ & $54.1^{\pm0.6\#}\textcolor{black}{(51.9)}$ & $43.5^{\pm0.2}\textcolor{black}{(41.0)}$ & $\mathbf{59.3}^{\pm1.4\#}\textcolor{black}{(55.7)}$ & $\underline{58.1}^{\pm1.6\#}\textcolor{black}{(55.2)}$ & $55.0^{\pm0.4}\textcolor{black}{(52.1)}$ \\ \midrule
MELD &  $45.6^{\pm0.4}\textcolor{black}{(24.3)}$ & $45.3^{\pm0.6}\textcolor{black}{(24.0)}$ & $41.9^{\pm0.5}\textcolor{black}{(23.1)}$ & $47.6^{\pm0.3\#}\textcolor{black}{(27.4)}$ & $43.2^{\pm0.2}\textcolor{black}{(23.3)}$ & $\mathbf{53.3}^{\pm0.7}\textcolor{black}{(33.4)}$  & $\underline{52.6}^{\pm0.4}\textcolor{black}{(32.8)}$ & $48.1^{\pm0.8}\textcolor{black}{(28.8)}$ \\ \midrule
CMU-MOSI &  $64.1^{\pm0.8}\textcolor{black}{(64.2)}$ & $62.5^{\pm0.6}\textcolor{black}{(62.5)}$ & $59.7^{\pm0.4}\textcolor{black}{(58.9)}$ & $66.5^{\pm0.6}\textcolor{black}{(65.9)}$ & $\underline{74.6}^{\pm0.3}\textcolor{black}{(73.9)}$ & $\mathbf{78.0}^{\pm0.7}\textcolor{black}{(77.0)}$ & $72.8^{\pm1.0}\textcolor{black}{(72.0)}$ & $66.7^{\pm1.0}\textcolor{black}{(66.2)}$ \\ \midrule
DAIC-WOZ &  $63.2^{\pm1.5}\textcolor{black}{(61.5)}$ & $65.9^{\pm2.0}\textcolor{black}{(61.9)}$ & $\underline{67.8}^{\pm1.4}\textcolor{black}{(65.7)}$ & $61.6^{\pm0.7}\textcolor{black}{(61.0)}$ & $64.3^{\pm0.4}\textcolor{black}{(63.7)}$ & $62.6^{\pm3.4}\textcolor{black}{(60.4)}$ & $64.7^{\pm3.0}\textcolor{black}{(61.1)}$ & $\mathbf{68.5}^{\pm2.1}\textcolor{black}{(67.1)}$ \\ \midrule
RAVDESS &  $50.5^{\pm3.6}\textcolor{black}{(49.1)}$ & $\underline{53.5}^{\pm1.1}\textcolor{black}{(55.7)}$ & $38.5^{\pm5.2}\textcolor{black}{(40.8)}$ & $48.5^{\pm1.0}\textcolor{black}{(51.0)}$ & $11.8^{\pm2.0}\textcolor{black}{(10.8)}$ & $50.2^{\pm1.3}\textcolor{black}{(54.2)}$ & $51.9^{\pm3.6}\textcolor{black}{(53.4)}$ & $\mathbf{60.1}^{\pm1.6}\textcolor{black}{(62.0)}$ \\ \midrule
CaFE & $66.6^{\pm2.6}\textcolor{black}{(69.0)}$ &$66.5^{\pm4.5}\textcolor{black}{(69.1)}$ & $48.8^{\pm4.3}\textcolor{black}{(51.0)}$ & $59.3^{\pm3.8}\textcolor{black}{(62.6)}$ & $5.7^{\pm1.4}\textcolor{black}{(7.1)}$ & $59.9^{\pm2.0}\textcolor{black}{(62.8)}$ & $\underline{69.8}^{\pm3.3}\textcolor{black}{(71.4)}$ & $\mathbf{77.0}^{\pm1.5}\textcolor{black}{(78.1)}$ \\ \midrule
EmoDB & $66.5^{\pm4.8}\textcolor{black}{(68.5)}$ & $66.9^{\pm3.9}\textcolor{black}{(68.2)}$&  $48.9^{\pm3.1}\textcolor{black}{(49.9)}$ &  $64.4^{\pm2.6}\textcolor{black}{(66.7)}$ & $10.2^{\pm2.0}\textcolor{black}{(12.6)}$ & $\underline{82.8}^{\pm2.9}\textcolor{black}{(85.3)}$ & $82.2^{\pm4.0}\textcolor{black}{(84.1)}$ & $\mathbf{83.4}^{\pm2.0}\textcolor{black}{(83.9)}$ \\ \midrule \midrule
Avg. & $59.3\textcolor{black}{(56.5)}$ & $59.5\textcolor{black}{(57.0)}$ & $51.8\textcolor{black}{(49.5)}$ &  $58.7\textcolor{black}{(56.9)}$ & $39.1\textcolor{black}{(36.7)}$ & $65.2\textcolor{black}{(63.2)}$& $\underline{65.6}\textcolor{black}{(63.1)}$ & $\mathbf{66.0}\textcolor{black}{(63.5)}$ \\ \bottomrule
\end{tabular}}
\vspace{-0.1in}
\end{table}

\subsubsection{Performance of CARE}\label{sec:careresults}
The results on the $8$ downstream datasets using representations from the proposed CARE model are shown in Table~\ref{tab:results-cat-oth}.  \textcolor{black}{These baseline models are categorized into two groups based on the number of parameters used during inference: base models (parameter size $<200$M), and large models ($>500$M), which also include LLM based models.} The following observations are made for each category:

\noindent \textbf{Base models:} We compare HuBERT \cite{hsu2021hubert}, WavLM \cite{chen2022wavlm}, data2vec \cite{baevski2022data2vec} and emotion2vec~\cite{ma2024emotion2vec} representations as the baseline models in this category. Among these baseline systems, the emotion2vec is also pre-trained on IEMOCAP and MELD datasets, partially explaining the improved results seen on the downstream tasks on these datasets.
While CARE performs similar to emotion2vec on CMU-MOSI, it improves over all the base-sized models on other datasets. On the average, the proposed CARE achieves a relative improvement of $15.6$\% over the best baseline model (HuBERT). 

\noindent \textbf{Large models:} SONAR~\cite{duquenne2023sonar} is selected as the speech encoder in this category. For the six English-based datasets, the pre-trained English speech encoder\footnote{\url{https://dl.fbaipublicfiles.com/SONAR/spenc.eng.pt}} is used, while the French and German speech encoders are utilized for the CaFE and EmoDB datasets, respectively. Similar to the CARE backend, the layer representations from the SONAR encoder are linearly combined and the classification head is trained on the downstream task.
 Although SONAR has nearly four times the parameter size of CARE, our proposed model outperforms SONAR across all datasets except CMU-MOSI. 

 \noindent \textbf{LLM based models:} Two versions of SALMONN~\cite{tang2023salmonn} ($7$B and $13$B)\footnote{\url{https://huggingface.co/tsinghua-ee/SALMONN}} are considered as examples of LLM-based models. 
        These are typically applied in a zero-shot setting; however, due to variability in emotion classes across datasets, their zero-shot performance is inconsistent. E.g. while SALMONN-$13$B model achieves $68.75\%$ weighted F1-score on the IEMOCAP-4 dataset (on which it is trained), it achieves only $24.06\%$ for MELD. Thus, for fair comparison, the same framework used in CARE and other baseline models is followed for the LLM based evaluations as well. The internal representations from all layers ($41$ layers for SALMONN $13$B and $33$ layers for SALMONN $7$B) are aggregated using a convex combination, and the classification head (similar to CARE) is trained for each downstream dataset. 
    Similar to emotion2vec, SALMONN includes IEMOCAP in its pre-training, leading to superior performance on IEMOCAP-4 and IEMOCAP-6 compared to CARE. The larger model size and extensive pre-training data allows SALMONN to outperform CARE by $10\%$ and $34\%$ (relative improvements) on the MELD and CMU-MOSI datasets, respectively. 
      However, on the remaining four tasks, CARE surpasses the SALMONN models, achieving relative improvements of $17\%$ and $24\%$ on the RAVDESS-song and CaFE datasets, respectively. Notably, though music datasets are used to pre-train SALMONN, CARE emerges as the best model on the RAVDESS-Song dataset.

\noindent{\textbf{Key takeaways:}}
1) On average, CARE emerges as the top-performing model across the eight datasets, surpassing even the SALMONN $13$B model, which has nearly $80$ times more parameters. 
Although LLM-based models show strengths in in-domain emotion recognition datasets, their performance declines on out-of-domain tasks, indicating limited generalizability across diverse tasks and multilingual emotional speech. 2) CARE’s advantage over speech SSL models like WavLM, HuBERT, and data2vec is expected, given that these models are trained on non-emotional data (see  Sec.~\ref{sec:cont} for a related experiment). 3) Notably, CARE outperforms the multilingual SONAR model on CaFE and EmoDB datasets although it is trained on English speech only. This showcases the generalizability of our pre-training technique to out-of-domain tasks in SER.

\subsubsection{Emotional Attribute Prediction} \label{sec:care_improv}
The   emotion recognition can be posed as a regression problem, where valence, arousal and dominance of a particular utterance are predicted~\cite{parthasarathy2017jointly}. We use the MSP-IMPROV~\cite{busso2016msp} for this purpose. We refer the reader to Sec.~\ref{sec:setting_datasets} for details about this dataset.\par
We use the concordance correlation coefficient (CCC) as the metric. Denoting the mean, variance of ground truth by $\mu_g$, $\sigma_g^2$  and predicted scores by $\mu_p$, $\sigma_p^2$, the CCC is defined as 
\begin{equation}\label{eq:ccc}
    CCC = \frac{2\rho\sigma_g\sigma_p}{\sigma_p^{2} + \sigma_{g}^{2} + (\mu_g-\mu_p)^2}
\end{equation}
In Eq.~\ref{eq:ccc}, $\rho$ is the Pearson's correlation coefficient between the ground truth and the predicted scores. For training the downstream model, the representations from CARE and other models are aggregated similar to the categorical datasets. This is followed by a two-layer regression head with $256$ as the hidden dimension and $3$ as the output dimension ($1$ for each of the three attributes). The objective is to increase the CCC between the ground truth and the predicted values for each of the dimensions of valence, arousal and dominance. 
The results for this dataset along with other baseline models are shown in Table.\ref{tab:results-dim-oth}. We note that for this task, the CARE embeddings achieve the best results in terms of the valence and dominance attributes, while the performance  on arousal is marginally better for the SALMONN-$7$B model.
\begin{table}[t!]
\centering
\caption{Results for the MSP-IMPROV dataset. CCC stands for Concordance Correlation Coefficient while V, A, D stand for valence, arousal and dominance respectively.}\label{tab:results-dim-oth}
    \resizebox{0.5\columnwidth}{!}{%
\begin{tabular}{@{}l|l|l|l@{}}
\toprule
Method & CCC-V & CCC-A & CCC-D \\ \midrule
WavLM-base \cite{chen2022wavlm} & $0.51$ & $0.64$ & $0.47$ \\ \midrule
 emotion2vec \cite{ma2024emotion2vec} & $0.5$ & $0.61$ & $0.49$ \\ \midrule
 SALMONN-$7$B \cite{tang2023salmonn} & $0.53$ & $\mathbf{0.67}$ & $\underline{0.52}$ \\ \midrule
 SALMONN-$13$B \cite{tang2023salmonn} & $\underline{0.56}$ & $0.65$ & $0.51$ \\ \midrule
 CARE & $\mathbf{0.57}$ & $\underline{0.66}$ & $\mathbf{0.53}$ \\ \bottomrule
\end{tabular}}
\end{table}

\subsection{Discussion}
\subsubsection{Comparison with Baselines}
\begin{table*}[t!]
\centering
\caption{Baseline results on the different downstream datasets in terms of weighted F1-score. All numbers are averaged over $5$ random initializations of the downstream network. We also show the results of the different components of CARE in this table.  }\label{tab:results-cat}
    \resizebox{\columnwidth}{!}{%
\begin{tabular}{@{}l|c|c|c|c||c|c|c@{}}
\toprule
\multicolumn{1}{l|}{\multirow{2}{*}{Datasets}} 
& PASE+ \cite{ravanelli2020multi} 
& \textcolor{black}{Whisper~\cite{radford2023robust}}
& \begin{tabular}[c]{@{}c@{}}Whisper \cite{radford2023robust}+\\ RoBERTa \cite{liu2019roberta}\end{tabular} 
& \begin{tabular}[c]{@{}c@{}}\textcolor{black}{Teacher-}\\\textcolor{black}{fusion}\end{tabular}
& \begin{tabular}[c]{@{}c@{}}Semantic +\\ Common Enc.\end{tabular}  
& \begin{tabular}[c]{@{}c@{}}Acoustic +\\ Common Enc.\end{tabular} 
& CARE \\ 
 & \textbf{Params}:$8$M 
 & \textcolor{black}{\textbf{Params}:$800$M} 
 & \textbf{Params}:$1.6$B 
 & \textcolor{black}{\textbf{Params}:$1.6$B}
 & \textbf{Params}:$110$M 
 & \textbf{Params}:$94$M 
 & \textbf{Params}:$160$M \\ 
\midrule
IEMOCAP-4 \cite{busso2008iemocap} & $56.68$ & \textcolor{black}{$56.40$} & $61.97$ & \textcolor{black}{$69.49$} & $66.44$ & $65.91$ & $\mathbf{69.39}$ \\ 
\midrule
IEMOCAP-6 \cite{busso2008iemocap} & $41.38$ & \textcolor{black}{$40.62$} & $49.28$ & \textcolor{black}{$56.61$} & $53.05$ & $52.09$ & $\mathbf{55.02}$ \\ 
\midrule
MELD \cite{poria2019meld} & $35.86$ & \textcolor{black}{$40.11$} & $\mathbf{49.29}$ & \textcolor{black}{$49.72$} & $47.37$ & $46.98$ & $48.05$ \\ 
\midrule
CMU-MOSI \cite{zadeh2016mosi} & $50.69$ & \textcolor{black}{$55.60$} & $\mathbf{75.14}$ & \textcolor{black}{$74.12$} & $64.23$ & $64.17$ & $66.74$ \\ 
\midrule
DAIC-WOZ \cite{gratch2014distress} & $66.84$ & \textcolor{black}{$62.08$} & $64.04$ & \textcolor{black}{$67.63$} & $66.32$ & $66.89$ & $\mathbf{68.49}$ \\ 
\midrule
RAVDESS-Song \cite{livingstone2018ryerson} & $46.05$ & \textcolor{black}{$34.20$} & $9.58$ & \textcolor{black}{$48.48$} & $55.23$ & $56.17$ & $\mathbf{60.11}$ \\ 
\midrule
CaFE \cite{gournay2018canadian} & $52.86$ & \textcolor{black}{$19.22$} & $13.59$ & \textcolor{black}{$53.42$} & $69.23$ & $71.62$ & $\mathbf{76.98}$ \\ 
\midrule
EmoDB \cite{burkhardt2005database} & $62.59$ & \textcolor{black}{$24.77$} & $14.98$ & \textcolor{black}{$66.75$} & $75.42$ & $78.63$ & $\mathbf{83.41}$ \\ 
\midrule \midrule
Avg. & $51.62$ & \textcolor{black}{$41.63$} & $42.23$ & \textcolor{black}{$60.78$} & $62.16$ & $62.81$ & $\mathbf{66.02}$ \\ 
\bottomrule
\end{tabular}}
\vspace{-0.1in}
\end{table*}

\textcolor{black}{Four baseline systems  (Table~\ref{tab:results-cat}) are considered:-} \\
 \textbf{PASE+:} For each downstream dataset, PASE+ features are extracted and a classification network is trained to predict the emotion class of each utterance similar to CARE. The total number of parameters used during inference is $8$M. \\
 \noindent\textcolor{black}{\textbf{Whisper:} For each downstream dataset, the representations from the $33$ encoder layers of the Whisper-large-v3 model~\cite{radford2023robust} are  linearly combined with learnable weights. A two-layer classification head is trained on top of these representations for the task of emotion recognition. The total number of parameters used during inference is $800$M.}\\
 \noindent \textbf{Whisper+RoBERTa:}   The transcripts are generated using the Whisper-large-v3 model and subsequently processed by a pre-trained RoBERTa model. The internal representations from RoBERTa are linearly combined by learnable weights, followed by training a two-layer classification head. This has a total of $1.6$B parameters in use during inference.\\
 \noindent \textcolor{black}{\textbf{Teacher-fusion:}   The PASE+ and Whisper+RoBERTa representations are concatenated and a two-layer classification head is trained for each downstream dataset. This baseline also has a total of $1.6$B parameters during inference.}\\
 \textbf{Key takeaways:} 1) The performance of CARE surpasses that of the acoustic supervisory signal by $29.76\%$ (relative) on average across the $8$ datasets. This improvement can be attributed to the larger parameter size of CARE compared to the PASE+ model. \textcolor{black}{2) CARE is seen to outperform Whisper and Whisper+RoBERTa systems by $41.79\%$ and $41.18\%$ in relative terms. This indicates that, although the Whisper-based baselines are much larger in size, the combination of the acoustic and semantic information in CARE results in effective emotion recognition. 3) On MELD and CMU-MOSI, CARE is outperformed by the Whisper+RoBERTa baseline. For these datasets, text-based models are known to significantly outperform speech-only systems~\cite{dutta2023hcam,lian2022smin}. 
 In the Whisper+RoBERTa setup, the RoBERTa model is fine-tuned on transcripts generated by Whisper-large-v3 ($1.6$B sized model). In contrast, CARE is a smaller model ($160$M), and does not use directly use the ASR transcripts during inference. To further elucidate the fairness in model-size, 
 we replace Whisper-large-v3 with a Whisper-base model for the ASR, followed by the RoBERTa modeling. Then, the performance drops from $49.29\%$ to $46.02\%$ on MELD and from $75.14\%$ to $71.91\%$ on CMU-MOSI. This underscores the importance of accurate transcriptions and large model capacity in settings where the textual information is emotion rich.} \textcolor{black}{4) While the teacher-fusion baseline is competitive for a number of datasets involving English speech, CARE outperforms this baseline on average by $5.24\%$ absolute. This also motivates why CARE was pre-trained using knowledge distillation as it outperforms the fusion baseline with only $10\%$ of the parameters.}
 \subsubsection{Importance of the Two Encoders}
We present the performance of CARE when we use only one of acoustic and semantic encoders along with the common encoder for the downstream datasets in Table~\ref{tab:results-cat}. For evaluating the combination of the semantic and common encoders, we use the $768$-dimensional representations from the convolutional feature extractor, the common encoder, and the semantic encoder, excluding outputs from the acoustic encoder. Similarly, the semantic encoder representations are disregarded during the evaluation of the acoustic-common encoder combination. Note that, while CARE has more number of parameters ($160$M) as compared to models like WavLM or emotion2vec, both these combinations have similar number of parameters during inference. While the semantic-common encoder combination has an inference time parameter size of $110$M, the acoustic-common encoder has a total of $94$M parameters during evaluation on each downstream dataset.\\
\noindent{\textbf{Key takeaways:}} 1) The combination of the acoustic and common encoder representations outperforms the best performing SSL model (HuBERT) by $8.08\%$ (relative) on average for the $8$ datasets (Table~\ref{tab:results-cat-oth}). Given the similar parameter count, this performance suggests an advantage of our pre-training approach. 2) For the three out-of-domain datasets, the acoustic-common combination fares better than its semantic counterpart. 
3) Across all datasets, the combination of both encoders in CARE yields the highest performance, suggesting that while the individual encoder performances are comparable, they capture distinct characteristics of the speech signal.

\subsubsection{Modifications in the Semantic Encoder} 
To evaluate the suitability of our design choices for the semantic encoder, we made three architectural modifications:
\begin{enumerate}
\item \textbf{CARE-No init.:} Removing the convolutional adapters, the transformer layers in the semantic encoder are initialized randomly (instead of pre-trained RoBERTa weights). 
\item \textbf{CARE-Trans.:} Removing the convolutional adapters while the RoBERTa transformer layers are  updated.   
\item \textbf{CARE-FT:} Keeping the convolutional adapters, we update all the parameters (conv. adapters and transformer weights) in the semantic encoder.
\end{enumerate}
The results for these modifications are shown in Table~\ref{tab:results-cat-ada}. \\
\noindent{\textbf{Key takeaways:}} 1) Initializing the transformer weights with RoBERTa is essential for CARE's performance. Random initialization of the semantic encoder leads to performance drops across all five datasets, suggesting that a randomly initialized semantic encoder struggles to regress to the semantic supervisory signal. 2) Removing convolutional adapters (in CARE-Trans) negatively impacts performance, highlighting the necessity of our convolution-based adaptation technique for aligning speech representations with RoBERTa's transformer layers. 3) Updating the transformer layers in the semantic encoder decreases performance. Since RoBERTa is pre-trained on text, fine-tuning with speech data degrades its effectiveness.
\begin{table}[t!]
\centering
\caption{Results on the downstream datasets (weighted F1-score) with modifications in the semantic encoder.}\label{tab:results-cat-ada}
    \resizebox{0.4\columnwidth}{!}{%
\begin{tabular}{@{}l|ll@{}}
\toprule
Dataset & \multicolumn{1}{l|}{Method} & WF1 \\ \midrule
\multicolumn{1}{l|}{\multirow{3}{*}{IEMOCAP(4-class)}} & \multicolumn{1}{l|}{CARE-No init.} & $65.71$ \\
\multicolumn{1}{l|}{} & \multicolumn{1}{l|}{CARE-Trans.} & $65.89$ \\
\multicolumn{1}{l|}{} & \multicolumn{1}{l|}{CARE-FT} & $68.16$ \\
\multicolumn{1}{l|}{} & \multicolumn{1}{l|}{CARE} & $\mathbf{69.39}$ \\ \midrule
\multicolumn{1}{l|}{\multirow{3}{*}{IEMOCAP(6-class)}} & \multicolumn{1}{l|}{CARE-No init.} & $50.96$ \\
\multicolumn{1}{l|}{} & \multicolumn{1}{l|}{CARE-Trans.} & $51.19$ \\
\multicolumn{1}{l|}{} & \multicolumn{1}{l|}{CARE-FT} & $52.37$ \\
\multicolumn{1}{l|}{} & \multicolumn{1}{l|}{CARE} & $\mathbf{55.02}$ \\ \midrule
\multicolumn{1}{l|}{\multirow{3}{*}{MELD}} & \multicolumn{1}{l|}{CARE-No init.} & $45.16$ \\
\multicolumn{1}{l|}{} & \multicolumn{1}{l|}{CARE-Trans.} & $46.57$ \\
\multicolumn{1}{l|}{} & \multicolumn{1}{l|}{CARE-FT} & $47.93$ \\
\multicolumn{1}{l|}{} & \multicolumn{1}{l|}{CARE} & $\mathbf{48.05}$ \\ \midrule
\multicolumn{1}{l|}{\multirow{3}{*}{CMU-MOSI}} & \multicolumn{1}{l|}{CARE-No init.} & $62.16$ \\
\multicolumn{1}{l|}{} & \multicolumn{1}{l|}{CARE-Trans.} & $65.97$ \\
\multicolumn{1}{l|}{} & \multicolumn{1}{l|}{CARE-FT} & $64.27$ \\
\multicolumn{1}{l|}{} & \multicolumn{1}{l|}{CARE} & $\mathbf{66.74}$ \\ \midrule
\multicolumn{1}{l|}{\multirow{3}{*}{DAIC-WOZ}} & \multicolumn{1}{l|}{CARE-No init.} & $64.57$ \\
\multicolumn{1}{l|}{} & \multicolumn{1}{l|}{CARE-Trans.} & $65.93$ \\
\multicolumn{1}{l|}{} & \multicolumn{1}{l|}{CARE-FT} & $67.43$ \\
\multicolumn{1}{l|}{} & \multicolumn{1}{l|}{CARE} & $\mathbf{68.49}$ \\ \bottomrule
\end{tabular}}
\vspace{-0.1in}
\end{table}
\begin{table}[t!]
\centering
\caption{Results on the downstream datasets (weighted F1-score) with different initializations of the acoustic  encoders.}\label{tab:results-ssl}
    \resizebox{0.6\columnwidth}{!}{%

\begin{tabular}{@{}l|c|c|c|c}
\toprule
Datasets & Random init. & HuBERT init. & Data2vec init. &  WavLM init. \\ \midrule
IEMOCAP-4 & $66.72$ &$67.65$ &$66.76$ & $\mathbf{69.39}$  \\ \midrule
IEMOCAP-6 & $51.47$ &$51.40$ &$52.98$  & $\mathbf{55.02}$ \\ \midrule
MELD & $46.41$ &$46.97$ &$47.19$ & $\mathbf{48.05}$  \\ \midrule
CMU-MOSI & $65.07$ &$66.26$ &$\mathbf{68.14}$  & $66.74$ \\ \midrule
DAIC-WOZ & $67.56$ &$65.19$ &$66.59$ & $\mathbf{68.49}$  \\ \midrule
RAVDESS-Song & $57.81$ &$\mathbf{60.30}$ & $55.09$ & $60.11$ \\ \midrule
CaFE & $73.83$ &$72.23$ & $62.46$ & $\mathbf{76.98}$  \\ \midrule
EmoDB & $78.61$ &$\mathbf{86.51}$& $77.19$ & $83.41$ \\ \midrule \midrule
Avg. &$63.44$ &$64.56$ & $62.05$ & $66.02$ \\ \bottomrule

\end{tabular}}
\end{table}

\subsubsection{Initialization of Acoustic and Common Encoders}
As indicated in Section~\ref{sec:pretraining_method_care}, the common and the acoustic encoders of CARE are initialized with the WavLM-base model weights. We present the results of our method when this initialization is modified to i) random, ii) HuBERT-base~\cite{hsu2021hubert} or iii) data2vec-base~\cite{baevski2022data2vec} (Table~\ref{tab:results-ssl}). \\
\noindent{\textbf{Key takeaways:}} 1) The model's performance decreases with data2vec initialization, likely due to data2vec’s lower baseline performance compared to HuBERT and WavLM (see Table~\ref{tab:results-cat-oth}). An exception is the CMU-MOSI dataset, where this initialization improves over the WavLM initialized model by $4.21\%$ (relative). 2) The HuBERT-initialized model performs best on the RAVDESS-Song and Emo-DB datasets. Notably, HuBERT outperforms WavLM for these two out-of-domain datasets  (Table~\ref{tab:results-cat-oth}). 3) Initialization impacts the acoustic and common encoders less than the semantic encoder, as the latter requires alignment with text representations.

\subsubsection{Choice of Acoustic Targets}

\begin{figure}
    \centering
    \includegraphics[width=0.7\textwidth,trim={2cm 7cm 5cm 3.5cm},clip]{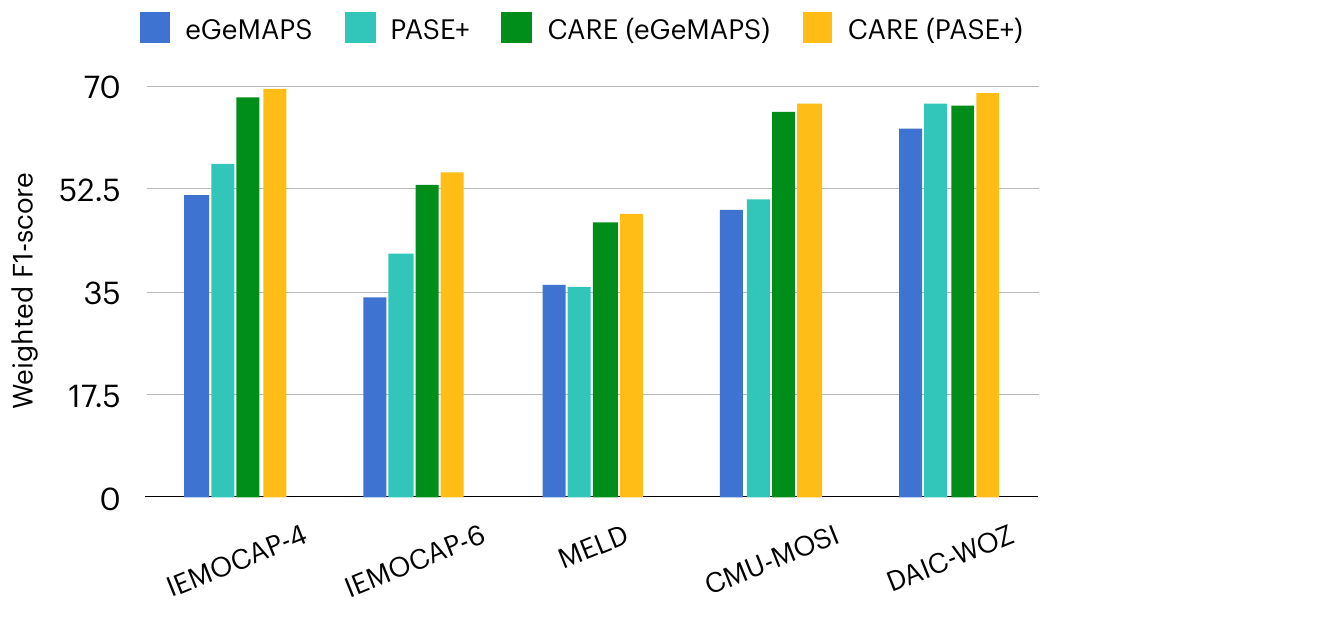}
    \caption{Performance of CARE when different acoustic targets are used.  The model with eGeMAPS as features is trained similarly to that of the PASE+ baseline. All numbers are shown as the average of $5$ random initializations.}
    \label{fig:acoustic}
\end{figure}

We run an experiment where the acoustic encoder is trained with targets based on eGeMAPS~\cite{eyben2015geneva} features extracted from the openSMILE toolkit~\cite{eyben2010opensmile}. The  PASE+ targets of the acoustic encoder of the CARE model is replaced by the eGeMAPS features. The performance of this model, called CARE (eGeMAPS), is shown in Fig.~\ref{fig:acoustic}. \\
\noindent{\textbf{Key takeaway:}} The baseline model using eGeMAPS input features performs worse than the baseline with PASE+ features, as expected, since eGeMAPS are handcrafted. Consequently, the average performance of CARE with eGeMAPS is also lower than that of CARE with PASE+ targets.

\begin{figure}
    \centering
    \includegraphics[width=0.7\textwidth,trim={2cm 7cm 5cm 3.5cm},clip]{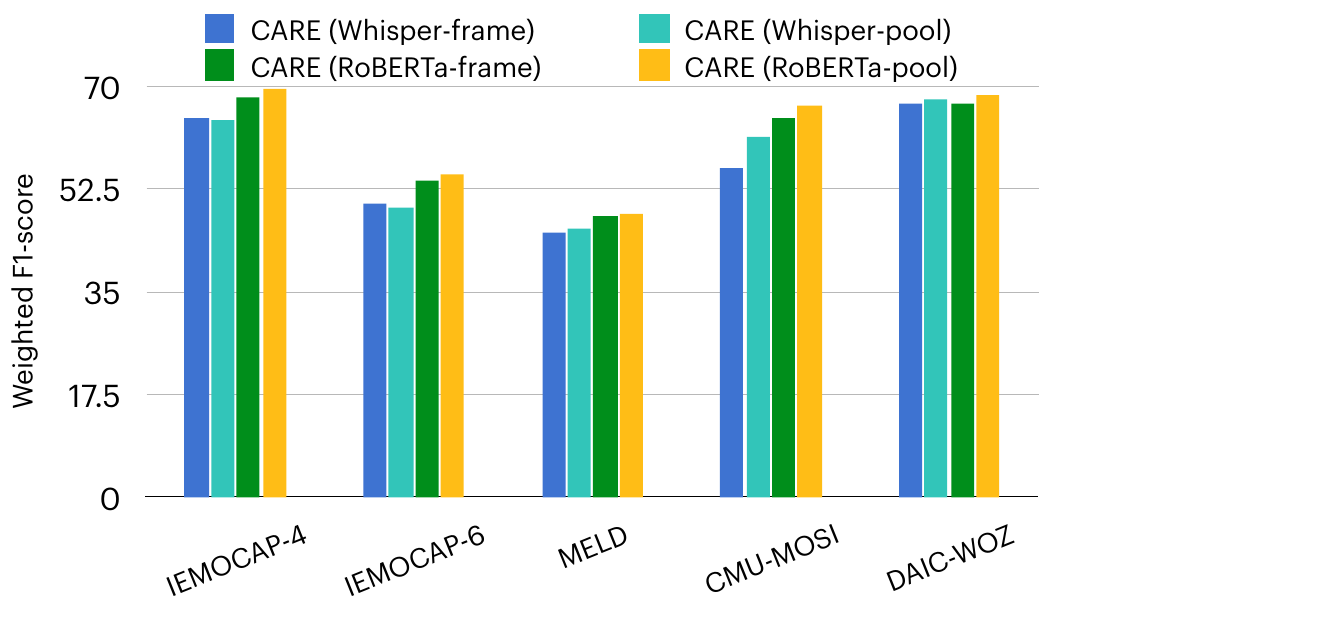}

    \caption{\textcolor{black}{Performance of CARE when different semantic targets are used. All numbers are shown as the average of $5$ random initializations.}}
    \label{fig:semantic}
\end{figure}

\textcolor{black}{\subsubsection{Choice of Semantic Targets}}\label{sec:sem_targets}
\textcolor{black}{
We conduct an experiment where the Whisper encoder representations serve as supervisory signals for the semantic encoder. We explore two variants of this: 1) We pool the Whisper representations to serve as semantic targets while pre-training. This model is called CARE (Whisper-pool). 2) We pre-train a model with the frame-level representations of Whisper as the targets. We call this model CARE (Whisper-frame). 3) We also use the frame level alignments between speech and the RoBERTa tokens and use the frame-level RoBERTa representations as the semantic targets. We call this model CARE (RoBERTa-frame). The comparative performances of the different variants along with the proposed model-CARE (RoBERTa-pool) are shown in Fig.~\ref{fig:semantic}.}\\
\noindent \textcolor{black}{
\noindent{\textbf{Key takeaways:}} 1) The performance of CARE (RoBERTa-pool) is seen to be superior to both variants trained with Whisper encoded representations.
2) The performance of the systems when the semantic encoder is trained with the pooled   targets is observed to be better than those trained with frame-level representations. }

\begin{figure}[ht!]
    \centering    \includegraphics[width=0.7\textwidth,trim={1.9cm 7.0cm 1cm 3.7cm},clip]{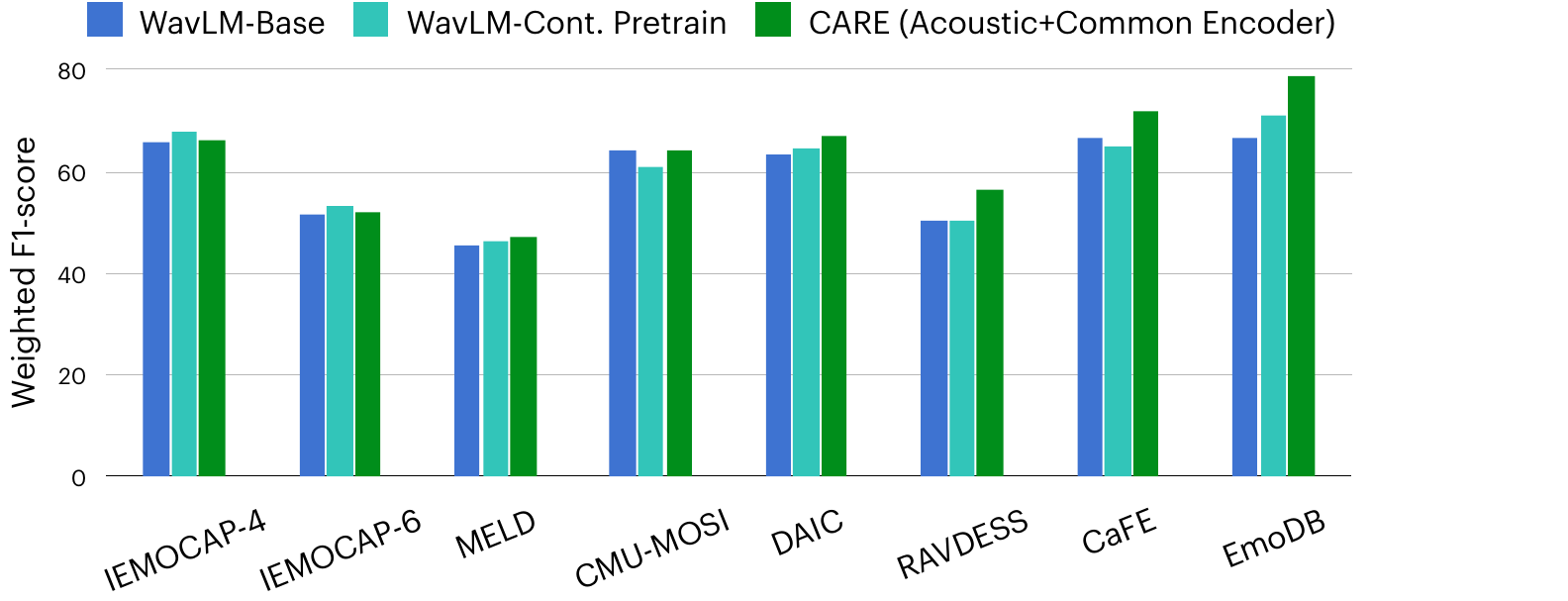}
    \vspace{-0.1in}
    \caption{Comparison of the performance when WavLM is continually pre-trained on MSP-PODCAST. The performance of the combination of acoustic and common encoders of CARE is shown for reference. Here, RAVDESS refers to the RAVDESS-Song dataset.}
    \label{fig:init}
\end{figure}

\subsubsection{Continued Pre-training of WavLM}\label{sec:cont}

Since all self-supervised learning (SSL) models are trained on neutral data, their ability to accurately discern emotions from speech signals is typically limited. The emotion recognition performance of these SSL models when pre-trained on emotion datasets thus becomes crucial. To explore the impact of pre-training setup in the proposed CARE, we continued the pre-training of the publicly available WavLM-base model,  using the MSP-PODCAST dataset. This was done following the WavLM pre-training procedure, with masked language modeling loss, for an additional $200,000$ steps (similar to the CARE). The results of this experiment are shown in Fig.~\ref{fig:init}, wherein the performance of the continually pre-trained WavLM model is denoted by {WavLM-Cont. Pretrain}. The performance of the combination of the common and acoustic encoders is also shown for comparison.

\noindent{\textbf{Key takeaway:}} Continued pre-training improves WavLM-base performance on certain downstream tasks, like IEMOCAP. However, except for IEMOCAP, WavLM performs worse than CARE's acoustic-common encoder combination. As all the models in Fig.~\ref{fig:init} have the same size ($94$M) during inference, this experiment highlights the benefits of our proposed distillation-based pre-training.

\subsection{Summary of Pre-training for Speech Emotion Recognition}
\begin{figure}
    \centering
    \includegraphics[width=0.6\textwidth,trim={3cm 4cm 7cm 3.5cm},clip]{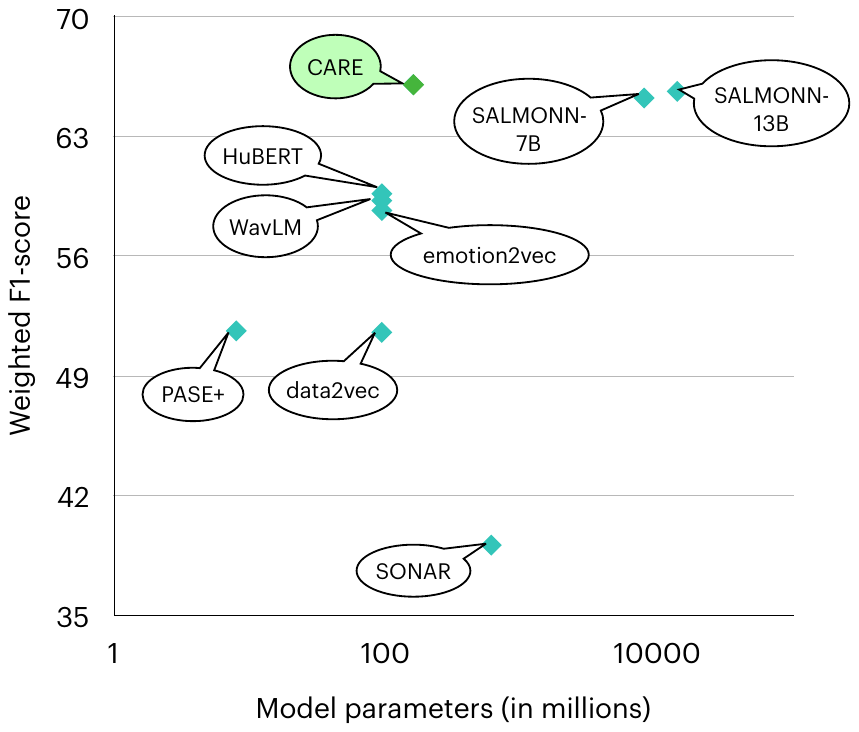}
    \caption{Scatter plot of  inference model size (parameters in millions) versus the SER performance (average weighted F1-score over $8$ datasets). CARE is seen to achieve a better trade-off compared to the existing solutions. For more details, refer Tables~\ref{tab:results-cat-oth} and ~\ref{tab:results-cat} and the associated discussions.}.
    \label{fig:compare}
\end{figure}

The landscape of various SER methods is summarized in Fig.~\ref{fig:compare}. In current modeling frameworks: models either prioritize efficiency with limited performance (those in the lower end of the x-axis), or focus on maximizing performance with increased memory and compute requirements (typically based on LLMs). Our proposed system, CARE, combines the computational efficiency of smaller models with the high performance of large-scale systems, thereby providing a superior trade-off between efficiency and performance. 

\section{Pre-training for Text Emotion Recognition}\label{sec:pretraining_ter}
We propose a pre-training methodology through a multi-modal approach that leverages both text and speech representations. Specifically, we introduce a strategy to improve emotion classification from text by leveraging unsupervised speech data with large-scale language models (LLMs). To this end, we first utilize a pre-trained ASR based on Whisper-large model \cite{radford2023robust} to  transcribe speech. The ``noisy'' speech transcripts are labeled with an LLM to automatically generate pseudo-labels of speech sentiments. These labeled text-transcripts are then used to fine-tune a RoBERTa text encoder model \cite{liu2019roberta} for sentiment classification, allowing it to capture nuanced emotional patterns in textual data. 
\subsection{Proposed Approach}\label{sec:meritsmodel}
\begin{figure*}
    \centering
    \includegraphics[width=0.9\textwidth,trim={2cm 8.5cm 2cm 6cm},clip]{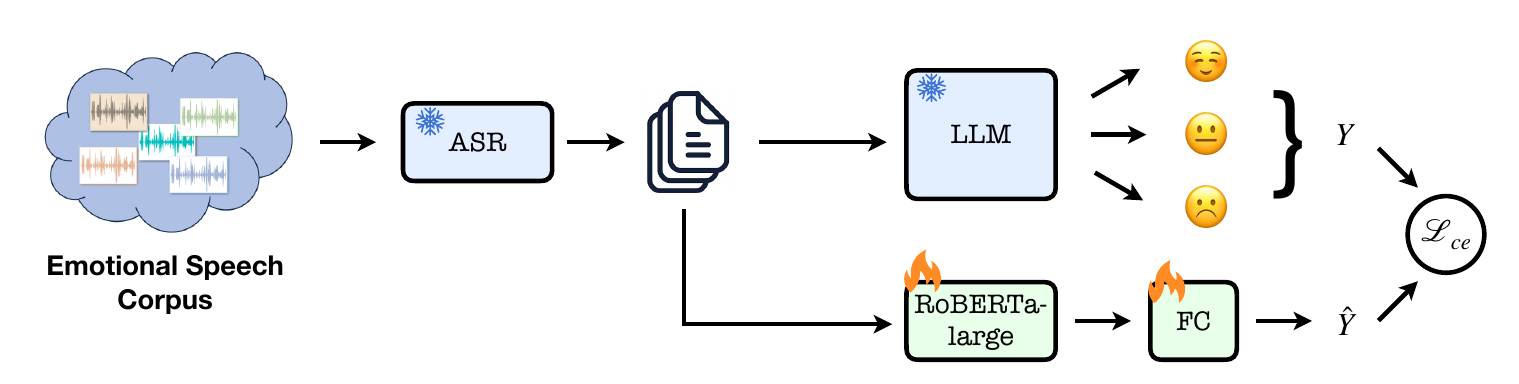}
    \caption{An ASR system is used to generate the transcripts for the pre-training data which are annotated by a large language model (LLM) as positive, negative or neutral sentiment. These ``silver'' labels with the text transcripts form the supervised training  dataset  for RoBERTa-large model.}
    \label{fig:ter_pretraining}
\end{figure*}

The text transcripts from the emotional speech corpus are first extracted from an ASR system (Whisper-large-v3 \cite{radford2023robust}). Generally, the word error rates on emotional speech are higher than neutral speech~\cite{li2023asr}.  With the ASR generated transcripts of the speech corpus, a large language model (LLM) is prompted to annotate the transcript as three classes, ``positive'', ``negative'' or ``neutral''.  The pre-trained RoBERTa-large model is fine-tuned to predict the pseudo-classes (from the LLM predictions). A block diagram of the proposed pre-training strategy is shown in Fig.~\ref{fig:ter_pretraining}
\subsection{Experiments and Results}
\subsubsection{Pre-training}\label{sec:merits_pre}
The MSP-PODCAST corpus~\cite{busso2025msp} is used for the task of pre-training. A total of $149,307$ speech-turns amounting to   $230$ hours of emotional speech data is used. Out of the total number of samples, $80\%$ of the data is randomly chosen as the training set while the remaining $20\%$ serves as the validation set. The Whisper-large-v3 model\footnote{\url{https://huggingface.co/openai/whisper-large-v3}} is used for generating the transcripts. These transcripts are annotated using the \texttt{GPT-3.5 Turbo}\footnote{\url{https://platform.openai.com/docs/models/gpt-3-5-turbo}} model.
\subsubsection{Downstream Tasks}\label{sec:merits_down}
Three datasets are used for evaluating the performance of our pre-training strategy - IEMOCAP~\cite{busso2008iemocap}, MELD~\cite{poria2019meld} and CMU-MOSI~\cite{zadeh2016mosi}. The details of these datasets are provided in Sec.~\ref{sec:setting_datasets}.
\subsubsection{Implementation Details}
Once the transcripts are generated by the Whisper model, the \texttt{GPT-3.5 Turbo} model is prompted as follows: \par
\begin{tcolorbox}[enhanced,width=\columnwidth,center upper,size=fbox,
    fontupper=\small\bfseries,
    colframe=red!50!black,colback=yellow!10]
\texttt{You are a sentiment classification bot. Given the [sentence], classify as positive, negative or neutral sentiment. Please give the sentiment and no extra text as output.}
\end{tcolorbox}
\noindent
The RoBERTa-large model is pre-trained with the LLM generated labels ($3$ classes) for a total of $10$ epochs with a learning rate of $1e$-$4$ and a batch size of $32$.
\begin{figure}
    \centering
    \includegraphics[width=0.7\columnwidth,trim={8cm 7.5cm 5cm 5cm},clip]{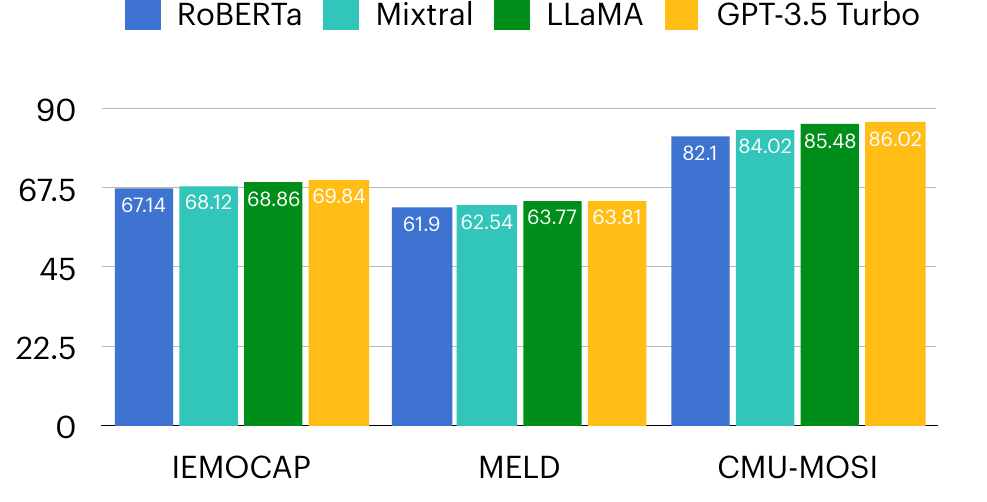}
    \caption{The performance of the   RoBERTa-large models on the different datasets. Different LLMs are used for generating pseudo emotion labels  from speech transcripts. The performance of pre-trained RoBERTa without any supervised fine-tuning is also reported.}
    \label{fig:llm}
\end{figure}
\subsubsection{Results}\label{sec:meritsresults}

We perform a downstream evaluation using the three choices of LLM. The impact of the different LLM annotation ability is shown in Fig.\ref{fig:llm}. The performance of the pre-trained RoBERTa model (without any LLM guidance) is also shown for reference. We notice that   the RoBERTa model fine-tuned with labels provided by \texttt{GPT-3.5 Turbo} achieves relative improvements of $8.22\%$, $5.01\%$ and $21.9\%$ for IEMOCAP, MELD and CMU-MOSI respectively over the pre-trained RoBERTa model.   The relative improvements achieved by \texttt{GPT-3.5 Turbo} over \texttt{Mixtral-8x7B-Instruct-v0.1} are $5.4\%$, $3.39\%$ and $12.51\%$ for IEMOCAP, MELD and CMU-MOSI respectively. The highest performance improvement in the CMU-MOSI dataset may be attributed to the  binary classification task in this dataset.
\subsection{Summary of Pre-training for Text Models}
In this section, we proposed a very simple way to improve the performance of text-based models for emotion recognition. By utilizing the prior knowledge of large language models (LLMs), we annotate emotionally salient spoken language. These emotion labels are then utilized to pre-train text models by supervised pre-training. The performance is seen to improve across multiple spoken language emotion recognition benchmarks.

\section{Chapter Summary}
\label{sec:chap3_summary}

This chapter investigated emotion-aware pre-training strategies for spoken language, motivated by the observation that conventional self-supervised objectives are primarily optimized for linguistic reconstruction and may not adequately capture affective cues essential for emotion understanding. We explored how acoustic and semantic information can be explicitly incorporated during pre-training to improve generalization across emotion recognition tasks.

First, we introduced CARE, a speech representation learning framework that integrates acoustic and semantic supervision during pre-training. Through extensive experiments and ablation studies, we demonstrated that CARE achieves a favourable balance between computational efficiency and emotion recognition performance, outperforming both lightweight models and large-scale systems across multiple benchmarks. These results highlight the importance of incorporating emotion-relevant inductive biases at the pre-training stage.

Second, we proposed a complementary approach for emotion-aware pre-training of text models derived from spoken language. By leveraging the prior knowledge of large language models to annotate transcribed speech with emotion labels, we enabled supervised pre-training of text representations without requiring manually annotated text corpora. This strategy resulted in consistent performance improvements across several emotion recognition benchmarks, demonstrating the effectiveness of speech-driven supervision for semantic emotion modeling.

Together, the methods presented in this chapter establish frameworks for emotion-aware pre-training for both the acoustic and semantic dimensions of spoken language.

\chapter{Multimodal Emotion Recognition from Spoken Language}
\label{chap:recognition}

This chapter focuses on multimodal emotion recognition from spoken language in two challenging settings: Emotion Recognition in Conversations (ERC) and emotion recognition under naturalistic recording conditions. ERC requires models to capture conversational context while remaining robust to interactions involving multiple speakers. Emotion recognition from speech recorded under naturalistic conditions further introduces challenges such as variability in speakers and recording environments, as well as highly skewed emotion distributions.

In ERC, emotions must be inferred not only from individual utterances but also from their conversational context, including interactions across speakers and temporal dependencies between utterances. Modeling such dependencies introduces challenges related to context aggregation and multimodal fusion. In this chapter, we investigate hierarchical approaches for capturing conversational structure and analyze different mechanisms for fusing acoustic and semantic information. We further propose a modular mixture-of-experts modeling paradigm to effectively combine these modalities for ERC.

Existing ERC datasets, however, do not reflect the complexity of naturalistic speech. As discussed in Sec.~\ref{sec:setting_datasets}, these datasets predominantly consist of acted emotions recorded in controlled laboratory environments, limiting their generalizability to spontaneous spoken language. The latter part of this chapter, therefore, outlines our efforts to improve emotion recognition from naturalistic speech, carried out as part of our participation in the \textit{Interspeech 2025 Challenge on Speech Emotion Recognition in Naturalistic Conditions}~\footnote{\url{https://lab-msp.com/MSP-Podcast_Competition/IS2025/}}. We evaluate several recent large language model-based architectures and investigate the use of loss functions tailored to class imbalance, achieving competitive performance on the challenge leaderboard.

\section{Emotion Recognition in Conversations (ERC)}

We begin by formalizing the problem of Emotion Recognition in Conversations (ERC). We then introduce two modeling frameworks for multimodal conversational emotion recognition: a hierarchical cross-attention model (HCAM), followed by a modular mixture-of-experts architecture.

The HCAM framework is designed to explicitly model conversational context while integrating acoustic and semantic information. The architecture follows a hierarchical training paradigm comprising three stages. In the first stage, unimodal predictors for speech and text are trained independently in a context-agnostic manner. In the second stage, conversational context is incorporated through contextual modeling layers, enabling each modality to capture temporal dependencies across utterances. In the final stage, the acoustic and textual streams are fused using a cross-attention-based co-attention module to produce multimodal predictions.

While HCAM effectively captures contextual and cross-modal interactions, it exhibits a key limitation common to feature-level fusion approaches: when one modality substantially outperforms the other, multimodal fusion does not consistently improve upon the stronger unimodal system. In such cases, the fused representation is dominated by the weaker modality, limiting the overall benefit of multimodal modeling.

To address this limitation, we propose \textbf{Mi}xture of \textbf{S}peech-\textbf{T}ext \textbf{E}xperts for \textbf{R}ecognition of \textbf{E}motions (MiSTER-E), a modular architecture that performs fusion at the decision level rather than at the feature level. MiSTER-E adopts a Mixture-of-Experts (MoE) formulation with three independently optimized branches: a speech-only contextual expert, a text-only contextual expert, and a multimodal expert.

Each unimodal expert models modality-specific conversational dynamics using temporal inception networks followed by recurrent layers, while the multimodal expert fuses speech and text representations through stacked cross-attention and self-attention modules. The outputs of all experts are combined via a gating network that computes a weighted sum of expert logits prior to softmax normalization. This logit-level fusion enables dynamic expert selection, allowing the model to emphasize unimodal or multimodal cues depending on their relative reliability.

To further encourage expert specialization and stabilize training, we incorporate contrastive losses and KL-divergence-based consistency constraints across experts. Additionally, each expert is supervised using focal loss to mitigate the effects of class imbalance. Empirically, this modality-decoupled MoE formulation demonstrates a distinct advantage in scenarios where one modality dominates the other, consistently outperforming both feature-level fusion approaches and individual unimodal baselines.
\begin{figure}
    \centering
    \includegraphics[width=0.9\textwidth,trim={0cm 5cm 0cm 5cm},clip]{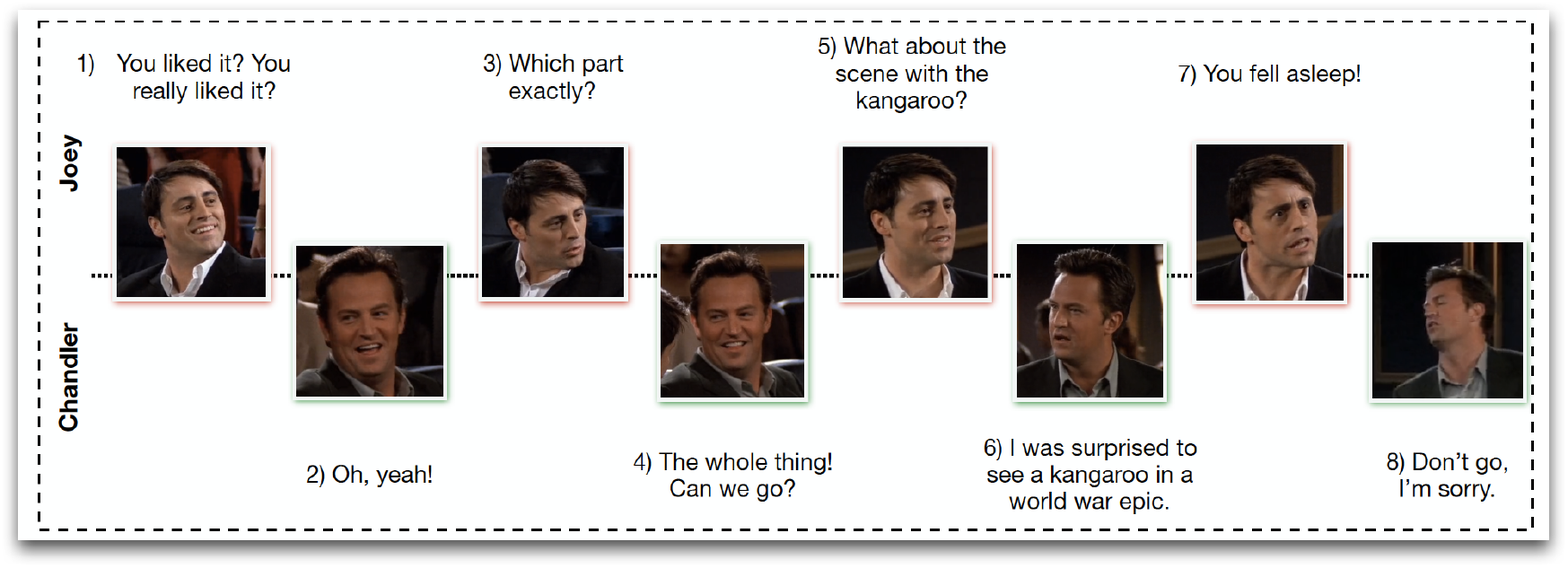}
    \caption{An example of a conversation from the ERC dataset MELD. This example shows how emotions change when multiple speakers interact in a conversation}.
    \label{fig:erc}
\end{figure}
\subsection{Problem Description}\label{sec:erc_prob}

Let $\mathcal{D}$ denote an Emotion Recognition in Conversations (ERC) dataset consisting of $P$ conversations,
$\mathcal{C} = \{\mathbf{c}_1, \mathbf{c}_2, \dots, \mathbf{c}_P\}$.
Each conversation $\mathbf{c}_i$ is composed of a sequence of utterances,
$\mathcal{U}_i = \{\mathbf{u}_{i1}, \mathbf{u}_{i2}, \dots, \mathbf{u}_{iN_i}\}$,
where $N_i$ denotes the number of utterances in conversation $\mathbf{c}_i$.
In this work, we consider two modalities for each utterance: speech and text. Accordingly, each utterance is represented as
$\mathbf{u}_{ik} = \{\mathbf{s}_{ik}, \mathbf{t}_{ik}\}$, for $k = 1, \dots, N_i$,
where $\mathbf{s}_{ik}$ denotes the speech signal and $\mathbf{t}_{ik}$ denotes the corresponding textual content.
Each utterance $\mathbf{u}_{ik}$ is associated with an emotion label $y_{ik} \in \mathcal{Y}$, where $\mathcal{Y}$ represents the set of emotion categories defined for dataset $\mathcal{D}$.
The task of ERC is to predict the sequence of emotion labels
$\mathbf{y}_i = \{y_{i1}, y_{i2}, \dots, y_{iN_i}\}$
given the sequence of utterances
$\{\mathbf{u}_{i1}, \mathbf{u}_{i2}, \dots, \mathbf{u}_{iN_i}\}$
for each conversation $\mathbf{c}_i$. This formulation naturally corresponds to a sequence-to-sequence classification problem, where contextual dependencies across utterances play a critical role.

An example conversation from the MELD dataset~\cite{poria2019meld} is illustrated in Fig.~\ref{fig:erc}. The conversation consists of eight utterances exchanged between two speakers, with each utterance annotated with a corresponding emotion label. This example highlights the importance of modeling both conversational context and speaker interactions in ERC. The illustration is adapted from Poria et al.~\cite{poria2017review}.

\subsection{Hierarchical modeling for Multimodal ERC}
\begin{figure*}
    \centering
    \includegraphics[width=\textwidth,trim={0cm 3cm 0cm 2cm},clip]{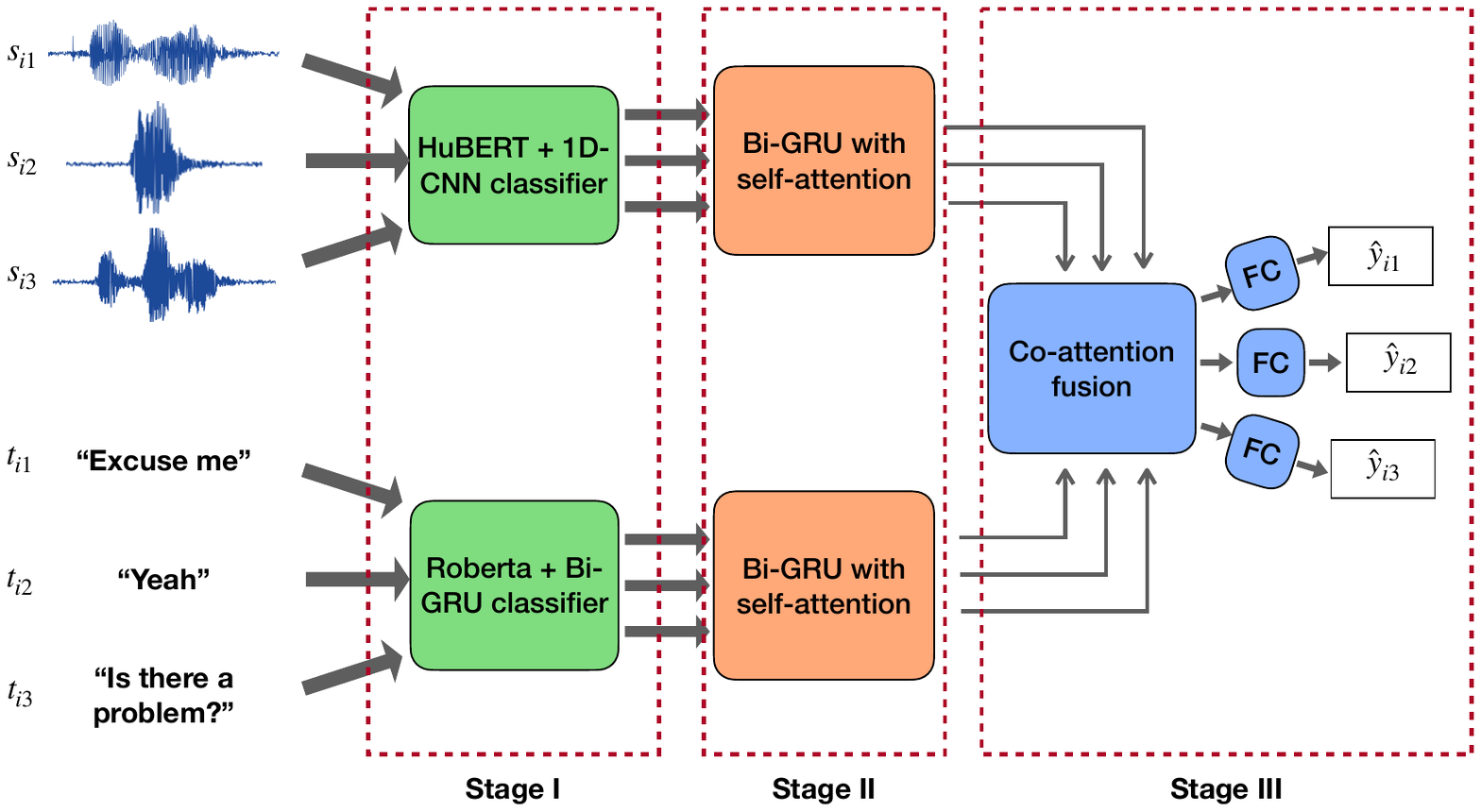}
    \caption{Block diagram of the proposed model. The three stages of training are also marked in the diagram.}
    \label{fig:hcam}
\end{figure*}

The block diagram of HCAM is shown in Fig. \ref{fig:hcam}. Our model has three distinct stages that are trained hierarchically, where each stage uses the pre-trained model parameters of the previous stage without fine-tuning.
\subsubsection{Background}
\textbf{Self and Cross Attention:} The self-attention mechanism, as proposed by Vaswani et. al~\cite{vaswani2017attention}, considers a sequence of length $N$, and of dimension of $d_k$. This sequence is converted into three matrices, $Q \in \mathcal{R}^{N \times d_k}$, $K \in \mathcal{R}^{N \times d_k}$ and $V\in \mathcal{R}^{N \times d_k}$. Self-attention is involved in computing the similarity between query representation (denoted by $Q$) with key representation (denoted by $K$). This similarity matrix, converted to a probability distribution by the softmax function, is then used to take a weighted sum of the value representations (denoted by $V$). Finally, the query matrix ($Q$) is added to this weighted sum followed by a layer normalization block.
    \begin{eqnarray}
    &Attention(Q, K, V) = softmax(\frac{QK^{T}}{\sqrt{d_k}})V\\\label{selfattn1}
    &Q^{'} = Q + Attention(Q, K, V)\\\label{selfattn2}
    &O = LayerNorm(Q^{'})\label{selfattn3}
\end{eqnarray}
\par
The cross-attention network is similar to the self-attention module with a key difference. Here, the query matrix and the key/value matrices are constructed from representations of different modalities. If we consider the query matrix from speech modality (denoted as $Q_S$), and the key/value matrices from the text modality (denoted as $K_T$/$V_T$), the cross-attention network operations are, 
\begin{eqnarray}
&Atten(Q_S, K_{T}, V_{T}) = softmax(\frac{Q_{A}K_{T}^{T}}{\sqrt{d_k}})V_{T}\label{crossatt1}\\
&Q_S^{'} = Q_S + Atten(Q_S, K_{T}, V_{T})\label{crossatt2}\\
&O_S = LayerNorm(Q_S^{'}).\label{crossatt3}
\end{eqnarray}

\subsubsection{Proposed Approach}\label{sec:hcammodel}
\noindent \textbf{HCAM Stage I:} In the first stage, we train utterance-level embedding extractors from speech and text. The models are trained to classify individual utterances without considering the inter-utterance conversational context. The models trained in this stage classify the individual utterances into the corresponding emotion classes based on cues present in either the speech signal or the text transcript. The pre-trained feature extractors are generally models with large computational requirements, and this constrains the number of utterances that can be processed in every iteration. The large number of utterances in a single conversation inhibits the fine-tuning of these feature extractors in previous works, as they process conversations as a whole. Our hierarchical modeling allows us to fine-tune the embeddings from the pre-trained feature extractors for improved emotion classification for each individual utterance in this training stage.
\begin{itemize}
    \item \textbf{Speech Embedding Extractor:} The speech is input to the HuBERT large model~\cite{hsu2021hubert}. Inspired by the strategy proposed in Pepino et. al.~\cite{pepino2021emotion}, we fine-tune the transformer layers in the HuBERT network while keeping the lower convolutional layers unchanged. The hidden layer outputs from the HuBERT model, for all the transformer layers, are summed at the frame-level, and passed through a 1-D CNN network. 
  Finally, the embeddings from the CNN network are average pooled over the utterance level to generate speech embeddings for the given utterance. The embeddings obtained at the output of the 1D-CNN network are considered for the contextual GRU layer.
  \item \textbf{Text Embedding Extractor:} The embedding extractor on the text data follows a similar architecture to that of the speech feature extraction. We obtain the word embeddings through a pre-trained RoBERTa model~\cite{liu2019roberta} and splice the last four hidden layer representations from the model. A bi-GRU layer allows the incorporation of intra-utterance context to generate utterance level embeddings.  Unlike the speech embedding  extraction, all the layers of the RoBERTa and bi-GRU model are fine-tuned. 
\end{itemize}

\begin{figure}
    \centering
    \includegraphics[width=0.7\columnwidth,trim={2cm 4cm 0cm 2cm},clip]{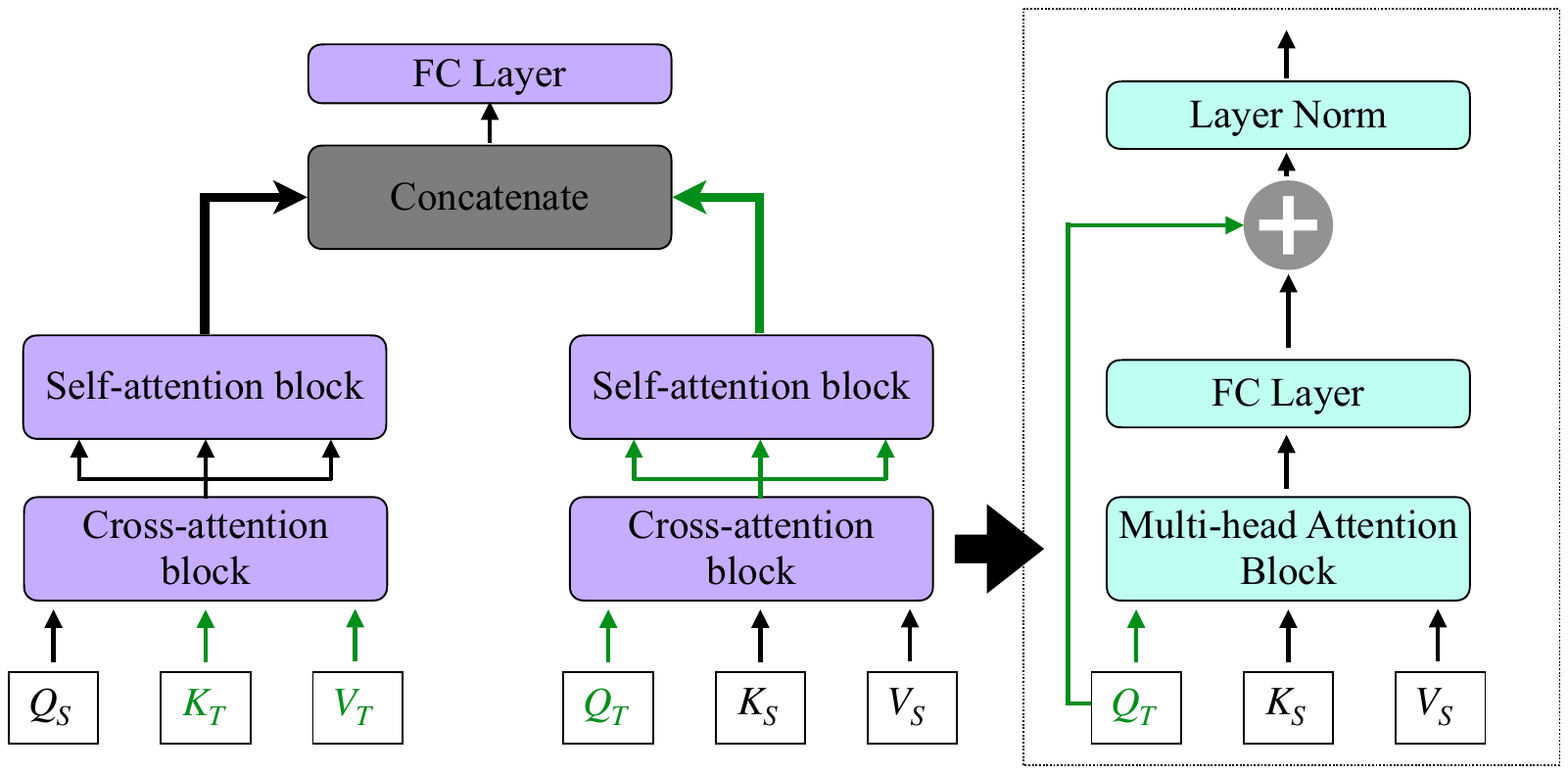}
    \caption{The co-attention network used in the proposed model. It consists of two sub-blocks - the cross-attention and the self-attention blocks.}
    \label{fig:attention}
\end{figure}

\noindent \textbf{HCAM Stage II:} The utterance-level embeddings obtained from the previous modeling stage are used in this stage, where we introduce inter-utterance context by means of a bidirectional gated recurrent unit (Bi-GRU) architecture with self-attention mechanism. The representations  extracted from models trained on each individual utterance in stage I are further enhanced with the conversational context information.
\begin{itemize}
    \item \textbf{Inter-utterance contextual GRU:} We propose a simple block, called the contextual-GRU, which takes into account the information from all utterances in the conversation.  
 While the bi-GRU itself can incorporate contextual information, it may not capture long term dependencies in the conversations that have a large number of utterances. For this reason, self-attention is used, which allows effective modeling of the long-term context. 
The output from the self-attention block is processed by a position wise feed forward layer with ReLU activation. 
\end{itemize}
\noindent \textbf{HCAM Stage III:} The third stage of the model consists of the effective fusion of the embeddings from the different modalities. 
The network architecture of the co-attention block, is shown in Fig. \ref{fig:attention}, and it consists of two sub-blocks, namely cross-attention and self-attention. 
In the co-attention network, we exploit the cross-attention between the modalities. 
There are two ways of performing the cross-attention between speech and text, one where the query representations are derived from the speech data, while the key and value representations are derived from the text data.
The other way is the reversal of the roles of the speech and text. The cross-modal embeddings at the output of the cross-attention blocks are further enriched by adding a self-attention block.  The two arms of cross-attention, as shown in Fig.~\ref{fig:attention}, are concatenated and this forms the multi-modal representation. 

The final representations, which combines the information from speech and text, is passed through a position-wise feed forward layer. 

\subsubsection{Loss Functions}
\noindent \textbf{Supervised Contrastive Loss:} We explore the supervised contrastive loss function, proposed by Khosla et.al~\cite{khosla2020supervised}, that encourages similarity between utterance representations from the same emotion class. Let us consider a mini-batch size of $\mathcal{B}$, where the utterance-level representations, derived from multiple conversations,  are denoted as $\{x_1, x_2, \dots, x_\mathcal{B}\}$. The features that appear in this loss formulation are normalized, that is, $||x_i|| = 1 \quad \forall i = \{1,2,\dots,\mathcal{B}\}$. We denote the corresponding labels as $\{y_1, y_2, \dots, y_\mathcal{B}\}$. Considering the sample $x_j$, the set of positive examples from the mini-batch is denoted by $P_j: \{i \in \mathcal{B} \quad s.t. \quad y_i = y_j\}$. The supervised contrastive loss is,
\begin{equation}\label{supconeq}
    L^{sup-con} = \sum_{j\in \mathcal{B}} \frac{-1}{|P_j|}\sum_{p \in P_j} log \frac{exp(x_j^T x_p/\tau)}{\sum \limits_{a\in \mathcal{B}}exp(x_j^T x_a/\tau)},
\end{equation}
where, $\tau$ is a hyper-parameter indicating the temperature of this loss.
Generally, contrastive losses have been used in representation learning tasks. The dependence of the contrastive loss on the ground truth labels enables one to use these losses in supervised settings too. In addition to the cross-entropy loss, this loss makes the system focus on the hard-to-classify samples. However, unlike in representation learning tasks, the datasets for ERC are not large enough to be able to classify the utterances by means of a contrastive loss alone. We therefore, use the supervised contrastive loss in combination with the cross-entropy loss for training our emotion recognition modules.\\
\noindent \textbf{Combined loss:} We use a convex combination of the the cross entropy loss and the supervised contrastive loss. The final loss, used to train all the stages of our model, is given by,
\begin{equation}\label{lossfinal}
    L^{tot} = \beta CE(y, \hat{y}) + (1-\beta)L^{sup-con},
\end{equation}
where $\beta$ is a hyper-parameter in the range of $[0,1]$. Here, $y$ and $\hat{y}$ denote the true label and the probability predicted by the model respectively.

\subsubsection{Experiments and Results}\label{sec:hcam_down}
\noindent \textbf{Datasets:} We evaluate our work on three widely used datasets, IEMOCAP~\cite{busso2008iemocap}, MELD~\cite{poria2019meld} and CMU-MOSI~\cite{zadeh2016mosi}. The details of these datasets are covered in Sec.~\ref{sec:setting_datasets}.\\
\noindent \textbf{Implementation Details:} The models are trained with Adam optimizer using a learning rate of $1e-5$ and a batch size of $32$ in MELD and IEMOCAP dataset, while the batch size is reduced to $8$ for the CMU-MOSI dataset. All the experiments reported in this work use $5$ random weight initialization choices. The mean performance using the random initializations are reported in all the experiments below. We run the different stages of our model for $100$ epochs as we found the validation performance to saturate within this limit. We also employ gradient clipping with a L2 norm of $0.25$ in all our implementation.

\begin{table}[t!]
\centering
\caption{Results on the different datasets in terms of weighted F1-score.}\label{tab:hcam_results}
\resizebox{0.7\columnwidth}{!}{%
\begin{tabular}{@{}l|cc|c|c@{}}
\toprule
\multirow{2}{*}{Modality (Training Stage)} & \multicolumn{2}{c|}{IEMOCAP} & \multirow{2}{*}{MELD} & \multirow{2}{*}{CMU-MOSI} \\ \cmidrule(lr){2-3}
 & \multicolumn{1}{c|}{$4$ way} & $6$ way &  &  \\ \midrule
Speech (Stage I) & \multicolumn{1}{c|}{$64.3\%$} & $55.6\%$ & $48.2\%$ & $64.4\%$ \\
Speech (Stage II) & \multicolumn{1}{c|}{$78.7\%$} & $65.7\%$ & $50.1\%$ & $67.4\%$ \\ \midrule
Text (Stage I) & \multicolumn{1}{c|}{$68.4\%$} & $53.8\%$ & $63.3\%$ & $84.3\%$ \\
Text (Stage II) & \multicolumn{1}{c|}{$81.4\%$} & $64.4\%$ & $65.6\%$ & $85.4\%$ \\ \midrule
Speech + Text (Stage III) & \multicolumn{1}{c|}{$85.9\%$} & $70.5\%$ & $65.8\%$ & $85.8\%$ \\ \bottomrule
\end{tabular}}
\end{table}
\noindent \textbf{Results:} We report the performance of the proposed model for each individual modality followed by the performance of the model on the multi-modal setting.  The key results on the three datasets are shown in Table \ref{tab:hcam_results}. The following are the observations from these results, 
\begin{enumerate}
    \item In the IEMOCAP dataset, the speech and the text modalities perform relatively similarly, while in the MELD and CMU-MOSI datasets, the text transcripts emerge to be the stronger modality.
    \item The context addition framework proves to be effective for all the three datasets. The proposed contextual GRU architecture with self-attention, though a simple architecture, leads to an improvement for both the modalities. For speech, the relative improvement over the stage I performance is $40.3\%$, $22.7\%$, $3.7\%$ and $8.4\%$ for IEMOCAP 4-way, IEMOCAP 6-way, MELD and CMU-MOSI datasets, respectively. Similarly, for the textual modality, we notice a relative improvement of $41.1\%$, $22.9\%$, $6.3\%$ and $7\%$ respectively. IEMOCAP has the largest number of utterances per conversation among the three datasets while MELD has the lowest. As the number of utterances increase in the datasets, the contextual modeling becomes more effective
\end{enumerate}
\noindent \textbf{Impact of Hierarchical modeling:} In order to understand the advantages of our hierarchical modeling approach, we modify our training paradigm where we combine stages II  and III of training.  The results for these modifications in the training paradigm are shown in Table \ref{hierarchy}. We note that, for all the four dataset settings, we achieve a better performance with the proposed hierarchical modeling with no change in the model architecture. These experiments highlight the benefits of a curriculum style design of the stages proposed in the HCAM framework.
\begin{table}[t!]
\caption{Weighted F1 score of our system with different training paradigms}\label{hierarchy}
\begin{center}
\resizebox{0.6\columnwidth}{!}{
    \begin{tabular}{@{}c|cc|c|c@{}}
\toprule
\multirow{2}{*}{\begin{tabular}[c]{@{}l@{}}Training Paradigm\end{tabular}} & \multicolumn{2}{c|}{IEMOCAP} & \multirow{2}{*}{MELD} & \multirow{2}{*}{CMU-MOSI} \\ \cmidrule(lr){2-3}
 & \multicolumn{1}{c|}{4 way} & 6 way &  &  \\ \midrule
Hierarchical & \multicolumn{1}{c|}{$85.9\%$} & $70.5\%$ & $65.8\%$ & $85.8\%$ \\
Non-hierarchical & \multicolumn{1}{c|}{$82.3\%$} & $68.5\%$ & $65\%$ & $84.4\%$ \\ \bottomrule
\end{tabular}}
\end{center}
\end{table}

\subsubsection{Summary and Limitations}
The hierarchical cross-attention model (HCAM) provides a structured framework for multimodal emotion recognition in conversations by decoupling unimodal representation learning, contextual modeling, and multimodal fusion. By introducing conversational context in a staged manner and employing cross-attention mechanisms for speech-text interaction, HCAM effectively captures both temporal dependencies and cross-modal correlations.

However, this architecture also exposes an important limitation. When there exists a significant performance disparity between modalities—such as when either speech or text is substantially more informative—the fused representation does not consistently outperform the strongest unimodal system (see performance of MELD and CMU-MOSI in Table~\ref{tab:hcam_results}). In such cases, multimodal fusion may fail to provide complementary gains and can even dilute modality-specific strengths.

This observation motivates the need for a more flexible fusion strategy that enables dynamic modality selection rather than enforcing uniform feature-level integration. In the next subsection, we address this limitation by introducing a modular mixture-of-experts framework that performs fusion at the decision level, allowing the model to adaptively leverage unimodal and multimodal experts based on conversational context.
\begin{figure*}[t!]
    \centering
\includegraphics[width=\linewidth, trim={0cm 2cm 0.5cm 2cm},clip]{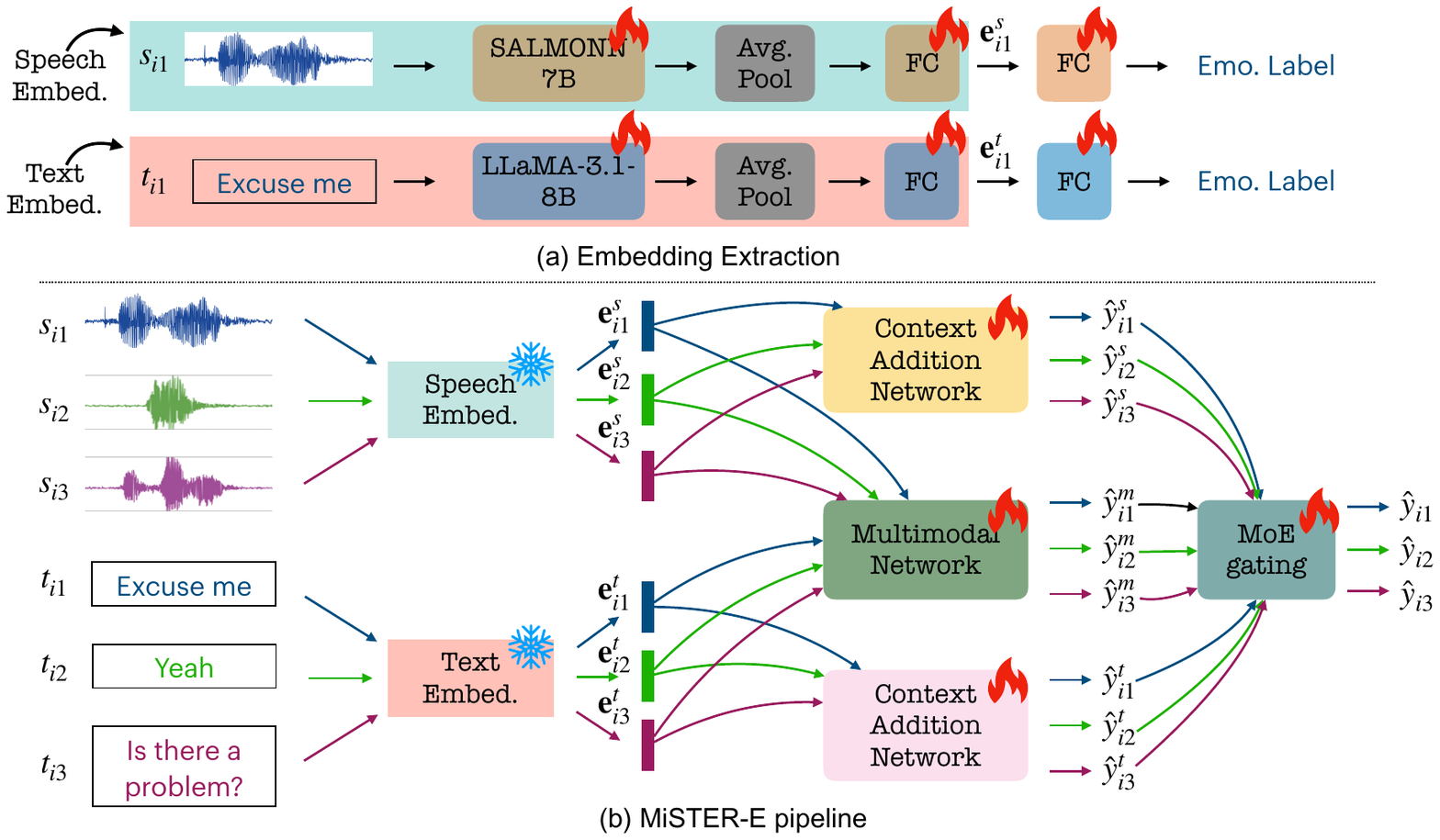}

    \caption{(a) Training of the unimodal feature extraction module (b) The entire pipeline of MiSTER-E. The speech and text embedding modules are frozen during training of the rest of the pipeline. Two context addition networks are trained for the two modalities along with a multimodal  network. Finally, a mixture of experts gating network is trained to predict the emotion category for each utterance.}
    \label{fig:model}

\end{figure*}
\subsection{Mixture-of-Experts modeling for ERC}
In this section, we introduce MiSTER-E~\cite{DUTTA2026101965}, a modular architecture that enforces a combination of experts at the logit (decision) level rather than at the feature level.    MiSTER-E adopts a Mixture-of-Experts (MoE) formulation with three independently optimized  branches: a speech-only contextual expert, a text-only contextual expert, and a multimodal expert. The uni-modal systems capture modality-specific conversational dynamics using temporal inception networks followed by  a recurrent layer, while the multimodal expert fuses speech and text features via cross-attention and self-attention layers. The outputs from these experts are combined using a  gating mechanism, which computes a weighted sum before the softmax normalization. This logit-level fusion enables dynamic expert selection, allowing the model to  combine complementary multimodal cues.
    To further stabilize expert specialization, we incorporate contrastive and KL-based consistency losses, while supervising each expert with focal loss to address class imbalance.
    In particular, when either modality (speech or text) has a dominant advantage over the other modality (text or speech), our modality decoupling followed by MoE based combination is shown to have a unique fusion advantage compared to other prior works. \\
A block diagram of our proposed method is shown in Fig.~\ref{fig:model}. Further, a summary of the different notations used for the problem formulation (Sec.~\ref{sec:erc_prob}) and model description is provided in Table~\ref{tab:notation}.
\begin{table}[t]
\centering
\resizebox{\textwidth}{!}{%
\begin{tabular}{l l}
\toprule
\textbf{\textcolor{black}{Symbol}} & \textbf{\textcolor{black}{Description}} \\
\midrule
\textcolor{black}{$\mathcal{D}$} & \textcolor{black}{Emotion Recognition in Conversations (ERC) dataset} \\
\textcolor{black}{$P$} & \textcolor{black}{Number of conversations in the dataset} \\
\textcolor{black}{$\mathcal{C}=\{\mathbf{c}_1,\mathbf{c}_2,\dots,\mathbf{c}_P\}$} & \textcolor{black}{Set of all conversations in $\mathcal{D}$} \\
\textcolor{black}{$\mathbf{c}_i$} & \textcolor{black}{The $i$-th conversation} \\
\textcolor{black}{$\mathcal{U}_i=(\mathbf{u}_{i1},\mathbf{u}_{i2},\dots,\mathbf{u}_{iN_i})$} & \textcolor{black}{Sequence of utterances in conversation $\mathbf{c}_i$} \\
\textcolor{black}{$N_i$ }& \textcolor{black}{Number of utterances in conversation $\mathbf{c}_i$} \\
\textcolor{black}{$\mathbf{u}_{ik}$} & \textcolor{black}{The $k$-th utterance in conversation $\mathbf{c}_i$} \\
\textcolor{black}{$\mathbf{s}_{ik}$ }& \textcolor{black}{Speech modality of utterance $\mathbf{u}_{ik}$} \\
\textcolor{black}{$\mathbf{t}_{ik}$ }& \textcolor{black}{Text modality of utterance $\mathbf{u}_{ik}$} \\ 
\textcolor{black}{$y_{ik}$ }& \textcolor{black}{Emotion label corresponding to utterance $\mathbf{u}_{ik}$} \\
\textcolor{black}{$\mathcal{Y}$} & \textcolor{black}{Set of emotion categories in the dataset} \\
\textcolor{black}{$\mathbf{y}_i=(y_{i1},y_{i2},\dots,y_{iN_i})$ }& \textcolor{black}{Sequence of emotion labels for conversation $\mathbf{c}_i$} \\
\midrule
\textcolor{black}{$\mathbf{e}_{ik}^s,\mathbf{e}_{ik}^t$} & \textcolor{black}{Speech and text embeddings for utterance $\mathbf{u}_{ik}$} \\
\textcolor{black}{$\hat{\mathbf{y}}_{ik}^s,\hat{\mathbf{y}}_{ik}^t,\hat{\mathbf{y}}_{ik}^m$ }& \textcolor{black}{Pre-softmax logits from speech, text, and multimodal experts} \\
\textcolor{black}{$\hat{\mathbf{y}}_{ik}$ }& \textcolor{black}{Final fused logit after mixture-of-experts gating} \\
\textcolor{black}{$\boldsymbol{\beta}_{ik}$ }& \textcolor{black}{Mixture-of-experts  weights for utterance $\mathbf{u}_{ik}$ }\\
\textcolor{black}{$\hat{\mathbf{p}}_{ik}^s,\hat{\mathbf{p}}_{ik}^t,\hat{\mathbf{p}}_{ik}^m$} & \textcolor{black}{Class probability distributions for speech, text and multimodal experts }\\
\midrule
\textcolor{black}{$\mathcal{L}_{\text{CAN}}^i$ }& \textcolor{black}{Focal loss for unimodal context addition networks} \\
\textcolor{black}{$\mathcal{L}_{\text{multi}}^i$ }& \textcolor{black}{Multimodal classification and contrastive loss} \\
\textcolor{black}{$\mathcal{L}_{\text{moe}}^i$ }& \textcolor{black}{MoE gating loss with KL-divergence consistency regularization }\\
\textcolor{black}{$\mathcal{L}_{\text{tot}}$ }& \textcolor{black}{Overall training objective} \\
\bottomrule
\end{tabular}}
\caption{\textcolor{black}{Notations used in the problem formulation and model description.}}
\label{tab:notation}
\end{table}
\subsubsection{Proposed Approach}\label{sec:mistere}
\noindent \textbf{Unimodal Feature Extraction:} While large language models (LLMs) excel in text generation, their use in emotion recognition has been limited. Prior work typically uses them in autoregressive mode or as frozen feature extractors~\citep{behnamghaderllm2vec}.\\
In this work, we fine-tune an LLM—specifically, \texttt{LLaMA-3.1-8B}—to act as a text encoder rather than a generator. Each utterance transcript $\mathbf{t}_{ik}$ is tokenized and processed through the LLM. Token-level hidden states are then pooled and passed through a two-layer feedforward classifier trained with task-specific supervision. To preserve the pretraining knowledge while enabling efficient adaptation, we apply LoRA~\citep{hu2022lora} to fine-tune the weights.
We denote the resulting text embedding as:
\begin{equation}
    \mathbf{e}_{ik}^{t} = \texttt{Text-Embed}(\textbf{t}_{ik})
\end{equation}
where $\mathbf{e}_{ik}^t$ is extracted from the first fully connected layer of the classifier. \\
For speech, we adopt a similar approach using \texttt{SALMONN\allowbreak-7B}~\citep{tang2023salmonn}, a speech large language model (SLLM) comprising a speech encoder, Q-former~\citep{li2023blip}, and an LLM backbone.
For ERC, we fine-tune this model by updating the Q-former, the LLM, and a classification head with LoRA. This allows the system to learn emotionally salient acoustic patterns while retaining the semantic features of the LLM backbone. Given a speech signal $\mathbf{s}_{ik}$, we extract its representation as:
\begin{equation}
    \mathbf{e}_{ik}^{s} = \texttt{Speech-Embed}(\textbf{s}_{ik})
\end{equation}
where the embedding is taken from the first fully connected layer of the speech classification head. \\
\noindent \textbf{Conversational Modeling:} For a conversation $\textbf{c}_i$, the text and speech embedding sequences are denoted by $\mathbf{E^t_i}=(\mathbf{e}_{i1}^{t},\dots,\mathbf{e}_{iN_i}^{t})$ and $\mathbf{E^s_i}=(\mathbf{e}_{i1}^{s},\dots,\mathbf{e}_{iN_i}^{s})$ respectively.
\begin{figure*}[t!]
    \centering
\includegraphics[width=0.75\textwidth,trim={2cm 2.3cm 1cm 3cm},clip]{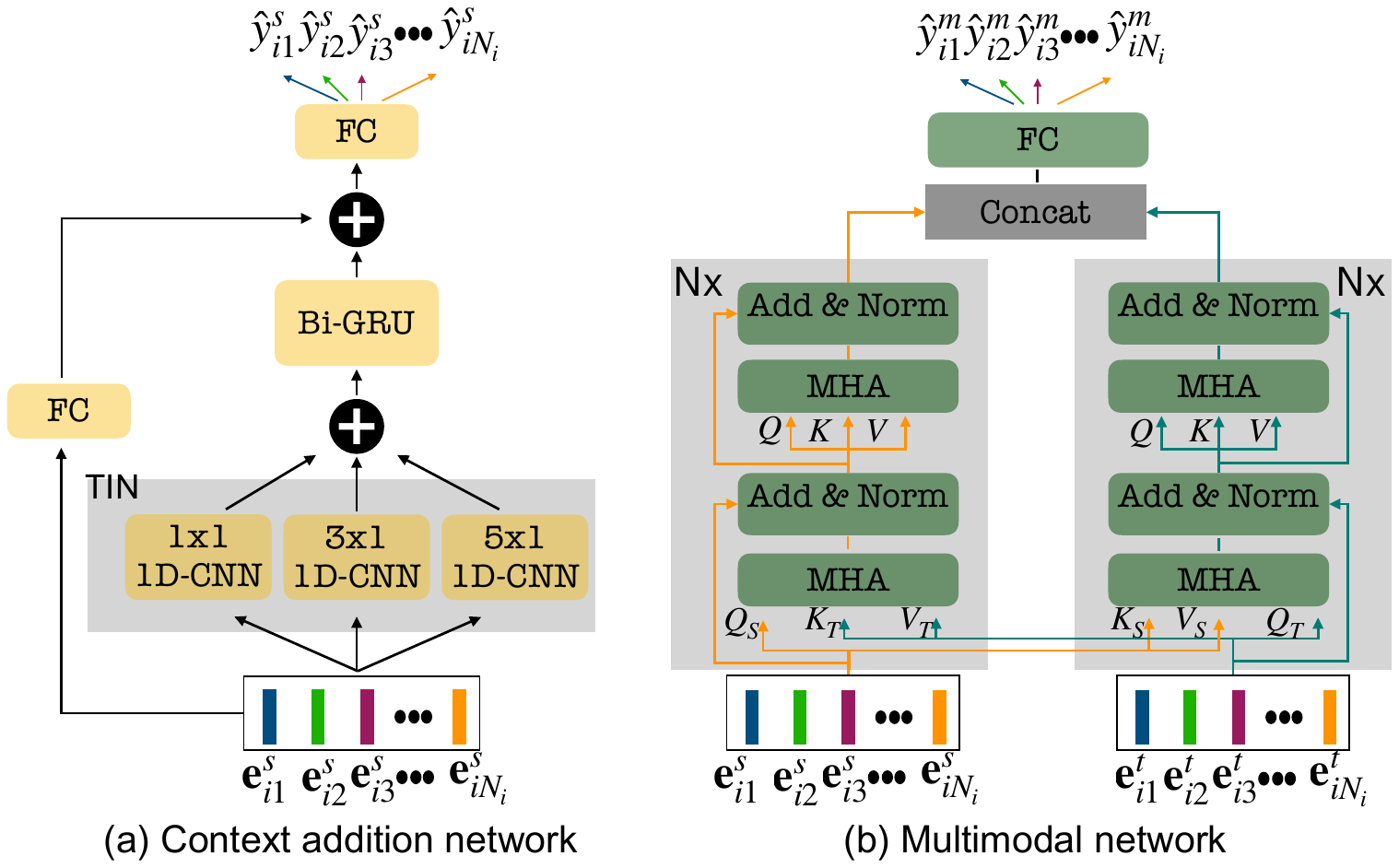}

    \caption{(a) The context addition network (for the speech modality) and (b) the multimodal network used in MiSTER-E. The inputs to both the blocks are derived from the uni-modal feature extractor modules. \texttt{TIN} stands for Temporal Inception Network, \texttt{MHA} stands for multi-head attention.}
    \label{fig:subparts}
\end{figure*}
As shown in Figure~\ref{fig:model}, the fine-tuned speech and text embedding extractors are frozen. We now discuss the blocks used for modeling the conversational context in ERC.
\begin{itemize}
    \item \textbf{Context Addition Network:} To make utterance representations context-aware, we introduce a Context Addition Network (\texttt{CAN}) that enhances both text and speech embeddings with conversational context. At its core is a Temporal Inception Network (\texttt{TIN}), which applies 1D convolutions with kernel sizes of $1$, $3$, and $5$ to capture short-range dependencies—simulating varying receptive fields across local utterance neighbourhoods. \textcolor{black}{Inspired by the Inception architecture~\citep{szegedy2015going} originally developed for image classification, TIN enables multi-scale temporal context modeling within conversations.}\\
However, emotional signals in dialogue often evolve over longer durations. To capture such global dependencies, we append a Bi-GRU layer, allowing the model to integrate information across the entire conversational span.
A residual connection links the original embedding with its context-enhanced version, enabling additive refinement while preserving the semantic grounding of the base LLM features.
 Finally, a FC layer is used to map each utterance $\textbf{u}_{ik}$ to its corresponding emotion category $y_{ik}$.
 We train two similar networks for the speech and text modalities, respectively. These operations are denoted as:
\begin{eqnarray}\label{eq:can_s}
    \mathbf{\hat{y}^s_i} = (\hat{y}_{i1}^s, \dots,\hat{y}_{iN_i}^s)=\texttt{CAN}(\mathbf{E_i^s})\\ 
    \mathbf{\hat{y}^t_i} =(\hat{y}_{i1}^t, \dots,\hat{y}_{iN_i}^t)=\texttt{CAN}(\mathbf{E_i^t})
\end{eqnarray}
A schematic of the speech-side \texttt{CAN} is shown in Fig.~\ref{fig:subparts}(a).

\item \textbf{Multimodal Network:} To integrate information across modalities, we design a fusion module based on cross-attention between speech and text. This mechanism allows the model to align semantic cues from text with affective signals from audio.
The speech and text embeddings, $\mathbf{E_i^s}$ and $\mathbf{E_i^t}$, are first projected into query, key, and value spaces. Cross-attention is then applied bidirectionally:
\begin{eqnarray}
    \mathbf{M^{t\rightarrow s}_i} = \texttt{LN}(\texttt{FC}(\mathbf{E_i^s})+\texttt{MHA}(\mathbf{Q_i^s},\mathbf{K_i^t},\mathbf{V_i^t})) \\
    \mathbf{M^{s\rightarrow t}_i} = \texttt{LN}(\texttt{FC}(\mathbf{E_i^t})+\texttt{MHA}(\mathbf{Q_i^t},\mathbf{K_i^s},\mathbf{V_i^s}))
\end{eqnarray}
Here, \texttt{MHA} and \texttt{LN} refer to multi-head attention and layer normalization, respectively.\\
While this enables inter-modal alignment, it overlooks cues from the conversational context that are essential for ERC. To address this, we apply modality-specific self-attention layers over the aligned representations, capturing temporal context across the conversation: 
\begin{eqnarray}\label{eq:multspeech}
    \mathbf{M^{s}_i} = \texttt{Self-attn.}( \mathbf{M^{t\rightarrow s}_i}) \\ 
 \mathbf{M^{t}_i} = \texttt{Self-attn.}(\mathbf{M^{s\rightarrow t}_i})
\end{eqnarray}
The outputs $\mathbf{M^{s}_i}$ and $\mathbf{M^{t}_i}$ are concatenated and passed through a fully connected (FC) layer:
\begin{equation}\label{eq:mult}
    \mathbf{\hat{y}^m_i} =(\hat{y}^m_{i1},\hat{y}^m_{i2}, \dots,\hat{y}^m_{iN_i})=\texttt{FC}([\mathbf{M^{s}_i};\mathbf{M^{t}_i}])
\end{equation}
An overview is presented in Fig.~\ref{fig:subparts}(b).
\end{itemize}

\noindent \textbf{Mixture-of-Experts Gating:} \textcolor{black}{We obtain pre-softmax logits from the three experts for each utterance $u_{ik}$:
the speech expert ($\mathbf{\hat{y}^s_{ik}}$), the text expert ($\mathbf{\hat{y}^t_{ik}}$),
and the multimodal expert ($\mathbf{\hat{y}^m_{ik}}$).} \\
 \textcolor{black}{To combine the logits, we dynamically fuse the experts' logits using a gating mechanism that learns to weigh them.  To compute these weights, we concatenate the logits and feed them into a fully connected (FC) layer:}
\begin{equation}\label{eq:moegate}
\textcolor{black}{\mathbf{g}_{ik} = \texttt{FC}([\mathbf{\hat{y}^s_{ik}}; \mathbf{\hat{y}^t_{ik}}; \mathbf{\hat{y}^m_{ik}}])}
\end{equation}
\textcolor{black}{The gating network outputs $\mathbf{g}_{ik} \in \mathbb{R}^{3}$, which is transformed via a softmax operation to produce adaptive mixture weights ${\boldsymbol \beta}_{ik}$.}

\textcolor{black}{The final logit is computed as a weighted sum of the expert predictions:}

\begin{equation}\label{eq:final_pred}
\textcolor{black}{\mathbf{\hat{y}_{ik}} = \beta_{ik}^s \cdot \mathbf{\hat{y}^s_{ik}} + \beta_{ik}^t \cdot \mathbf{\hat{y}^t_{ik}} + \beta_{ik}^m \cdot \mathbf{\hat{y}^m_{ik}}}
\end{equation} 
 
 \textcolor{black}{A softmax operation is applied to $\mathbf{\hat{y}_{ik}}$ to obtain the final predictive distribution for utterance $u_{ik}$.
This gating network is trained end-to-end, empowering the model to flexibly integrate modalities by emphasizing the most reliable expert.}\\
\subsubsection{Model Training}
\noindent \textbf{Loss Function:} ERC is typically characterized by severe class imbalance, where  rare emotion classes are often misclassified~\citep{poria2019emotion}. To address this, we employ the focal loss~\citep{lin2017focal}, which modulates the contribution of each training example based on its complexity,  reducing the relative loss for well-classified examples while focusing on hard (possibly minority class) instances.
In MiSTER-E, this loss is applied during the training of the uni-modal embedding extractors  for text and speech as well as their respective context addition networks (CANs). The loss for the speech and text CAN networks is given by:
\begin{equation}
\mathcal{L}_{\text{CAN}}^i = \sum_{k=1}^{N_i} \texttt{FL}(\hat{y}_{ik}^s, y_{ik}) + \sum_{k=1}^{N_i} \texttt{FL}(\hat{y}_{ik}^t, y_{ik})
\end{equation}
Here, $\texttt{FL}(\cdot)$ represents the focal loss function.
Denoting the predicted probability for the sample with logits $\hat{y}_{ik}^\cdot$ as $\hat{p}_{ik}^\cdot$, the focal loss   is given as,
\begin{equation}\label{eq:focal}
\texttt{FL}(\hat{y}_{ik}^\cdot, y_{ik}) = -\sum_{c=1}^{\mathcal{Y}}y_{ik}^c(1-\hat{p}_{ik}^{\cdot c})^\gamma \log (\hat{p}_{ik}^{\cdot c})
\end{equation}
where $y_{ik}^c=1$ for the true class and $0$ otherwise.  $\hat{p}_{ik}^{\cdot c}$ is the predicted probability for class $c$ for the sample with logits $\hat{y}_{ik}^{\cdot}$ ($\cdot$ is replaced by $s$, $t$ or $m$) and $\gamma$ is the hyperparameter associated with the focal loss.\\
\noindent \textbf{Multimodal Contrastive Loss:} In many cases, speech and text provide complementary signals about a speaker’s emotion. To exploit this, we incorporate a supervised contrastive loss that structures the joint representation space based on emotion labels, drawing together utterances with shared emotional intent across modalities.  
Consider the multimodal speech and text representations denoted by $\mathbf{M_i^s}=\{\mathbf{m_{i1}^s}, \mathbf{m_{i2}^s}, \mathbf{m_{i3}^s},\dots,\mathbf{m_{iN_i}^s}\}$ and $\mathbf{M_i^t}=\{\mathbf{m_{i1}^t}, \mathbf{m_{i2}^t}, \mathbf{m_{i3}^t},\dots,\mathbf{m_{iN_i}^t}\}$ respectively. These embeddings  are batched to get, 
\begin{equation}
    \mathbf{Z_i}=\{\mathbf{\tilde{m}_{i1}^s},\dots,\mathbf{\tilde{m}_{iN_i}^s},\mathbf{\tilde{m}_{i1}^t},\dots,\mathbf{\tilde{m}_{iN_i}^t}\}
\end{equation}
where \textcolor{black}{$\mathbf{\tilde{m}_{ik}^\cdot}=\frac{\mathbf{m_{ik}^\cdot}}{||\mathbf{m_{ik}^\cdot}||}$} denotes the normalized embeddings. 
Let $\mathbf{z}_a \in \mathbf{Z_i}$ for $a \in \{1,2,\dots,2N_i\}$, and let $y_a$ be the emotion label associated with $\mathbf{z}_a$.
The contrastive loss   is given by:
\begin{equation}\label{eq:con}
\mathcal{L}^\text{i}_{\text{con}} =
  \sum_{a=1}^{2N_i} \frac{-1}{|P(a)|}
\sum_{p \in P(a)} \log \frac{\exp\left(\mathbf{z}_a^\top \mathbf{z}_p / \tau\right)}{\sum\limits_{\substack{q=1 \\ \ q \ne a}}^{2N_i} \exp\left(\mathbf{z}_a^\top \mathbf{z}_q / \tau\right)}
\end{equation}
where $\tau$ is the temperature and $P(a)=\{p \in \{1,2,\dots,2N_i\} \backslash \{a\}|y_p=y_a\}$ is the set of positives for anchor $\mathbf{z}_a$. This objective pulls together utterances with the same emotion class across both speech and text, while pushing apart other samples, thereby guiding the model to learn emotionally coherent, modality-invariant representations.\\
\noindent \textbf{Total multimodal loss:} The final loss used to train the multimodal model combines the classification and contrastive objectives:
\begin{equation}\label{eq:tot_mult}
\mathcal{L}_{\text{multi}}^i = \sum_{k=1}^{N_i} \texttt{FL}(\hat{y}_{ik}^m, y_{ik}) + \lambda \mathcal{L}_{\text{con}}^i
\end{equation}
where $\hat{y}_{ik}^m$ denotes the multimodal prediction, and $\lambda$ controls the contribution of the contrastive loss. \\
\noindent \textbf{MoE Gating Loss:} To train the mixture-of-experts (MoE) gating network, we combine two objectives: (i) focal loss for emotion classification, and (ii) a regularization term to promote consistency among the expert predictions. Specifically, we enforce similarity in the predicted distributions of the three experts—speech-only, text-only, and multimodal—using the Kullback-Leibler (KL) divergence.
The total loss for the MoE layer is:
\begin{equation}\label{eq:kl}
\mathcal{L}_{\text{moe}}^i = \sum_{k=1}^{N_i} \texttt{FL}(\hat{y}_{ik}, y_{ik}) + \alpha \cdot \mathcal{L}_{\text{KL}}^i
\end{equation}
where $\hat{y}_{ik}$ is defined in Eq.~\ref{eq:final_pred}, $\alpha$ controls the strength of the consistency regularization, and the KL term is given by:
\begin{equation}
\textcolor{black}{\mathcal{L}_{\text{KL}}^i = \sum_{k=1}^{N_i} \Big[ \texttt{KL}(\hat{p}_{ik}^m || \hat{p}_{ik}^s) +
\texttt{KL}(\hat{p}_{ik}^m || \hat{p}_{ik}^t) \Big]}
\end{equation}
\textcolor{black}{Here, $\hat{p}_{ik}^{\cdot}$ denotes the class probability distribution obtained by applying a softmax function to the corresponding expert logits $\hat{y}_{ik}^{\cdot}$. The KL consistency loss encourages the unimodal speech and text experts to produce predictive distributions that are aligned with the multimodal expert. Since the multimodal expert jointly models both modalities, we treat its output distribution as a stronger supervisory signal and therefore, employ a directional KL divergence in this manner.
}\\
\noindent \textbf{Total Loss:} The context addition networks, the multimodal network, and the MoE gating layer are trained together. The total loss is:
\begin{equation}\label{eq:all}
 \mathcal{L}_{\text{tot}} = \sum_{i=1}^{P} \bigg[
\mathcal{L}_{\text{CAN}}^i + \mathcal{L}_{\text{moe}}^i \\
+ \mathcal{L}_{\text{multi}}^i \bigg]
\end{equation}

\subsubsection{Experiments}\label{sec:misterd_down}
\noindent \textbf{Datasets:} As in the previous approach, we conduct experiments on IEMOCAP, MELD and CMU-MOSI. For this experiment, we only evaluate the $6$-class problem in IEMOCAP as that has become more widely used in recent work.\\
\noindent \textbf{Implementation Details:} The two unimodal feature extractors are trained using LoRA with a rank of $8$ and the scaling parameter of $32$ with a dropout of $0.1$. Both models are trained with a batch size of $8$ and a learning rate of $1e-5$, with the focal loss. The hidden dimension in the FC (Sec.~\ref{sec:mistere}) is set to $2048$. For the rest of the MiSTER-E pipeline, we use a batch size of $8$ for IEMOCAP, MOSI and $32$ for MELD. The learning rate is set to $1e-5$ for all datasets. For MELD and MOSI, $\lambda$ (Eq.~\ref{eq:tot_mult}) is set to $1$ , while for IEMOCAP, it is set to $2$. The consistency regularization term,
$\alpha$ (Eq.~\ref{eq:kl}), is set to $0.1$ for IEMOCAP and MOSI, while it is kept at $1e-3$ for MELD (Sec.~\ref{sec:mistere}).
We report the weighted F1-score on the test data as the performance metric with $3$ random initializations. \\
All our experiments are run using a NVIDIA RTX A6000  GPU card with Pytorch 2.7.0 and CUDA 12.6. The number of trainable parameters in our model amounts to $97$M ($8$M each for the LLM/SLLM LoRA training and $81$M for the conversational modeling). Training the LLM takes about $10$ minutes per epoch, while the SLLM takes about $20$ minutes per epoch. The rest of the model, following feature encoders takes approximately $10$ minutes for $100$ epochs. We mention further hyperparameter choices for both datasets below:
\begin{itemize}
    \item The BiGRU used in the \texttt{CAN} network (Sec.~\ref{sec:mistere}) has a hidden dimension of $512$ and we use $3$ layers for IEMOCAP, MOSI, and $2$ for MELD. We use a dropout of $0.2$ between each fully connected layer for regularization.
    \item For the multimodal fusion network, we use $4$ layers with hidden dimension of $120$ and $4$ attention heads. We use a dropout regularization of $0.5$ in the multimodal fusion network. A fusion network with the same parameters is used for both the datasets. The temperature parameter for the contrastive loss, used for the fusion network (Eq.~\ref{eq:con}), is kept at $1$ for IEMOCAP, MOSI, and $0.05$ for MELD.
    \item While training the context addition networks, the multimodal network, and the MoE gating network, we use gradient clipping with norm $1.0$ for both the datasets and train for $100$ epochs. While training the LLM and the SLLM feature encoders, we run for $50$ epochs for both datasets \footnote{Code for MiSTER-E is available at \url{https://github.com/iiscleap/MiSTER-E}.}.
\end{itemize}
\noindent \textbf{Comparison with prior work:} Table~\ref{tab:method_comparison} presents a comparison of the proposed approach with several baseline methods. The competing approaches span a broad spectrum of modeling paradigms, including graph-based frameworks~\cite{ai2024gcn,shou2024efficient}, a state-space formulation~\cite{shou2024revisiting}, and a mixture-of-experts (MoE) based model~\cite{zhang2025moe}. As observed, the proposed method establishes new state-of-the-art performance on all the three datasets, attaining a weighted F1 score of $\mathbf{70.9}\%$ on IEMOCAP, $\mathbf{69.5}\%$ on MELD, \textcolor{black}{and $\mathbf{87.9\%}$ on CMU-MOSI}.\\
It is also pertinent to highlight that, for IEMOCAP, the absence of a standardized validation set has often resulted in prior studies performing model selection directly on the test set. Such a practice can artificially inflate performance metrics and limit reproducibility. To mitigate this issue, all experiments in this work (for which official implementations were available) were conducted using our own train, validation, and test splits. Consequently, certain IEMOCAP results reported in Table~\ref{tab:method_comparison} may differ from those originally published.

\begin{figure*}[t!]
    \centering
    \begin{subfigure}[t]{0.48\textwidth}
        \centering
        \includegraphics[width=\linewidth,trim={1cm 6cm 13cm 2cm},clip]{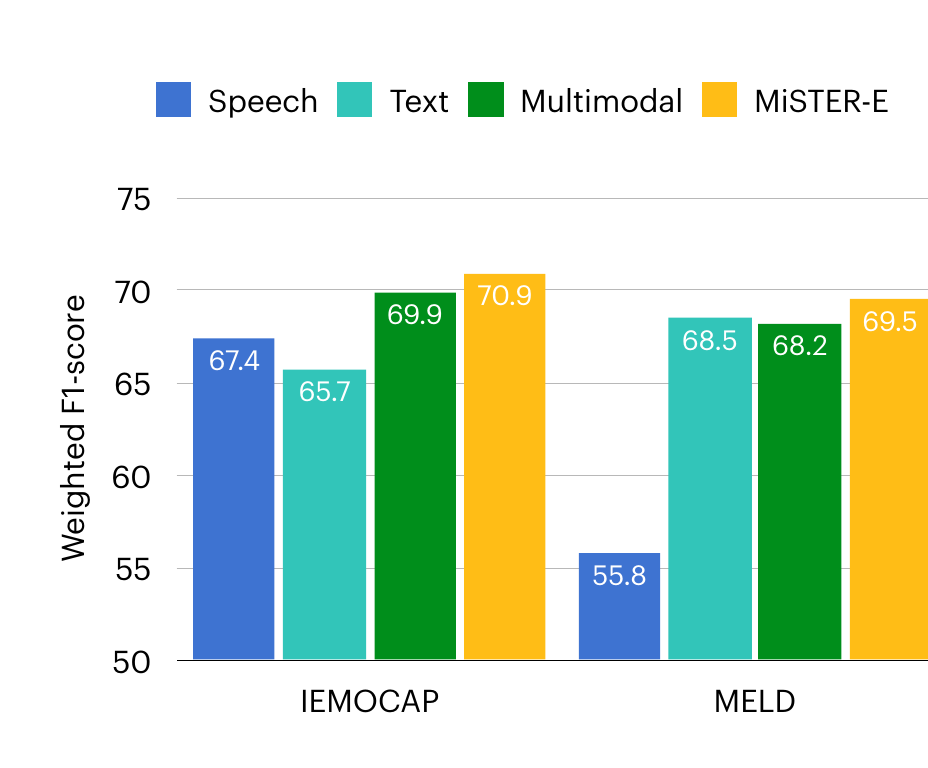}
        \caption{ Performance of MiSTER-E with changes in the MoE gating strategy for IEMOCAP and MELD and }
        \label{fig:moe}
    \end{subfigure}\hfill
    \begin{subfigure}[t]{0.5\textwidth}
        \centering
        \raisebox{1.35\height}{%
        \resizebox{\linewidth}{!}{%
        \begin{tabular}{@{}l|l|c|c@{}}
\toprule
\textbf{Modality} & \textbf{Model} & \textbf{IEMOCAP} & \textbf{MELD} \\
\midrule
\multirow{3}{*}{\texttt{Text}} 
& \texttt{Gemma-7B} & $48.2\%$ & $58.1\%$ \\
& \texttt{Qwen2.5-7B} & $53.1\%$ & $65.5\%$ \\
& \texttt{LLaMA-3.1-8B} & $\mathbf{55.3\%}$ & $\mathbf{67.1\%}$ \\
\midrule \midrule
\multirow{3}{*}{\texttt{Speech}} 
& \texttt{SALMONN-7B} & $\mathbf{59.7\%}$  & $54.3\%$ \\
& \texttt{Qwen2-Audio-7B} & $59.2\%$ & $\mathbf{54.9\%}$ \\
& \texttt{SALMONN-13B} & $58.7\%$ & $53.2\%$ \\
\bottomrule
\end{tabular}}}
        \caption{Performance of different SLLMs and LLMs on the two datasets.}
        \label{tab:others}
    \end{subfigure}
    \caption{Comparison of model performance on IEMOCAP and MELD. 
(a) Effect of different MoE gating strategies in MiSTER-E. 
(b) Performance of unimodal SLLMs and LLMs across the two datasets.}
\end{figure*}

\begin{table*}[t!]
\centering
\resizebox{0.75\textwidth}{!}{%
\begin{tabular}{@{}l|c|c|c@{}}
\toprule
\textbf{Method} 
& \textbf{IEMOCAP ($6$-class)} 
& \textbf{MELD ($7$-class)} 
& \textcolor{black}{\textbf{MOSI ($2$-class)}} \\
\midrule
SMIN~\citep{lian2022smin}$^{\#\#}$ & $\underline{70.5}\%$ & $63.7\%$ & \textcolor{black}{$79.7\%$} \\
MultiEmo~\citep{shi2023multiemo}$^{\#}$ & $66.9\%^{\pm2.0}$ & $65.3\%^{\pm0.5}$ & \textcolor{black}{$83.7\%^{\pm0.3}$} \\
HCAM~\citep{dutta2023hcam} (Sec.~\ref{sec:hcammodel}) & $\underline{70.5}\%$ & $65.8\%$ & \textcolor{black}{$\underline{85.8\%}$} \\
DF-ERC~\citep{li2023revisiting} & $69.5\%$ & $64.5\%$ & \textcolor{black}{--} \\
DER-GCN~\citep{ai2024gcn} & $64.7\%$ & $62.6\%$ & \textcolor{black}{--} \\
ELR-GNN~\citep{shou2024efficient} & $64.4\%$ & $63.2\%$ & \textcolor{black}{--} \\
Mamba-like-model~\citep{shou2024revisiting} & $70.2\%$ & $65.6\%$ & \textcolor{black}{--} \\
TelME~\citep{yun2024telme} & $69.3\%$ & $\underline{67.2\%}$ & \textcolor{black}{--} \\
CFN-ESA~\citep{li2024cfn} & $68.7\%$ & $\underline{67.2}\%$ & \textcolor{black}{--} \\
MMGAT-EMO~\citep{zhang2025moe}$^{\#}$ & $65.5\%^{\pm0.6}$ & $66.1\%^{\pm 0.4}$ & \textcolor{black}{$84.3\%^{\pm0.4}$} \\
\midrule
MiSTER-E 
& $\mathbf{70.9\%}^{\pm0.2}$ 
& $\mathbf{69.5\%}^{\pm0.3}$ 
& \textcolor{black}{$\mathbf{87.9\%}^{\pm 0.2}$} \\
\bottomrule
\end{tabular}}
\caption{Comparison of different methods on IEMOCAP, MELD, and MOSI datasets using weighted-F1 scores. All reported results are for speech-text systems. $^{\#\#}$ We report the numbers when no external emotional data is used for training SMIN. $^{\#}$ We ran the public implementations provided by the authors under our experimental settings. Superscripted results denote the mean and standard deviation over $3$ random initializations, where applicable.}
\label{tab:method_comparison}
\end{table*}
\noindent \textbf{Is modularity crucial?}: To evaluate the importance of architectural modularity, we design a monolithic variant of our model that eliminates the expert-based structure. In this baseline, the contextual representations from each modality are directly passed into the fusion module, bypassing the separation between modality-specific and multimodal experts. This modification leads to a notable degradation in performance, with the weighted F1-score dropping from $70.9\%$ to $67.8\%$ on the IEMOCAP dataset and from $69.5\%$ to $67.9\%$ on MELD. These results underscore a key insight: simply fusing contextual representations in a unified stream is insufficient. Instead, explicitly decoupling context modeling from cross-modal fusion, as implemented in MiSTER-E, enables each component to specialize and contribute more effectively to the final prediction.\\
\noindent \textbf{Are LLM  embeddings the panacea?}: An attentive reader might enquire whether the gains of our approach stem solely from the use of LLM/SLLM features. To isolate the impact of our architectural design from that of the input representations, we conduct controlled experiments where we retrain several prior ERC models using the same LLM and SLLM features employed in our proposed method. Since these baselines were originally optimized for different feature types, we perform our own hyperparameter tuning for each model to ensure a fair comparison (see Table~\ref{tab:our_feats}). Our findings show that while all baselines benefit to some extent from the improved representations, the performance of MiSTER-E improves over previously published baselines. This highlights a key finding: the gains achieved by our framework are not solely attributable to powerful embeddings, but instead emerge from the modular architectural choices—particularly the separation of context modeling and cross-modal fusion, and the use of expert-level supervision and gating.\\
\textcolor{black}{To further assess whether the gains of MiSTER-E are attributable to architectural design rather than encoder scale, we construct two variants of MiSTER-E by replacing the LLM/SLLM encoders with more conventional pretrained representations: RoBERTa for text and \{OpenSMILE, wav2vec\} features for speech. Despite the substantial reduction in model capacity, both variants of MiSTER-E consistently outperform all baseline methods that use the same feature representations (Table~\ref{tab:our_feats}).
\begin{table*}[t!]
\centering
\resizebox{\textwidth}{!}{%
\begin{tabular}{@{}l|l|c|c|c| c@{}}
\toprule
\textbf{Method} & \textbf{\# Params} & \textbf{Text Feats.} & \textbf{Speech Feats.} & \textbf{IEMOCAP} & \textbf{MELD} \\
\midrule
MultiEmo~\citep{shi2023multiemo} & $\approx450$M & RoBERTa & OpenSMILE & $66.9\%$ & $65.3\%$ \\
~~+~~LLM/SLLM  &$\approx14$B & LLaMA & SALMONN & $69.5\%$ & $68.2\%$ \\ \midrule
HCAM~\citep{dutta2023hcam} & $\approx750$M & RoBERTa & wav2vec & $70.5\%$ & $65.8\%$\\
~~+~~LLM/SLLM &$\approx14$B & LLaMA & SALMONN & $70.6\%$ & $\underline{68.4}\%$\\ \midrule
MMGAT-EMO~\citep{zhang2025moe} & $\approx450$M & EmoBERTa & OpenSMILE & $65.5\%$ & $66.1\%$ \\ 
~~+~~LLM/SLLM &$\approx14$B & LLaMA & SALMONN & $68.6\%$ & $67.7\%$\\ \midrule \midrule
\textcolor{black}{MiSTER-E~~w/o~~LLM/SLLM } &\textcolor{black}{$\approx450$M}& \textcolor{black}{RoBERTa} & \textcolor{black}{OpenSMILE} & \textcolor{black}{$69.5\%$} & \textcolor{black}{$66.4\%$} \\
\textcolor{black}{MiSTER-E~~w/o~~LLM/SLLM }&\textcolor{black}{$\approx750$M}& \textcolor{black}{RoBERTa} & \textcolor{black}{wav2vec} & \textcolor{black}{$\mathbf{71.1}\%$} & \textcolor{black}{$\underline{68.4\%}$}\\
MiSTER-E &$\approx14$B & LLaMA & SALMONN & $\underline{70.9\%}$ & $\mathbf{69.5\%}$\\
\bottomrule
\end{tabular}}
\caption{Comparison of some of the baseline methods \textcolor{black}{when re-designed with and without LLM features. The results show that the proposal is applicable even for non-LLM representations like OPENSMILE and wav2vec. } }
\label{tab:our_feats}
\end{table*}
 Notably, on IEMOCAP, MiSTER-E with RoBERTa and wav2vec features achieves performance comparable to—and in some cases higher than—the corresponding LLM/SLLM-based configuration, suggesting improved generalization under reduced model capacity. These experiments demonstrate that the advantages of our approach are not tied to any specific set of pretrained embeddings, but rather stem from the modeling design that is proposed in this work.}\\
\noindent \textbf{Is decision-level MoE gating crucial?}: To assess the importance of our decision-level Mixture-of-Experts (MoE) gating strategy, we conduct ablation studies by comparing MiSTER-E against two architectural variants. In the first variant, termed \texttt{feat-MoE}, we shift the fusion point to the feature space—that is, the modality-specific and multimodal representations are combined before classification, rather than fusing their predictions. This design tests whether earlier integration is sufficient. In the second variant, called \texttt{No-Loss-MoE}, we eliminate the expert-level supervision altogether. Here, the model is trained solely on the final gated prediction using focal loss, omitting the individual loss signals for the audio, text, and multimodal experts.
As shown in Fig.~\ref{fig:moe}, both variants degrade performance: \texttt{feat-MoE} results in a $2.6\%$ drop on IEMOCAP and $0.6\%$ on MELD, while \texttt{No-Loss-MoE} leads to similar degradation. These findings reinforce two key design principles behind MiSTER-E: (1) decision-level fusion is more effective than early feature fusion, likely because it preserves modality-specific discriminative cues; and (2) expert-specific supervision is essential for encouraging each expert to specialize in its respective modality or fusion pattern.\\
\noindent \textbf{Performance of Other LLMs/SLLMs}
We experiment with other speech and language large models (SLLMs and LLMs) for encoding speech and text, respectively. All models are trained using the same hyperparameters as \texttt{LLaMA-3.1-8B} and \texttt{SALMONN - 7B}. For the text modality, we experiment with \texttt{Qwen2.5-7B} and \texttt{Gemma-7B}, while for speech, we use \texttt{Qwen2-Audio-7B} and \texttt{SALMONN-13B}. 

As shown in Table~\ref{tab:others}, the performance of the speech encoders is comparable across models, with \texttt{SALMONN-7B} achieving the best result on IEMOCAP, and \texttt{Qwen2-Audio-7B} performing best on MELD. The larger \texttt{SALMONN-13B} underperforms compared to its smaller counterparts. For text, \texttt{LLaMA-3.1-8B} outperforms both \texttt{Gemma-7B} and \texttt{Qwen2.5-7B} on both datasets.
\subsubsection{Discussion}
\begin{figure*}[t!]
    \centering
\includegraphics[width=0.9\textwidth,trim={1cm 7cm 1cm 4cm},clip]{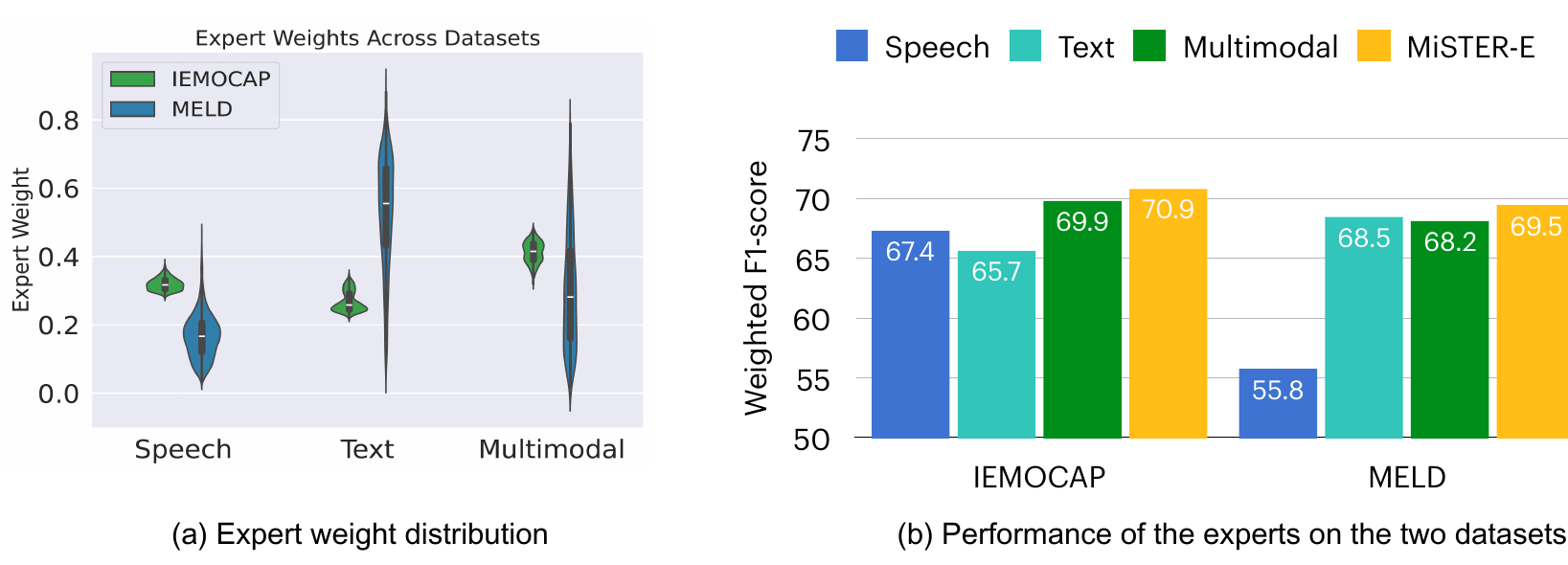}

    \caption{(a) Distribution of weights for the experts for the different datasets and (b) The performance of the experts and MiSTER-E for the two datasets}
    \label{fig:comb_2}
\end{figure*}

\begin{figure*}[t!]
    \centering
    \begin{subfigure}[b]{0.48\textwidth}
        \centering
        \includegraphics[width=\linewidth,trim={0cm 1cm 0cm 0cm},clip]{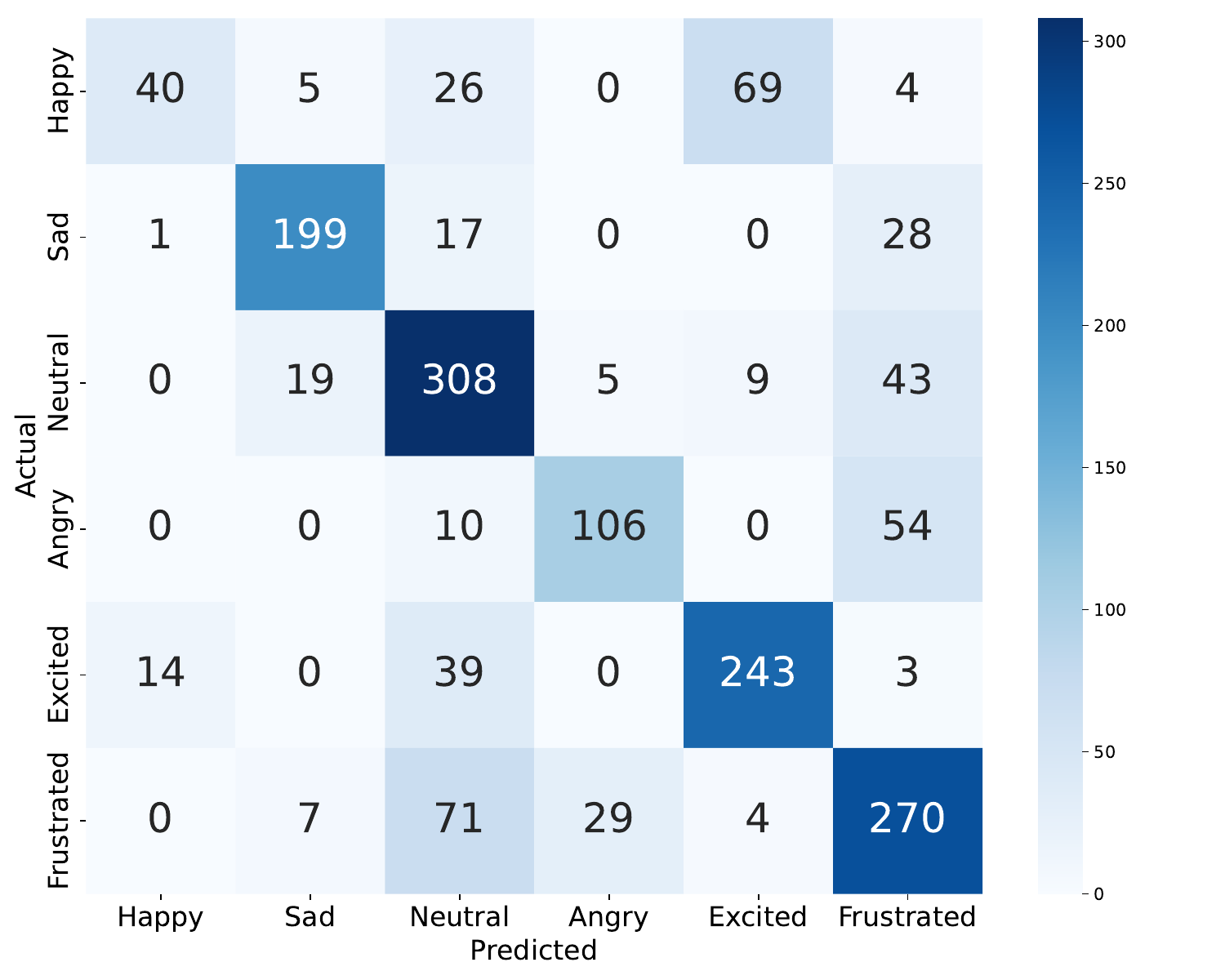}
        \caption{Confusion matrix of our model on the IEMOCAP dataset.}
        \label{fig:heatmap}
    \end{subfigure}\hfill
    \begin{subfigure}[b]{0.45\textwidth}
        \centering
        \resizebox{\linewidth}{!}{%
        \begin{tabular}{@{}l|c|l|c@{}}
        \toprule
        \multicolumn{2}{c|}{\textbf{IEMOCAP}} & \multicolumn{2}{c}{\textbf{MELD}} \\
        \midrule
        \textbf{Category} & \textbf{F1} & \textbf{Category}  & \textbf{F1}\\
        \midrule
        Angry & $68.4\%$ & Angry & $62.5\%$\\
        Excited & $77.9\%$ & Disgust & $42.9\%$ \\
        Frustrated & $69.0\%$ & Fear &  $30.6\%$\\
        Happy & $40.2\%$ & Joy & $65.4\%$ \\
        Neutral & $80.2\%$ & Neutral & $81.3\%$\\
        Sad & $83.8\%$ & Sad & $50.3\%$\\
        - & - & Surprise & $61.5\%$\\
        \bottomrule
        \end{tabular}}
        \caption{Per-class F1-scores for IEMOCAP and MELD.}
        \label{tab:class}
    \end{subfigure}
    \caption{(a) Confusion matrix for IEMOCAP. (b) Class-wise F1-scores on IEMOCAP and MELD, showing variation across different emotions.}
    \label{fig:heatmap_table}
\end{figure*}

\begin{figure*}[ht!]
    \centering
\includegraphics[width=\textwidth,trim={0cm 0cm 0cm 0cm},clip]{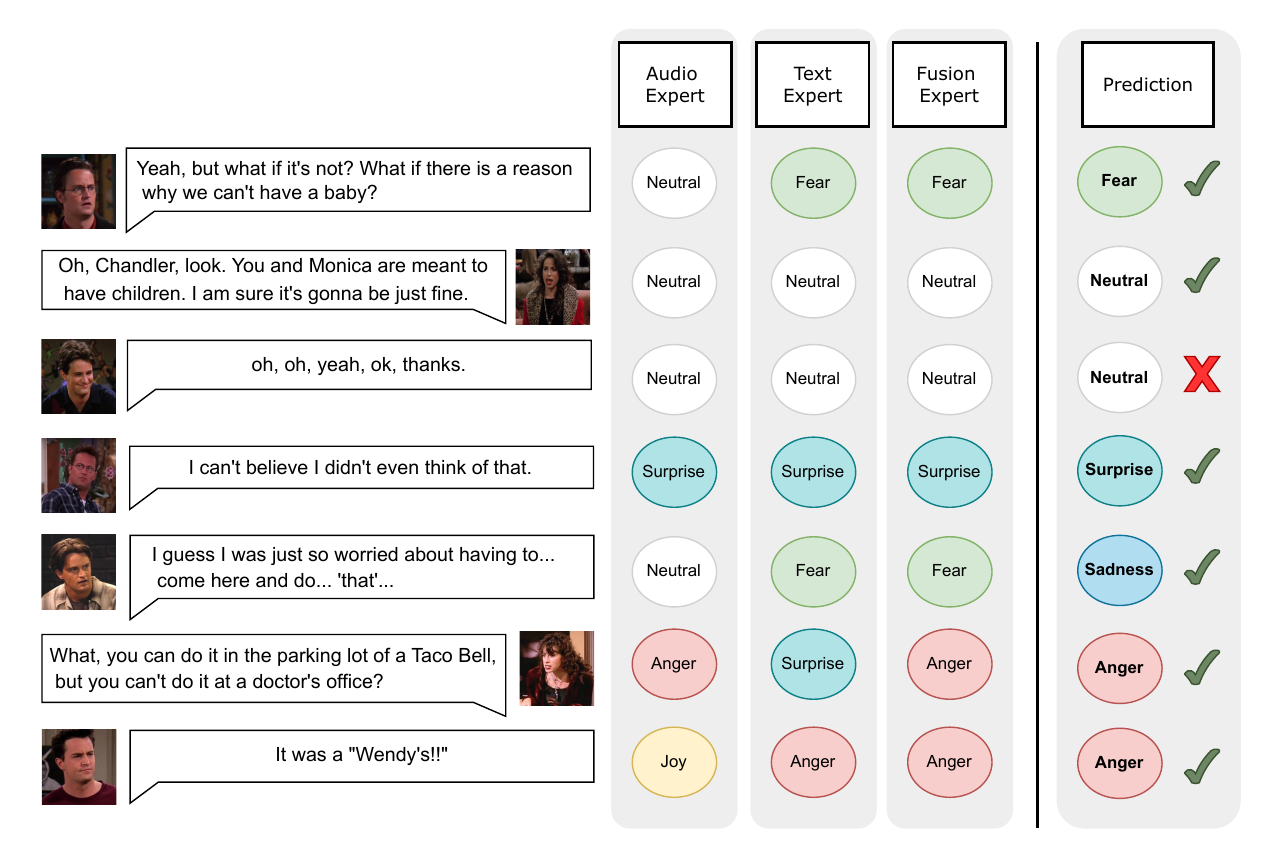}

    \caption{A case study from MELD where there are emotion shifts throughout the conversation. Out of the $7$ utterances in the conversation, MiSTER-E correctly predicts $6$.}
    \label{fig:case1}

\end{figure*}

\noindent \textbf{Expert Behaviour and Modality Imbalance:} We analyse the behaviour of the mixture-of-experts (MoE) gating mechanism on IEMOCAP and MELD to understand how the model adapts to dataset-specific modality characteristics. As shown in Fig.~\ref{fig:comb_2}(a), the gating network exhibits clear dataset-dependent preferences: it assigns higher weight to the multimodal expert for IEMOCAP, while favouring the text expert for MELD. This behaviour reflects the relative reliability of modalities in the two datasets.\\
These trends are consistent with the performance of individual experts in Fig.~\ref{fig:comb_2}(b). On MELD, the speech expert performs substantially worse than the text expert, and the multimodal expert slightly underperforms the text expert due to the difficulty of aligning modalities with highly unequal predictive power. In contrast, on IEMOCAP, where speech and text modalities exhibit comparable performance, the multimodal expert effectively fuses both sources and outperforms the best unimodal expert by $2.5\%$. Across both datasets, MiSTER-E consistently outperforms the strongest individual expert by approximately $1\%$, demonstrating the benefit of adaptive, per-utterance expert weighting.\\
The effect of expert consistency regularization aligns with these observations. 
For IEMOCAP, a moderate consistency strength ($\alpha=0.1$) improves performance from $70.4\%$ to $70.9\%$, while stronger regularization degrades performance. For MELD, where modality imbalance is pronounced, a small value ($\alpha=10^{-3}$) yields a modest gain of $0.1\%$, whereas larger values significantly reduce performance. \\
\noindent \textbf{Class-wise Performance of MiSTER-E}
The class-wise performance of our proposed method for the two datasets is shown in Table~\ref{tab:class}. We note that for IEMOCAP the worst performing class is ``happy'' - likely because it is the least frequent class and it is most confused with the ``excited'' category (Fig.~\ref{fig:heatmap}). Similarly, for MELD, the model performs most poorly on the ``fear'' category - again the least frequent class. 
However, we note that for both datasets, the model performs quite well on most emotion classes, thereby leading to an improved overall performance.\\
\noindent \textbf{Importance of the Contrastive Loss:} On MELD, incorporating the contrastive objective improves performance from $68.2\%$ ($\lambda=0$) to $69.5\%$ at $\lambda=1$, while on IEMOCAP performance increases from $70.2\%$ to $70.9\%$ at $\lambda=2$. These gains indicate that contrastive learning helps align emotion-relevant speech and text representations beyond what is achieved by focal loss alone. However, further increasing $\lambda$ leads to degraded performance on both datasets, suggesting that the contrastive loss serves as a beneficial auxiliary signal rather than a standalone supervisory objective.\\
\noindent \textbf{Case Study:} We provide a case study from the MELD dataset in Fig.~\ref{fig:case1}. A number of key points that we wish to highlight from this example are as follows:
\begin{itemize}
    \item MiSTER-E does not fail when there are emotion shifts. E.g., the conversation has $6$ emotions in $7$ utterances. Although, the model incorrectly predicts the third utterance to be neutral (instead of joy), it is able to predict the emotion shift from surprise to sadness to anger.
    \item Another interesting point is the output of the fifth utterance. None of the experts predict sadness class, yet the MoE strategy correctly marks the utterance as sad. This indicates the utility of the MoE gating strategy in our proposed method and differentiates it from static ensembling techniques.
\end{itemize}
\subsubsection{Summary and Limitations of MiSTER-E}
We proposed MiSTER-E, a modular framework for ERC that explicitly separates contextual modeling from multimodal fusion. Leveraging LLM-based representations for both speech and text, we model context in the conversations using a temporal inception block followed by a Bi-GRU, and perform modality fusion via an attention-based network. A Mixture-of-Experts (MoE) gate adaptively integrates decisions from context-aware and multimodal experts. MiSTER-E achieves new state-of-the-art performance on IEMOCAP, MELD, and CMU-MOSI, demonstrating the efficacy of modular context-fusion and decision-level gating. \\
While MiSTER-E achieves strong performance across multiple benchmarks, several limitations remain,  
(i) the use of large LLM/SLLM encoders introduces non-trivial computational and memory overhead, which may limit applicability in low-resource or real-time settings, despite the use of parameter-efficient fine-tuning. We have partially addressed this concern in Table~\ref{tab:our_feats}, where the proposed approach is shown to be beneficial even for non-LLM based features,  
(ii) our evaluation is primarily conducted on benchmark datasets consisting of scripted or semi-scripted dialogues (e.g., TV shows and acted conversations), and performance under domain shift to spontaneous, real-world conversational settings remains an open question, (iii) emotion inference models are known to be susceptible to dataset biases related to demographic factors, language use, and annotation subjectivity and we have not performed any bias/fairness analysis in this study, and, (iv) throughout this work,  we intentionally avoid explicit speaker identity modeling, however, the utilization of speaker metadata may improve the modeling of interpersonal dynamics.
\subsection{Summary of modeling for ERC}
In this section, we addressed the problem of multimodal Emotion Recognition in Conversations (ERC), where emotion inference requires modeling both conversational context and multimodal cues across multiple speakers. We proposed two techniques: hierarchical modeling and a mixture-of-experts model for this task. Together, these contributions advance ERC modeling by emphasizing modularity, robustness to modality imbalance, and flexible multimodal integration. However, existing ERC benchmarks predominantly consist of acted or semi-controlled conversational speech, which limits their ability to reflect the variability encountered in real-world recordings.\\
In the next section, we move beyond conversational settings and study emotion recognition from spoken language under naturalistic recording conditions, where challenges such as spontaneous speech, environmental noise, speaker diversity, and highly skewed emotion distributions become central.
\section{Emotion Recognition under Naturalistic Conditions}
Despite substantial progress in speech emotion recognition, most existing systems are developed and evaluated on datasets collected under controlled or semi-controlled conditions, often involving acted emotions and clean recording environments. In contrast, real-world spoken language is highly variable, spontaneous, and noisy, with emotions expressed subtly and unevenly across speakers, and recording conditions. These factors make emotion recognition from naturalistic speech significantly more challenging than laboratory-style benchmarks.

Emotion recognition under naturalistic conditions introduces several challenges that are not adequately addressed by existing SER frameworks. These include large inter-speaker variability, diverse acoustic environments, spontaneous speech patterns, and highly skewed emotion distributions, where neutral or weakly expressed emotions dominate. Moreover, models trained on acted datasets often fail to generalize to such settings, leading to substantial performance degradation when deployed in real-world applications.

To systematically study emotion recognition in such realistic scenarios, this thesis leverages the Speech Emotion Recognition in Naturalistic Conditions Challenge at Interspeech 2025, which is based on the MSP-Podcast corpus~\cite{busso2025msp}. This challenge provides a large-scale, real-world benchmark with spontaneous speech collected from diverse podcast sources and annotated under realistic conditions. We use this setting as a rigorous testbed to evaluate modern speech and language models for emotion recognition under naturalistic constraints.

\subsection{Proposed Approach}\label{sec:abhinayamodel}
\begin{figure}
    \centering
    \includegraphics[width=0.7\columnwidth,trim={0cm 5.5cm 0cm 1cm},clip]{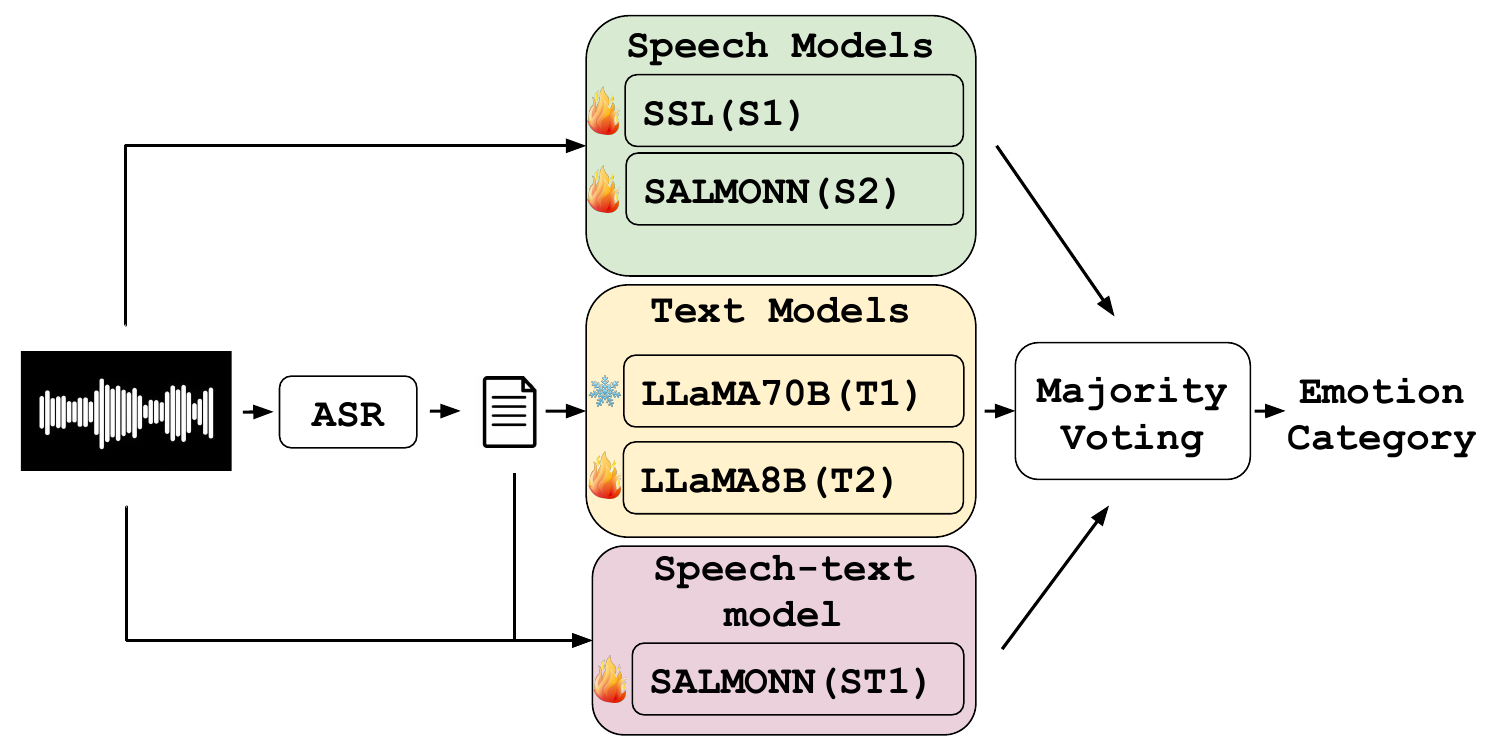}
    \caption{Schematic of the different  components of the Abhinaya SER system. We use three types of models - speech-only (\textbf{S1}, \textbf{S2}), text-only (\textbf{T1}, \textbf{T2}) and speech-text (\textbf{ST1}). Only \textbf{T1} is used in a zero-shot setting. The text used by the models are ASR transcripts   generated by Whisper  }.
    \label{fig:abhinaya_model}
\end{figure}

We now describe the details of ``Abhinaya'' (meaning - art of expression in \textit{Sanskrit}) system~\cite{dutta2025abhinaya}, an SER model, which was developed as part of this challenge.  The Abhinaya system directly addresses the challenge’s key issues—class imbalance, multimodal modeling, and data variability—through an ensemble of speech-only, text-only, and speech-text joint models. Rather than building these components from scratch, we leverage SSL models (WavLM-Large~\cite{chen2022wavlm}) and speech large language models (SLLMs) (SALMONN~\cite{tang2023salmonn}) as speech feature encoders. Given the potential benefits of multimodal information, we transcribe speech using Whisper-large-v3 automatic speech recognition (ASR) system  \cite{radford2023robust} and process the transcripts with LLMs. Further, we adopt an approach for the multimodal model, where speech and text are concatenated at the input and processed jointly using an SLLM. 
To mitigate class imbalance, we explore various loss functions for fine-tuning LLMs and SLLMs for SER. The schematic of the Abhinaya system~\footnote{Code:\url{https://github.com/iiscleap/ABHINAYA}} is shown in Fig.~\ref{fig:abhinaya_model}. 
\subsubsection{Speech-only models}
\noindent \textbf{Wav-LM SSL  (S1):} The speech foundation model used in this sub-system is the   WavLM-large model \cite{chen2022wavlm}. The convolutional feature extractor is frozen, and transformer layers are fine-tuned.  Since WavLM generates frame-level representations, we apply attentive pooling~\cite{okabe18_interspeech} to aggregate these features before emotion classification. The attentive pooling layer outputs a weighted mean of the features along with a weighted standard deviation, where the weights are learned with an attention mechanism.\\
\noindent \textbf{SALMONN SLLM (S2):} The SALMONN speech large language model (SLLM) \cite{tang2023salmonn}   
consists of speech encoder - Whisper encoder \cite{radford2023robust} + BEATS encoder  \cite{chen2023beats}, with a text-based LLM (LLaMA \cite{touvron2023llama}). The speech encoder and LLaMA weights remain frozen, while only the  Q-former~\cite{li2023blip} and the low-rank adapters (LoRA) in the LLaMA module are trained for audio and speech tasks  \cite{tang2023salmonn}. 
The SALMONN model is available in two sizes - $7$B and $13$B depending on the size of the back-end text LLM. \\
In our setting, we use all the layers of the SALMONN model  except the final text-token layer. 
 The representations from the final LLaMA transformer layer are passed through the attentive statistics pooling layer, and the model is fine-tuned for the emotion classification task with a softmax head. 
 During fine-tuning, we update only the Q-Former and LoRA weights, along with the classification head.
 \subsubsection{Text-only models}
 In the Naturalistic Conditions Challenge, the reference text transcripts were provided only for the training and validation subsets, while the evaluation data lacked ground-truth transcripts. To ensure consistency, we used ASR transcripts generated by Whisper-Large v3~\cite{radford2023robust} for our model development during training and testing. We explore two choices for text-only models.\\
 \noindent \textbf{Zero-shot LLaMA (T1):} We use the \texttt{LLaMA-3.3-70B-Instruct} model~\cite{dubey2024llama} in zero shot setting for this task. The model was prompted to classify emotions into predefined categories. \\
 \noindent \textbf{Fine-tuned smaller text LLM (T2):} Prior studies~\cite{zhang2024sentiment, dutta2025llm} indicate that LLMs often struggle with fine-grained emotion recognition in zero-shot scenario. To address this, we fine-tune the \texttt{LLaMA-3.1-8B} model~\cite{dubey2024llama} using LoRA for emotion recognition.
Similar to the SALMONN fine-tuning setup, we extract last-layer representations from LLaMA, apply attentive statistics pooling, and feed the utterance-level embeddings to an emotion classification head.
\subsubsection{Speech-text joint model (ST1)}
Since speech large language models (SLLMs) primarily adapt LLMs for speech sequences, we explore their potential for joint speech-text modeling. For this purpose, we use the SALMONN-$7$B model. In addition to the speech sequence, we append the Whisper-generated text transcript before fine-tuning the  SALMONN  model using LoRA. As in speech/text only fine-tuning, we extract final-layer representations, apply attentive statistics pooling, and  train the classification head.
\subsubsection{Loss Functions}
The dataset is highly imbalanced with the frequency of the majority class (``neutral'') almost $26$ times that of the minority class (``fear'') in the training data.  To address this issue, we explore various loss functions:
\begin{itemize}
    \item \textbf{Weighted cross entropy  (WCE):} Denoting the total number of data points by $N$, the number of classes by $C$, the weight assigned for a particular class, $c \in [1,2,\dots,C]$, is given by $w^c = \frac{N}{N_c \times C}$, where there are $N_c$ samples belonging to class $c$ in the training data. The cross-entropy loss $l_{i}$ for sample $i$ is given by:
\begin{equation}\label{eq:wce}
    l_{i} = -\sum_{c=1}^{C}t_{ic}w^c\log(p_{ic})~~~;~~~ {\mathbb E} = \frac{1}{N}\sum_{i=1}^N l_{i}
\end{equation}
where $p_{ic}$ is the prediction probability of class $c$ for sample $i$. $t_{ic}$ is set to $1$ for the true class $c$ of sample $i$, else it is kept at $0$. The  loss $\mathbb E$ is defined as the weighted cross-entropy loss (WCE). 
\item \textbf{Weighted FoCal Loss (WFL):} To further emphasize hard-to-classify samples, we explore the focal loss~\cite{lin2017focal}, which adjusts the weight based on the model’s confidence. 
The weighted focal loss (WFL) is defined as:
\begin{equation}\label{eq:weightedfocal}
    l_{i} = -\sum_{c=1}^{C}t_{ic}w^c(1-p_{ic})^{\gamma}\log(p_{ic}), 
\end{equation}
and the total loss $\mathbb E$ is the same as in Eq.~\ref{eq:wce}. Here, $\gamma \geq 0$ is a hyper-parameter.
\item \textbf{Vector scaling (VS) loss:} The vector scaling (VS) loss adjusts predictions using class-dependent temperature scaling and bias correction. It modifies the pre-softmax model prediction, $z_{ic}$, as:
\begin{equation}\label{eq:adjust}
    \hat{z}_{ic} = (\frac{N_{c}}{N_{max}})^{\gamma}z_{ic} + \tau \log(\frac{N_c}{N})
\end{equation}
where $\gamma \geq 0$ and $\tau \geq 0$ are hyper-parameters, while $N_{max}$ is the number of samples in the most frequent class. 
The softmax $\hat{p}_{ic} = softmax(\hat{z}_{ic})$ is used in the weighted loss:
\begin{equation}\label{eq:vs}
    l_{i} = -\sum_{c=1}^{C}t_{ic}w^c\log(\hat{p}_{ic})
\end{equation}
where $w^c$ and $t_{ic}$ are defined as before. The total loss ($\mathbb E$) is the same as in Eq.~\ref{eq:wce}.
\end{itemize}
\subsubsection{Experiments}\label{sec:abhinaya_data}
\begin{figure}
    \centering
    \includegraphics[width=\columnwidth,trim={3cm 8.5cm 3cm 5cm},clip]{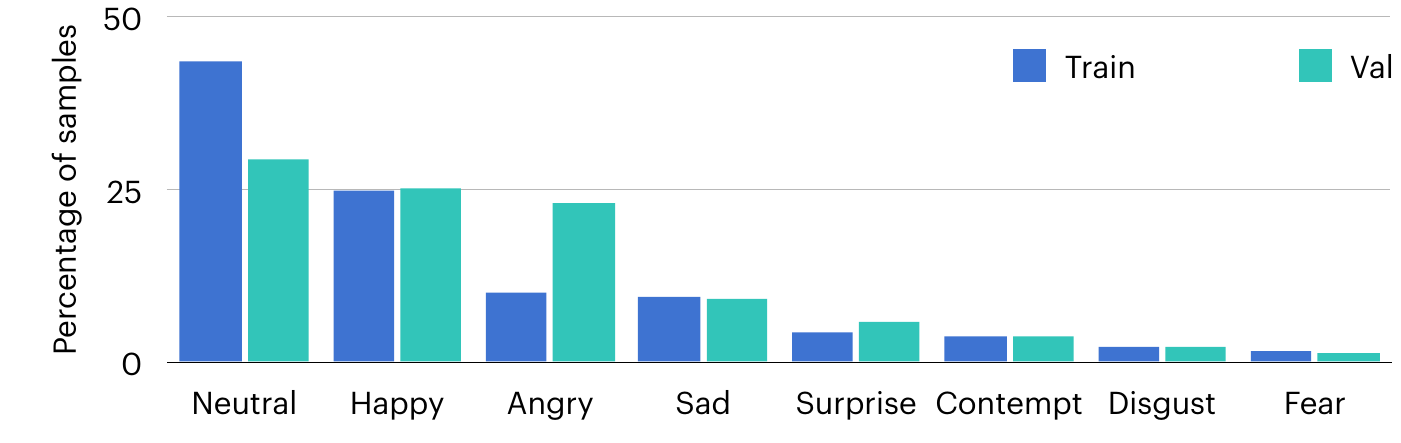}
    \caption{Distribution of the $8$ emotion categories in the training and validation data}.
    \label{fig:distribution}
\end{figure}

\begin{figure}[t]
    \centering
    \includegraphics[width=\columnwidth,trim={0cm 6.5cm 0cm 3cm},clip]{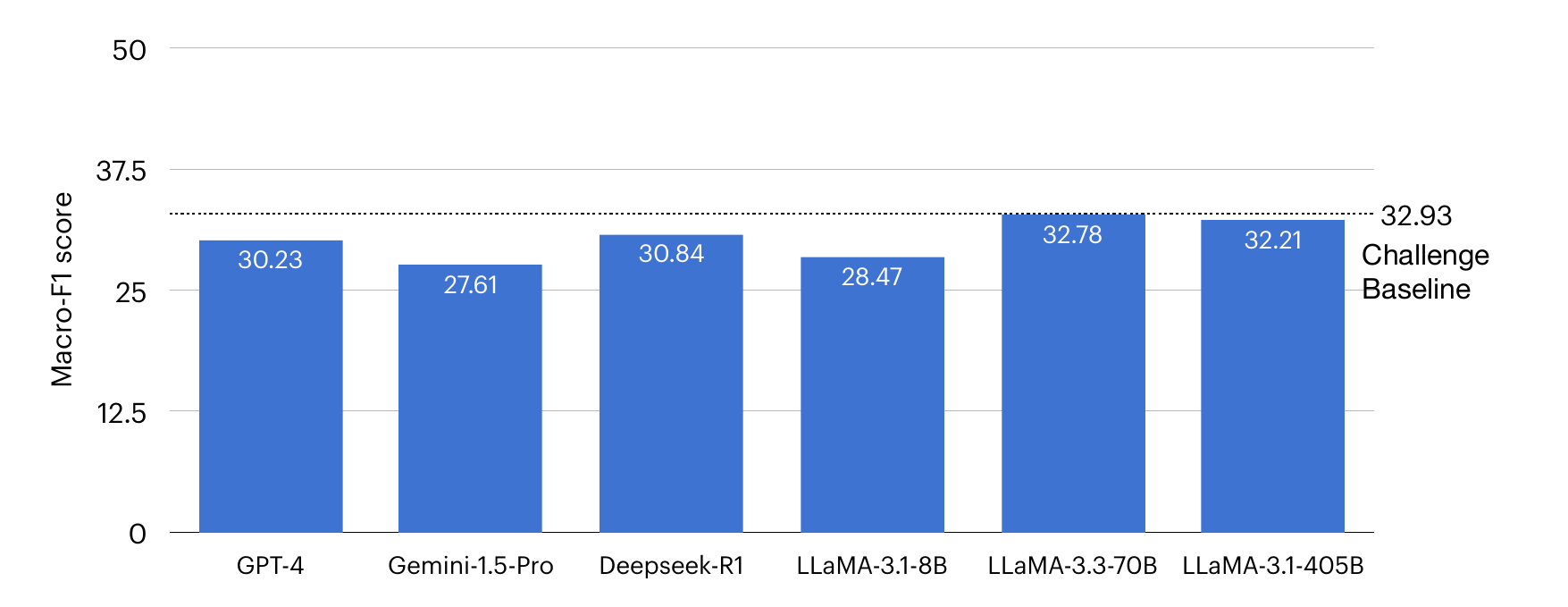}
    \caption{Validation macro F1-score (in $\%$) for different LLMs~\cite{dubey2024llama, guo2025deepseek, team2023gemini, achiam2023gpt} evaluated in zero-shot setting using the ASR transcripts. The LLaMA models considered are the Instruct versions. The baseline performance is also shown.}
    \label{fig:llms}
\end{figure}

\noindent \textbf{Dataset:} The MSP-PODCAST~\cite{busso2025msp} dataset serves as the basis for this challenge. The training, validation and the test splits contain $84260$, $31961$ and $3200$ speech files respectively. The dataset includes annotations for $8$ primary emotion categories: ``happy'', ``angry'', ``sad'', ``neutral'', ``surprise'', ``fear'', ``contempt'' and ``disgust''. Additionally, two other labels—``other'' (denoting emotions outside the predefined categories) and ``X'' (cases with no annotator consensus)—are present but excluded from emotion modeling. The training and the validation distribution for the $8$ categories are shown in Fig.~\ref{fig:distribution}. For the model selection, we construct a balanced validation set, using the same number of utterances ($326$) for each of the $8$ emotion classes. 
The test data labels in the challenge were not provided and the performance of the models was only available through the leaderboard. Unlike the training distribution, which shows a considerable imbalance, the test distribution is balanced across the $8$ emotion classes.\\
\noindent \textbf{Implementation Details:} Except for T1, which operates in a zero-shot manner, all models are fine-tuned using the AdamW optimizer~\cite{loshchilov2017decoupled} with a learning rate of $1e\text{-}5$. For models utilizing LoRA (S2, T2, and ST1), the LoRA parameters~\cite{hu2022lora} are set as $r=8$ and $\alpha=32$, with a dropout rate of $0.1$. Training configurations vary across models due to the diverse memory requirements:
\begin{itemize}
    \item For S1 and T2, the batch size is set to $8$ with gradient accumulation over $4$ steps.
    \item For SALMONN-based models (S2 and ST1), the batch size is set to $4$ with gradient accumulation over $8$ steps.
\end{itemize}
 In speech-only models (S1 and S2) and the speech-text model (ST1), audio samples are processed with a maximum duration of $10$ seconds. All four models are trained for $20$ epochs, with the checkpoint yielding the highest validation macro-F1 score selected for evaluation. Regarding loss functions, models trained with WFL (Eq.~\ref{eq:weightedfocal}) use $\gamma=2$, while those trained with VS loss (Eq.~\ref{eq:adjust}) have hyper-parameters  $\gamma=0.3$ and $\tau=1$. \\
\begin{table}[t]
\centering
\caption{The performance (in $\%$) of the different models on the balanced validation and test data. 
The performance of certain models on the test data are not reported as they were not submitted to the leaderboard.  
$*$ indicates the test score evaluated by the organizers after the official challenge deadline.}\label{tab:results}
\renewcommand{\arraystretch}{1.2}
 \resizebox{0.6\columnwidth}{!}{%
\begin{tabular}{@{}l|l|l|l|l@{}}
\toprule
Models & Modality & Loss & \begin{tabular}[c]{@{}l@{}}Val. F1 \\ (macro)\end{tabular} & \begin{tabular}[c]{@{}l@{}}Test F1\\ (macro)\end{tabular} \\ \midrule
Baseline~\cite{naini25_interspeech} & Speech & WCE & - & $32.93$ \\ \midrule
\textbf{S1} (WavLM-large-317M) & Speech & WFL & $34.43$ & $33$ \\
\textbf{S2} (SALMONN-13B) & Speech & WFL & $37.68$ & $35.34$ \\
\textbf{T1} (LLaMA-3.3-70B)& Text & - & $32.78$ & - \\
\textbf{T2} (LLaMA-3.1-8B) & Text & VS & $33.68$ & - \\
\textbf{ST1} (SALMONN-7B) & Speech-text & VS & $35.43$ & - \\ \midrule \midrule
ABHINAYA (4th place) & Speech-text & - & $\mathbf{43.01}$ & $41.81$ \\
ABHINAYA$^{*}$ (SoTA)  & Speech-text & - & $42.31$ & $\mathbf{44.02}$ \\ \bottomrule
\end{tabular}}
\end{table}
Once the outputs from the five models are available, majority voting is applied across the five classifiers, with S2 serving as the tiebreaker due to its superior validation performance. \\
\noindent \textbf{Performance of Abhinaya}: Table~\ref{tab:results} presents the validation and test set results. The key findings are summarized below:
\begin{itemize}
    \item \textbf{Fine-tuned SALMONN-13B (S2) achieves the best individual system performance}:  Note that, its zero-shot validation performance is only $18.63\%$, hence fine-tuning significantly improves the performance.
    \item \textbf{Fine-tuning a small model versus zero-shot evaluation of a large model}: The fine-tuned LLaMA-3.1-8B (T2) model outperforms the zero-shot LLaMA-3.3-70B model (T1).
    \item \textbf{Ensembling improves performance}: A majority vote ensemble of the five models achieves a $24.56\%$ relative gain over the best individual model (S2) on the test set.
    \item \textbf{Significant improvement over the challenge baseline}: Our system achieves a $33.68\%$ relative improvement over the baseline, demonstrating its effectiveness for naturalistic SER.
    \item \textbf{Post-challenge evaluation confirms state-of-the-art performance}: At the challenge submission deadline, the ST1 fine-tuning had  completed only $3$ epochs. We continued this model fine-tuning till completion ($20$ epochs) and the organizers performed a post-challenge evaluation of the Abhinaya system with this update.  The final system performance of $44.02\%$ surpasses the best published result in the challenge, thereby providing the SoTA results on this task.
\end{itemize}
\noindent \textbf{Evaluation of different LLMs for text-only models:} For the T1 zero-shot setting, we evaluate multiple large language models (LLMs) for the task of emotion recognition using a same prompt structure.  
The comparative performance on the validation set is presented in Fig.\ref{fig:llms}. The key observations are:
\begin{itemize}
    \item LLaMA-3.3-70B-Instruct achieves the best performance, outperforming all other LLMs considered here.
    \item While the zero-shot performance of LLaMA-3.1-8B-Instruct model is only $28.47\%$, the fine-tuning with VS loss improves the performance to $33.68\%$ (Table~\ref{tab:results}). 
    \item The other ``larger'' LLMs did not perform as well on the emotion recognition task with ASR transcripts.
\end{itemize}
\noindent \textbf{Choice of loss functions:} The impact of different loss functions is analyzed in Table~\ref{tab:loss}. The key observations are:
\begin{itemize}
    \item Speech-only models (S1, S2) benefit most from weighted focal loss, suggesting its effectiveness in handling class imbalance for SER. In contrast, VS loss is effective for the text-based (T2) and speech-text (ST1) models.
    \item The stronger zero-shot performance of T2 ($28.47\%$ versus $18.63\%$ for S2) suggests better initial separation between emotion classes. We hypothesize that this initial separation aids VS loss, which adjusts logits class-wise, whereas speech models with less discriminative initial representations benefit more from sample-wise loss weighting.
\end{itemize}
\begin{table}[t!]
\centering
\caption{The performance of the models with different loss functions. All results are macro F1-scores ($\%)$ on the balanced validation data. \textbf{Bold} indicates best loss function for each model.}\label{tab:loss}
\renewcommand{\arraystretch}{1.2}
 \resizebox{0.3\columnwidth}{!}{%
\begin{tabular}{@{}c|c|c|c@{}}
\toprule
Models & WCE & WFL & VS \\ \midrule
S1 & $33.07$ & $\mathbf{34.43}$ & $32.12$ \\
S2 & $36.34$ & $\mathbf{37.68}$ &  $33.17$\\
T2 & $29.79$ & $30.12$ & $\mathbf{33.68}$ \\ 
ST1 & $33.92$ & $34.73$ & $\mathbf{35.43}$ \\ \bottomrule
\end{tabular}}
\end{table}
\noindent \textbf{Speech-text versus speech-only SLLMs:} To assess the impact of joint speech-text fine-tuning, we fine-tune a SALMONN-$7$B model using speech inputs. This setup mirrors model S2 but replaces the SALMONN-$13$B model with the smaller $7$B variant for a direct comparison with ST1 model, which incorporates both speech and text. The speech-only model achieves $33.87\%$ validation macro-F1, while ST1 achieves $35.43\%$, demonstrating the benefits of multimodal fine-tuning. However, speech-text models incur higher computational costs due to the longer input sequences.\\
\noindent \textbf{Class-wise performance analysis:} Table~\ref{tab:class_abhinaya} presents macro-F1 scores per class for Abhinaya and its components. The speech models (S1, S2) perform poorly on the least frequent classes (see Fig.~\ref{fig:distribution}): ``fear'', ``contempt'', ``disgust''. In contrast, the fine-tuned text model (T2) and the speech-text model (ST1) show improved performance on these classes.  Interestingly, despite its weaker overall performance (Table~\ref{tab:results}), the zero-shot text model (T1) outperforms the speech-based models on the three rare classes and even outperforms the ensemble for the least frequent class, ``fear''. 
\begin{table}[t]
\centering
\caption{Class-wise macro F1-scores ($\%$) for the balanced validation data for Abhinaya and the different model components. \textbf{Bold} indicates best model for each emotion category.}\label{tab:class_abhinaya}
\renewcommand{\arraystretch}{1.2}
 \resizebox{0.6\columnwidth}{!}{%
\begin{tabular}{@{}l|c|c|c|c|c||c@{}}
\toprule
\begin{tabular}[c]{@{}l@{}}Emotion\\ Classes\end{tabular} & S1 & S2 & T1 & T2 & ST1 & ABHINAYA \\ \midrule
Angry & $49.74$ & $50.60$ & $32.48$ & $37.13$ & $50.73$ & $\mathbf{51.49}$ \\
Contempt & $20.02$ & $13.74$ & $22.37$ & $22.56$ & $\mathbf{25.62}$ & $22.87$ \\
Disgust & $18.91$ & $31.53$ & $33.33$ & $\mathbf{37.31}$ & $32.08$ & $36.13$ \\
Fear & $12.22$ & $16.79$ & $\mathbf{29.22}$ & $27.74$ & $29.09$ & $26.41$ \\
Happy & $55.93$ & $\mathbf{62.94}$ & $40.00$ & $43.33$ & $50.88$ & $62.83$ \\
Neutral & $33.13$ & $37.06$ & $31.24$ & $20.45$ & $16.37$ & $\mathbf{40.29}$ \\
Sad & $50.60$ & $51.47$ & $40.69$ & $47.47$ & $46.86$ & $\mathbf{55.62}$ \\
Surprise & $34.94$ & $37.31$ & $32.93$ & $33.48$ & $31.83$ & $\mathbf{42.81}$ \\ \bottomrule
\end{tabular}}
\end{table}
The speech-text model (ST1) generally performs between the speech and text models across classes (except ``neutral''), while the ensemble achieves the best scores in four out of eight emotion categories.\\
\noindent \textbf{Majority voting with other combinations:} We evaluate model combinations using majority voting (Table~\ref{tab:majority}). The best-performing trio—S2 (speech), T2 (text), and ST1 (speech-text)—achieves $40.36\%$ validation macro-F1 and $39.74\%$ on the test set (Comb. I). Adding S1 improves validation performance by $0.63\%$ absolute (Comb. III), while ensembling all five models boosts it by nearly $1.5\%$ absolute (ABHINAYA). Since T1, T2, and ST1 perform better on the minority classes, ensembling all models enhances performance on these underrepresented categories compared to ensembling only S1, S2, T2, and ST1. The impact of ST1 is evident from the nearly $2\%$ absolute drop in Comb. IV when it is removed.  We also evaluate majority voting with the three models having the fewest parameters (S1, T2, and ST1) (Comb. II), using ST1 for tie-breaking. The validation F1 score drops to $39.07\%$, highlighting the need for larger models for the SER task.
\begin{table}[t]
\centering
\caption{Results (in $\%$) with majority voting from different combinations of models from Abhinaya.}\label{tab:majority}
\resizebox{0.55\columnwidth}{!}{%
\begin{tabular}{@{}l|c|c|c|c|c||c|c@{}}
\toprule
\begin{tabular}[c]{@{}l@{}}Model\\ ensemble\end{tabular} & S1 & S2 & T1 & T2 & ST1 & \begin{tabular}[c]{@{}l@{}}Val. F1\\ (macro)\end{tabular} & \begin{tabular}[c]{@{}l@{}}Test F1\\ (macro)\end{tabular} \\ \midrule
Comb. I & \XSolid & \Checkmark  & \XSolid & \Checkmark & \Checkmark & $40.36$ & $39.74$ \\ \midrule
Comb. II & \Checkmark & \XSolid  & \XSolid & \Checkmark & \Checkmark & $39.07$ & $-$ \\ \midrule
Comb. III & \Checkmark & \Checkmark & \XSolid & \Checkmark & \Checkmark  & $40.99$ & $-$ \\ \midrule 
Comb. IV & \Checkmark & \Checkmark & \Checkmark & \Checkmark & \XSolid  & $40.48$ & $-$ \\ \midrule \midrule
ABHINAYA & \Checkmark & \Checkmark  & \Checkmark  & \Checkmark  & \Checkmark  & $42.31$ & $44.02$ \\ \bottomrule
\end{tabular}}
\end{table}
\subsection{Summary of SER in Naturalistic Conditions}
In this section, we present Abhinaya, a system for speech emotion recognition (SER) developed as part of the Interspeech Naturalistic Speech Emotion 2025 challenge. Our approach combines five models leveraging large language models (LLMs) and speech large language models (SLLMs)—two speech-based, two text-based, and one multimodal model. To address class imbalance, we explore various loss functions, demonstrating their impact on model performance. Our model ensemble achieves the best published results on this task. Experimental analysis highlights   the importance of the different components and loss functions suitable for fine-tuning each model.
\section{Chapter Summary}

This chapter addressed the problem of multimodal emotion recognition from spoken language in two challenging: emotion recognition in conversations and emotion recognition under naturalistic recording conditions. In the ERC setting, we highlighted the importance of modeling conversational context and modality-specific dynamics, and proposed hierarchical and modular architectures that enable effective fusion of acoustic and semantic cues while remaining robust to modality imbalance. 

We further extended our investigation to naturalistic speech, where spontaneous emotional expression, environmental variability, and severe class imbalance pose significant challenges to existing SER systems. Using a large-scale real-world benchmark, we evaluated contemporary speech and language models under these conditions and analyzed their limitations and strengths. 

Together, the studies in this chapter demonstrate that robust emotion recognition requires principled modeling of both acoustic and semantic information. 

\newpage\null\thispagestyle{empty}\newpage
\chapter{Emotion Style Transfer in Speech-to-Speech systems}\label{chap:styletransfer}

Emotion understanding is a necessary but not sufficient condition for building emotionally intelligent speech systems. Beyond recognizing emotions in spoken language, many applications require the ability to explicitly control how emotions are expressed in speech, while preserving the underlying linguistic content and speaker identity. Such controllable generation is essential for expressive speech synthesis, personalized conversational agents, and data augmentation for emotion recognition systems.

In spoken language, emotional expression is realized through subtle acoustic variations that are tightly coupled with linguistic structure. Generating expressive speech therefore requires disentangling emotion-related acoustic patterns from semantic content and speaker-specific characteristics. This problem is particularly challenging in speech-to-speech settings, where no textual supervision is available and paired recordings of the same content across different emotions are scarce.

In this chapter, we study the problem of \emph{emotion style transfer} (EST), where the goal is to transfer the emotional speaking style from a reference speech signal to a source speech signal while preserving both the linguistic content and speaker identity of the source. Unlike expressive voice conversion, EST requires emotion to be manipulated independently of speaker identity, making it a more stringent test of acoustic-semantic disentanglement.

We explore a textless and non-parallel setting, where neither transcriptions nor parallel emotional recordings are available during training. By framing emotion style transfer as a disentanglement problem, we propose a framework that enables zero-shot transfer of emotional style across speakers and utterances. Extensive objective and subjective evaluations demonstrate the effectiveness of the proposed approach, and we further show how emotion style transfer can be leveraged as a data augmentation strategy to improve downstream emotion recognition performance~\cite{dutta2025audio}.
\begin{figure}[t!]
    \centering
\includegraphics[width=0.7\textwidth,trim={0cm 2cm 3.5cm 2cm},clip]{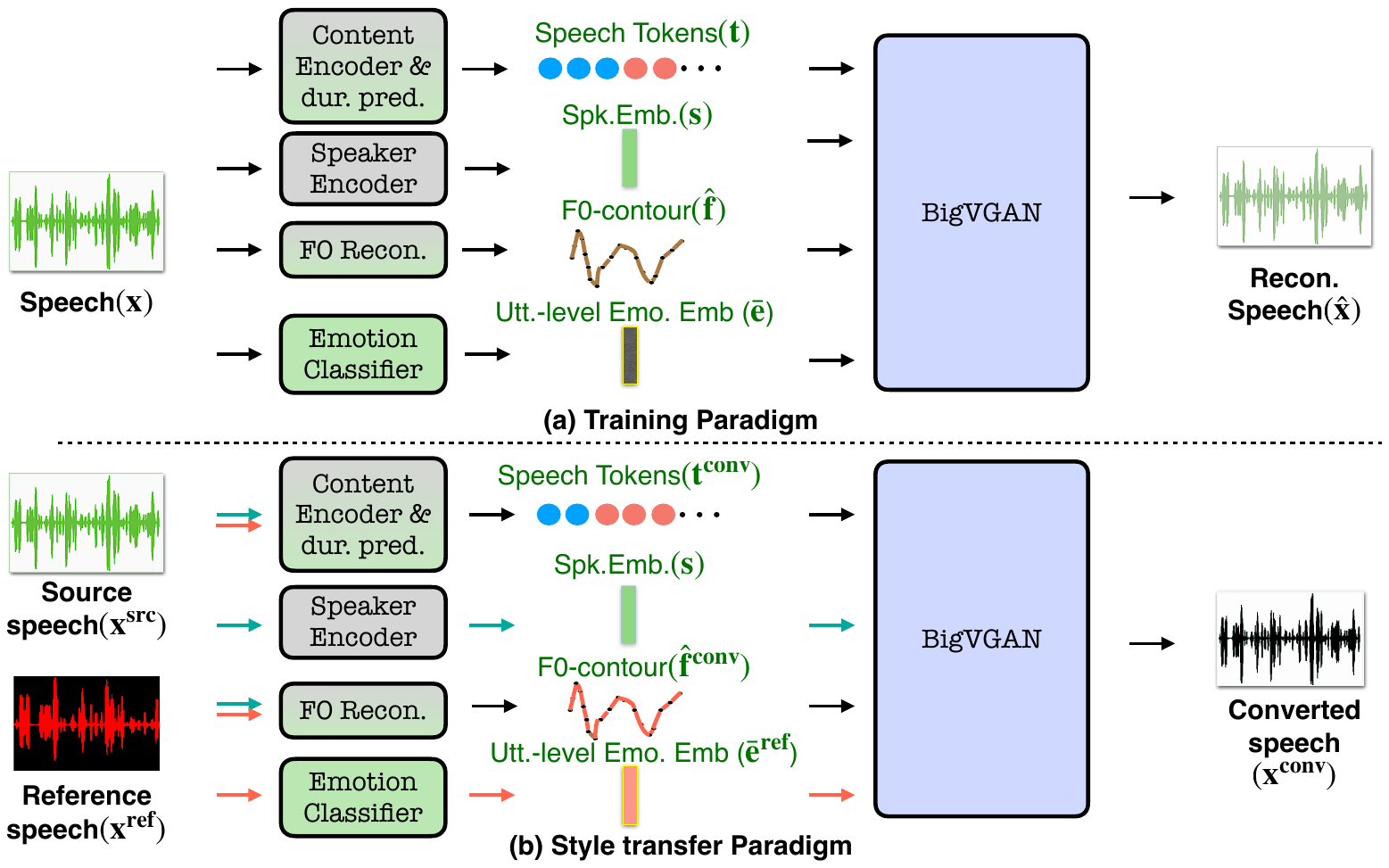}

    \caption{Overview of (a) S2S-ZEST training  and (b) style transfer paradigm. Emotion style factors are coloured differently from the rest. During style transfer, the source speech tokens are passed through the content and speaker encoder, while the duration predictor, F0 reconstruction module and emotion classifier modules receive their input from the reference speech.}
    \label{fig:overview}

\end{figure}
\section{Proposed Approach}\label{sec:est_method}

\subsection{Big picture}

Let an input speech recording be denoted as $\x = \{x_1,...,x_N\}$, where $N$ denotes the number of samples in the recording. Following a windowing operation with $N_w$ samples in each frame, we obtain $T$ frames (for non-overlapping frames, $T=N/N_w$). 
The windowed speech signal is   converted to a sequence of tokens using the content encoder (detailed in Sec.~\ref{sec:con}). 
The tokenizer generates $\{ t_1,... t_T \}$ discrete tokens, where $t_i \in \{1...V\}, $ is the cluster index and $V$ is the total vocabulary size of the tokenizer. 
The speaker information is captured by an utterance-level embedding $\mathbf{s}$, extracted using a pre-trained speaker encoder (Sec.~\ref{sec:spkr}).  
 The emotion information in the signal is extracted with an emotion classifier, which provides frame-level   embeddings ($\mathbf{E} = \{\mathbf{e}_1, \mathbf{e} _2, .. \mathbf{e} _T\}$) and an utterance-level pooled output ($\mathbf{\bar {e}}$).  The emotion embedding extraction is  detailed in Sec.~\ref{sec:sace}.
 
Using the basic components (token sequence, speaker and emotion embeddings), we train,
 \begin{itemize}
     \item Pitch contour reconstruction module, which reconstructs the frame-level F0 sequence, $\mathbf{\hat{f}} = \{f_1,...f_T\}$, of $\mathbf{x}$, described in Sec.~\ref{sec:pitchpred}. 
     \item Token duration predictor, which generates the duration $\mathbf{\hat{d}} = d_1,...,d_{T^{'}}$ of the de-duplicated tokens $t_1,..t_{T^{'}}$. Here, $T^{'}$ denotes the number of tokens after de-duplication\footnote{E.g. if tokens are $\{1,1,1,41,41,1,1,5,5,5,5,5\}$ the de-duplicated token sequence is $\{1,41,1,5\}$ with the durations being $\{3,2,2,5\}$.}. This is  detailed in Sec.~\ref{sec:dur}. 
 \end{itemize}
Thus, the encoder analysis pipeline extracts five components from the speech signal: i) de-duplicated tokens, ii) token durations, iii) speaker embedding, iv) emotion embedding, and v) pitch contour. With these components as input, we train a BigVGAN model for reconstructing the speech signal using the auto-encoding loss (Sec.~\ref{sec:synth}). A brief overview of the training procedure for S2S-ZEST is shown in Fig.~\ref{fig:overview} (a). 

 During style transfer (depicted in Fig.~\ref{fig:overview} (b)),
  the source speech tokens  are generated and de-duplicated to obtain unique tokens, $t_1,..,t_{T^{'}}$. 
 The duration predictor is fed with source speech content and reference speech style information, and this generates the token durations,  $\mathbf{\hat{d}}^{conv}$. The speaker embedding, $\mathbf{s}$,  is extracted from the source speech, while the emotion embedding, $\mathbf{E}^{ref}$, is extracted from the reference speech. The F0 contour, $\mathbf{\hat{f}}^{conv}$, is constructed using the source content and speaker along with the reference emotion embeddings. These representations are input to the synthesizer to generate the style-converted speech. The emotion style-transfer process is elaborated in Sec.~\ref{sec:emoconv}.

\subsection{Content Encoder}
\label{sec:con}

In order to encode speech content that is devoid of speaking-style or speaker information, previous work by Polyak~\cite{polyak2021speech} illustrated the benefits of discretization of self-supervised speech embeddings. 
 Separately, Niekerk et al.\cite{van2022comparison}  refined a HuBERT-base model to predict frame-level soft embeddings with a continued pre-training of the HuBERT back-bone.  
However, the continuous soft-HuBERT features make it difficult to control the speaking rate.
We use the pre-trained soft-HuBERT embeddings  \cite{van2022comparison} and train a k-means clustering model to generate discrete tokens of speech. 
This preserves the intelligibility benefits of soft-HuBERT while allowing control over speech duration during synthesis. 
Denoting the output from the content encoder as $\mathbf{t}=\{t_1, t_2, \dots,t_T\}$, we have  $\mathbf{t} = \texttt{k-means}(\mathbf{H})$, where the soft-HuBERT model embeddings are denoted as, $\mathbf{H} = \{\mathbf{h}_1,...,\mathbf{h}_T\}$, i.e., $\mathbf{H} = \texttt{soft-HuBERT}(\x)$
\begin{figure}[t!]
    \centering
\includegraphics[width=0.65\textwidth,trim={3cm 11cm 6cm 3cm},clip]{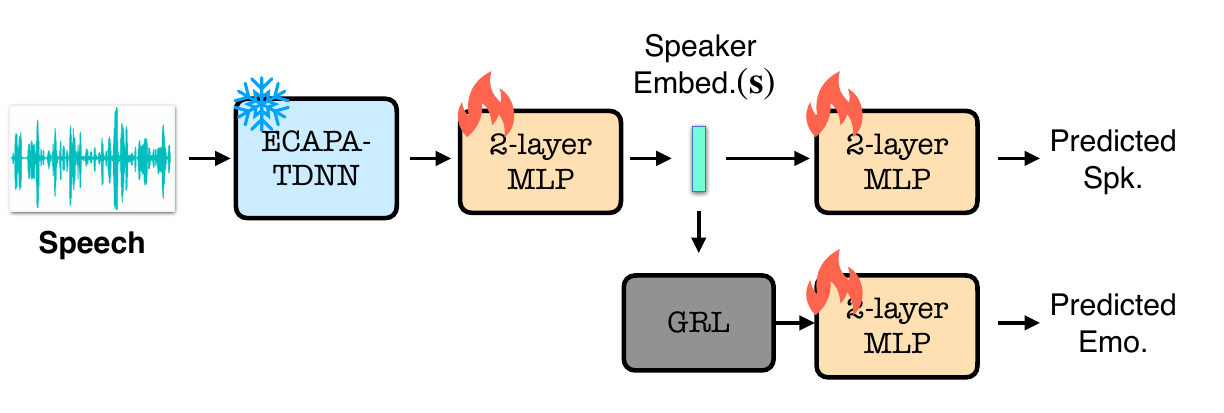}
    \caption{The model for extracting the speaker embedding. GRL stands for Gradient Reversal Layer. The blue block is kept frozen during training. }
    \label{fig:ease_model}
\end{figure}
\subsection{Speaker Encoder}
\label{sec:spkr}

To extract speaker embeddings, we employ a pre-trained speaker verification model. Specifically, we use ECAPA-TDNN~\cite{desplanques20_interspeech}, which generates x-vector embeddings by applying utterance-level pooling over frame-level representations.
Prior studies~\cite{pappagari2020x, shaheen23_interspeech} indicate that x-vectors also encode emotion information, which can interfere with style transfer. To mitigate this, inspired by Li et al.\cite{li2022cross}, we introduce two fully connected layers to the x-vector model and apply an emotion adversarial loss \cite{ganin2016domain}. The speaker embeddings in S2S-ZEST are trained using the following loss function:
\begin{equation}\label{eq:lossspkr}
    \mathrm{L}_{tot-spk} = \mathrm{L}_{ce}^{spk} - \lambda_{adv}^{emo} \mathrm{L}_{ce}^{emo}
\end{equation}
where $\mathrm{L}_{ce}^{spk}$ is the  speaker classification loss, and $\mathrm{L}_{ce}^{emo}$ is the emotion classification loss.  This is depicted in Fig.~\ref{fig:ease_model}. 

\subsection{Emotion Style Factors}
\label{sec:emo_pred}

\noindent \subsubsection{\textbf{Emotion classifier}}
\label{sec:sace}
We use an emotion classifier with the pre-trained HuBERT-base model, which comprises a convolutional feature extractor followed by $12$ transformer layers. 
For emotion classification, we fine-tune the transformer layers. 
The emotion embedding is the output of the last transformer layer, denoted as $\mathbf{E} = \texttt{Emo-embed}(\mathbf{x})$,
where $\mathbf{E} = {\{\mathbf{e}_1,...\mathbf{e}_T\}}, \mathbf{e}_i \in \mathrm{R}^{D}$,  with $D=768$. The frame-level embedding matrix $\mathbf{E}$ is average pooled over the temporal dimension to get the utterance-level emotion embedding $\mathbf{\bar{e}}$.  
 A softmax classification head is trained on $\mathbf{\bar{e}}$ for predicting the emotion class. We also require the  emotion embeddings to  be disentangled from the speaker attributes. To achieve this, we employ a speaker adversarial loss (similar to the speaker embedding extractor). The loss function is formulated as:
\begin{equation}\label{eq:lossemo}
    \mathrm{L}_{tot-emo} = \mathrm{L}_{ce}^{emo} - \lambda_{adv}^{spk} \mathrm{L}_{ce}^{spk}
\end{equation}
where $\mathrm{L}_{ce}^{emo}$ and $\mathrm{L}_{ce}^{spk}$ are the cross-entropy loss  functions for emotion and speaker classification,  respectively.
\begin{figure}[t!]
    \centering
\includegraphics[width=0.7\textwidth,trim={0cm 4cm 1cm 2cm},clip]{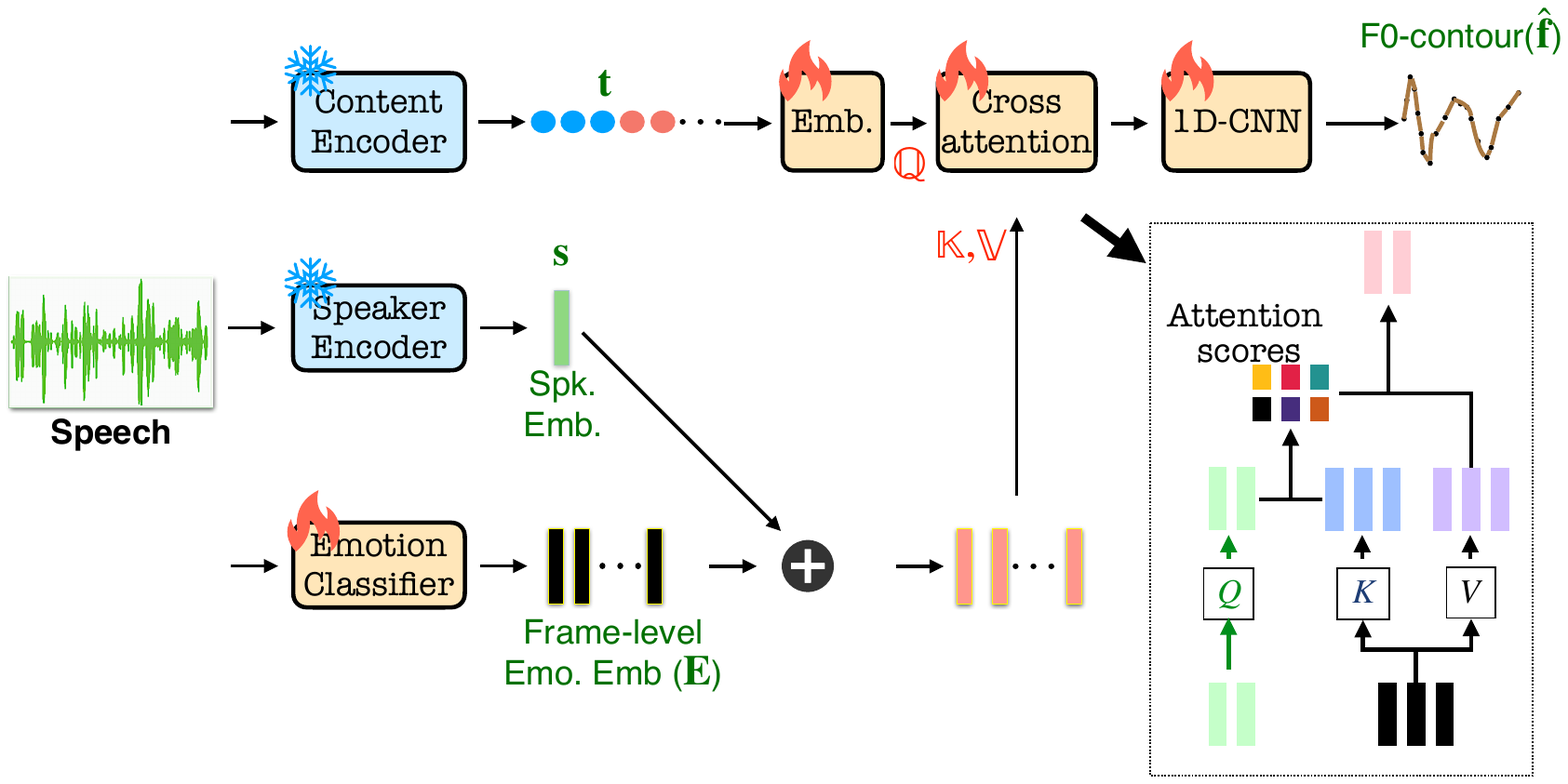}

     \caption{Pitch contour reconstruction module - Speaker embedding ($\mathbf{s}$) is added with frame-level emotion embeddings $\mathbf{E}$ and forms the key-value sequence while the source speech token embeddings ($\mathbf{C}$) form the query sequence. The frame-level outputs from the cross-attention block are passed through a position-wise feedforward network using 1D-CNNs to reconstruct the pitch contour ($\mathbf{\hat{f}}$). The cross-attention block architecture is also expanded for reference.}
    \label{fig:f0pred}

\end{figure}

\subsubsection{\textbf{Pitch Contour Reconstruction}}
\label{sec:pitchpred}
We propose a factored approach for pitch contour reconstruction, influenced by content, speaker, and emotion information. 
A learnable embedding layer provides the content representation $\mathbf{C}$ from the token sequence $\mathbf{t}$. 
As local pitch variations contribute to the emotional expressiveness~\cite{erro2009emotion,chen2024vesper}, we utilize the frame-level emotion embeddings $\mathbf{E}$ along with content representations $\mathbf{C}$ and speaker embedding $\mathbf{s}$.   
The pitch reconstruction module based on cross-attention and a $1$-D CNN network, is formulated as:
\begin{equation}\label{eq:f0}
    \mathbf{\hat{f}} = \texttt{1D-CNN}(\texttt{Attn}(\mathbf{C},~~ \mathbf{s}+\mathbf{E},~~\mathbf{s}+\mathbf{E} ) )
\end{equation}
In the above equation, the query representations $\mathbf{Q}$ come from the  content embeddings, while the speaker embedding $\mathbf{s}$ and the frame-level emotion embeddings $\mathbf{E}$ are added together to form the key ($\mathbf{K}$) and value ($\mathbf{V}$) representations. In the proposed  formulation, content embeddings are used as queries in the cross-attention model, while emotion and speaker embeddings are used as keys and values, respectively. This design allows the model to condition the pitch generation on  prosodic patterns of the target emotion and the source speaker identity, while anchoring the temporal alignment to the spoken source content.
The target pitch contour $\mathbf{f}=\{f_1, f_2,f_3,\dots,f_T\}$ is extracted using the YAAPT algorithm~\cite{kasi2002yet}, and the loss for pitch reconstruction module is defined as:
\begin{equation}\label{eq:lossf0}
    \mathrm{L}^{f0} = | |\mathbf{f}-\mathbf{\hat{f}}||_{L_1}
\end{equation}
Unvoiced frames identified by the YAAPT algorithm are assigned a value of zero and are included in the loss computation. This gives us a way to derive pitch contours based on content and speaker from one speech signal and emotion from another speech signal.  
The block diagram for pitch reconstruction is shown in Fig.~\ref{fig:f0pred}.

\subsubsection{\textbf{Duration Prediction}}
\label{sec:dur}
The discretization of self-supervised learning (SSL) features $\mathbf{H}$,   often results in repeated tokens~\cite{maimon2022speaking, lee2021direct, lee2021textless, kreuk2021textless}. To address this, we de-duplicate the tokens from the content encoder, $\mathbf{t}^{'} = \texttt{de-dup}(\mathbf{t})$.
During style transfer, we require the duration (speaking rate) of each de-duplicated source token conditioned on the speaker identity and the reference speech emotion~\cite{lee2021direct, maimon2022speaking}.  
To achieve this objective, we train the duration predictor module $\mathtt{D_{pred}}$ as:
\begin{equation}\label{eq:dur}
    \mathbf{\hat{d}} = \mathtt{D_{pred}}(\mathbf{\mathbf{t}^{'}}, \mathbf{s}, \mathbf{\bar{e}}) 
\end{equation}
The $\mathtt{D_{pred}}$ module consists of a learnable token-to-embedding layer (for inputs $\mathbf{t}^{'}$). The token embeddings are concatenated with utterance-level speaker embedding $\mathbf{s}$ and emotion embedding $\mathbf{\bar{e}}$, and processed with $1-D$ CNN layer to predict the duration of the de-duplicated tokens. Denoting the target and predicted durations by $\mathbf{d}=\{d_1, d_2,\dots, d_{T'}\}$ and $\mathbf{\hat{d}}=\{\hat{d_1}, \hat{d_2}, \hat{d_3}, \dots, \hat{d_{T'}}\}$ respectively, the loss for the duration predictor is defined as:
\begin{equation}\label{eq:lossdur}
    \mathrm{L}_{mse}^{dur} = \frac{1}{T'}||\mathbf{d} - \mathbf{\hat{d}} ||_{L_2} ^2
\end{equation}
\subsubsection{\textbf{Training Paradigm}}
\begin{figure}[t!]
    \centering
\includegraphics[width=0.7\textwidth,trim={1cm 5.5cm 2.5cm 3cm},clip]{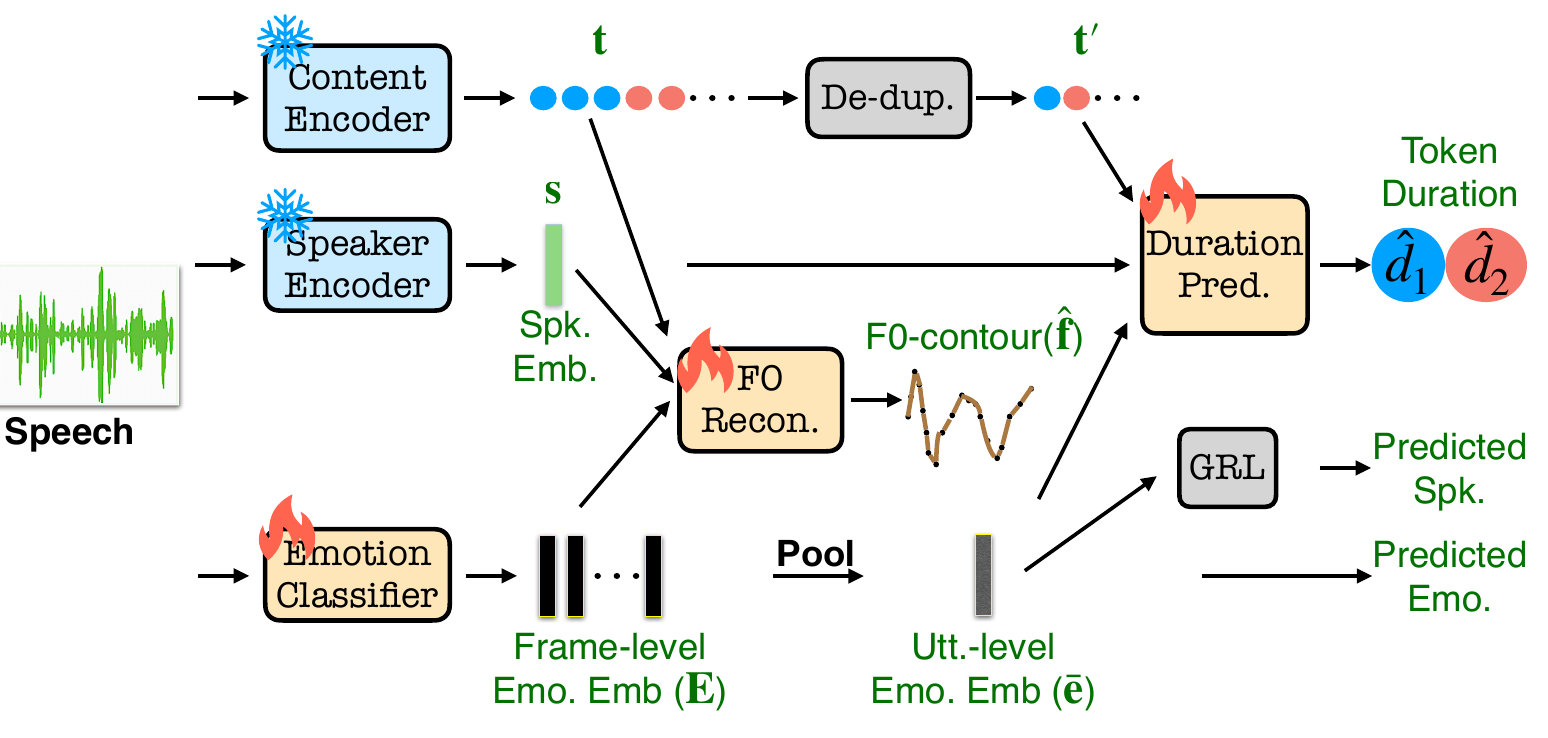}

    \caption{The different  factors that are derived from the speech in the analysis phase. The emotion classifier is trained with a speaker adversarial loss. The frame-level embeddings ($\mathbf{E}$), the speaker embedding ($\mathbf{s}$) and speech tokens ($\mathbf{t}$) are used to reconstruct the pitch contour ($\mathbf{\hat{f}}$). Further, the utterance-level emotion embedding ($\mathbf{\bar{e}}$) is used along with the de-duplicated tokens $\mathbf{t}^{'} = \{t_1,...,t_{T^{'}}\}$ to predict the duration of each of the tokens ($\mathbf{\hat{d}}$). All the blue blocks are kept frozen while the yellow blocks are trained. Grey blocks do not contain any learnable parameters.}
    \label{fig:emo_entire}

\end{figure}

The duration predictor, emotion classifier, and pitch contour reconstruction modules are trained jointly. The total loss  is:
\begin{equation}\label{eq:losstot}
    \mathrm{L}_{all} = \lambda_{e}\mathrm{L}_{tot-emo} + \lambda_{f}\mathrm{L}^{f0} + \lambda_{d}\mathrm{L}_{mse}^{dur}
\end{equation}
where $\lambda_{e}$, $\lambda_{f}$ and $\lambda_{d}$ are weighting coefficients.\\
The extraction of these components is shown in Fig.~\ref{fig:emo_entire}.
\subsection{Speech Reconstruction}
\label{sec:synth}
The token sequence, 
$\mathbf{t}$, 
is vectorized using a learnable embedding layer.  
Separately, the reconstructed pitch contour of the speech signal $\mathbf{\hat{f}}$ is vectorized through a learnable CNN-BiLSTM  network. These are concatenated with the vectorized token sequence, speaker embedding $\mathbf{s}$, and the utterance-level emotion embedding $\mathbf{\bar{e}}$ and are fed to the BigVGAN~\cite{leebigvgan} model for speech synthesis.

The BigVGAN network was proposed  as a universal vocoder that can generate high-fidelity raw speech waveform~\cite{leebigvgan}. The model was also shown to generalize well for various out-of-distribution scenarios without fine-tuning. 
The BigVGAN model upsamples the input sequence $N_w$ times using convolutional layers with residual connections.  
This network, called the generator, employs   \textit{Snake} activations ($f_\alpha(x) = x + \frac{1}{\alpha}\sin^2(\alpha x)$) for introducing the periodicity in speech signals~\cite{leebigvgan}. The discriminator network consists of two types - 1) Multi Period Discriminator (MPD)  to classify real from generated speech samples by focusing on  periodic structures of speech and 2) Multi Resolution Discriminator (MRD), where a number of discriminators are trained to separate real from generated speech samples by operating on spectrograms at different  resolutions. We use the same loss functions and hyper-parameters that are mentioned in Lee et al.~\cite{leebigvgan}.
Thus, the reconstructed speech signal $\mathbf{\hat{x}} = \{\hat{x_1}, \hat{x_2}, \hat{x_3}, \dots, \hat{x_N}\}$ is derived from the BigVGAN as, $
    \mathbf{\hat{x}} = \mathtt{BigVGAN}(\mathbf{t}, \mathbf{s}, \mathbf{\bar{e}}, \mathbf{\hat{f}})$
\begin{figure}[t!]
    \centering
\includegraphics[width=0.7\textwidth,trim={0cm 5cm 0cm 2cm},clip]{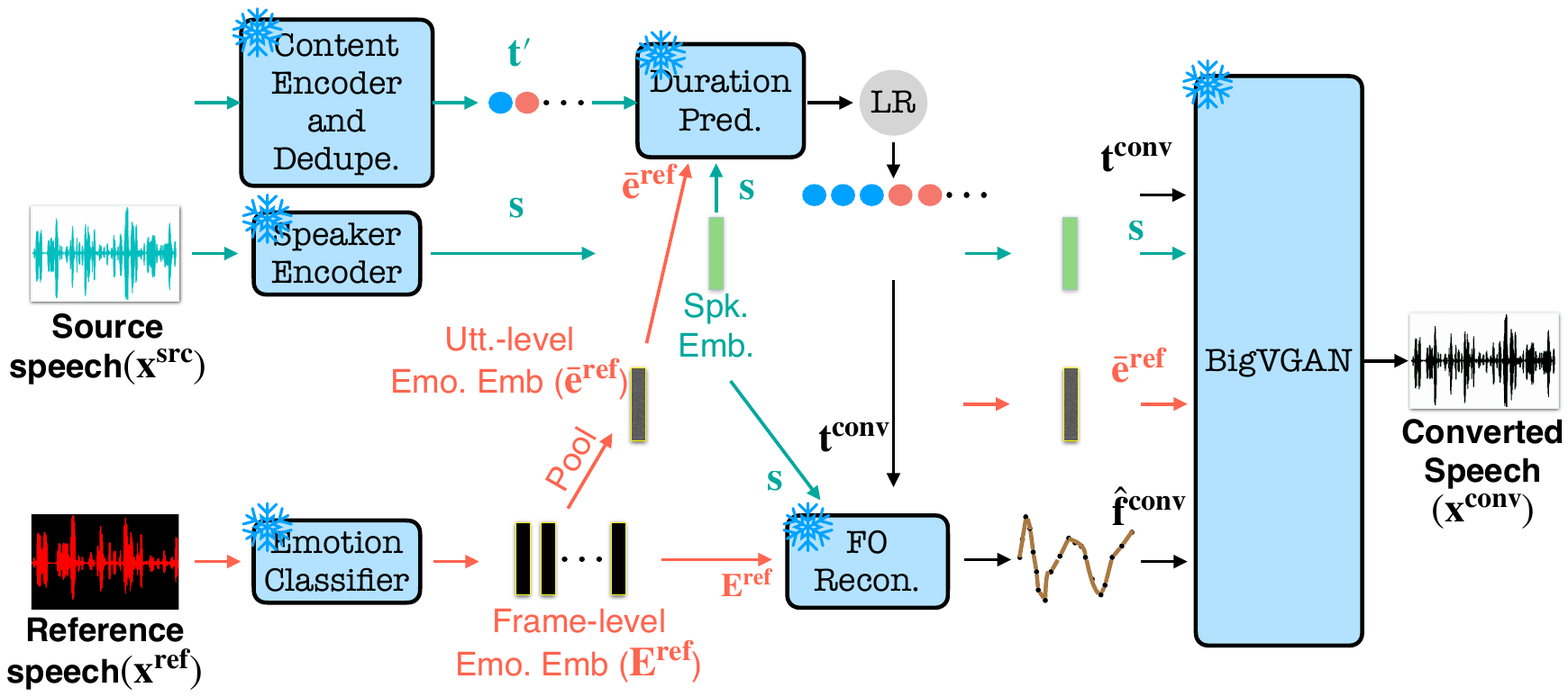}

    \caption{ Emotional style transfer - The frame-level ($\textbf{E}^{ref}$) and utterance-level ($\mathbf{\bar{e}}^{ref}$) embeddings are extracted from the reference speech. The duration prediction is performed using source tokens $\mathbf{t}^{'}$, speaker vector $\mathbf{s}$ and emotion embeddings
    $\mathbf{\bar{e}}^{ref}$. These predicted durations $\mathbf{\hat{d}}^{conv}$ are used to generate duplicated token sequence $\mathbf{t}^{conv}$. With this token sequence, $\mathbf{E}^{ref}$ and the speaker embedding $\mathbf{s}$, the $F_0$ contour is predicted, $\mathbf{\hat{f}}^{conv}$. Finally, the token sequence, speaker and emotion embeddings, and the predicted F0 contour are passed to the BigVGAN model to generate the converted speech.}
    \label{fig:conversion}

\end{figure}
\subsection{Emotion Style Transfer}
\label{sec:emoconv}
The emotion style transfer consists of the following steps:
\begin{itemize}
    \item \textbf{Emotion embeddings}: The frame-level and utterance-level emotion embeddings are extracted from the reference speech $\mathbf{x^{ref}}$, denoted as $\mathbf{E^{ref}}$ and $\mathbf{\bar{e}^{ref}}$.  
    \item \textbf{Token sequence}: We extract the de-duplicated token sequence $\mathbf{t}^{'}$ and the speaker embedding $\mathbf{s}$ from the source speech. Next, we predict the duration of each de-duplicated token using the emotion embedding $\mathbf{\bar{e}^{ref}}$ (similar to Eq.~\ref{eq:dur})  as:
    \begin{equation}\label{eq:dur_pred}
    \mathbf{\hat{d}}^{conv} = \mathtt{D_{pred}}(\mathbf{\mathbf{t}^{'}}, \mathbf{s}, \mathbf{\bar{e}^{ref}}) 
\end{equation} 
    The tokens in the sequence $\mathbf{t}^{'}$ are
    duplicated using $\mathbf{\hat{d}}^{conv}$,
    \begin{equation}
    \mathbf{t}^{conv} = \texttt{dup}(\mathbf{t}^{'},\mathbf{\hat{d}}^{conv}).     
    \end{equation} 
    
    \item \textbf{F0 Contour}: The token sequence $\mathbf{t}^{conv}$ is  vectorized to form $\mathbf{C}^{conv}$ and combined with     embeddings $\mathbf{s}$, $\mathbf{E^{ref}}$, to predict the pitch-contour (similar to Eq.~\ref{eq:f0}) as,
    \begin{equation}\label{eq:f0_pred}
    \mathbf{\hat{f}^{conv}} = \texttt{Attn}(\mathbf{C}^{conv},~~ \mathbf{s}+ \mathbf{E^{ref}},~~\mathbf{s}+ \mathbf{E^{ref}} ) 
\end{equation}

    \item \textbf{Speech Synthesis}: Finally, the converted speech is, 
    \begin{equation}
    \mathbf{x^{conv}} = \mathtt{BigVGAN}(\mathbf{t^{conv}}, \mathbf{s}, \mathbf{\bar{e}^{ref}}, \mathbf{\hat{f}^{conv}}) 
\end{equation}
    \end{itemize}
These steps are highlighted in Fig.~\ref{fig:conversion}.
\section{Experiments and Results}
\subsection{Datasets and Pre-training}
\label{est_dataset}
For the content encoder, we use the soft-HuBERT base model as proposed by Niekerk et al.~\cite{van2022comparison}, which was pre-trained on $960$ hours of the LibriSpeech dataset~\cite{panayotov2015librispeech}.  
We train a k-means clustering model on randomly selected $10\%$ of the training dataset, which forms the speech tokenizer.

The emotion embeddings are extracted from an emotion classifier model trained on the Emotional Speech Database (ESD)~\cite{zhou2021seen}. The ESD database consists of $350$ parallel utterances spoken by $10$ native English and $10$ native Chinese speakers and covers $5$ emotion categories (neutral, happy, angry, sad, and surprise). We only use the English training subset of the ESD dataset to build the emotion classifier model. We follow the dataset's predefined train-validation-test splits, using $300$ utterances per speaker per emotion for training. This results in a total of $15000$ training utterances, with $2500$ unseen utterances ($50$ per speaker per emotion) for validation.

The speaker encoder is initialized with the ECAPA-TDNN model~\cite{desplanques20_interspeech}, which was pre-trained on $2794$ hours and $7363$ speakers from the VoxCeleb dataset. 
We fine-tune the speaker encoder model using the emotion-adversarial loss (Eq.~\ref{eq:lossspkr}) on the ESD dataset. 

\textcolor{black}{Unless otherwise stated, all the modules are trained on the gender-balanced English training partition of ESD. To analyze the impact of training data diversity on generalization, we additionally train a variant (denoted as S2S-ZEST-diverse) using the original ESD data augmented with $100$ hours of LibriSpeech. For this variant, the speech tokenizer, speaker encoder, and BigVGAN model are trained on the augmented dataset. This variant is used solely to study scaling behavior under increased speaker variability.}

\begin{table*}[t]
\centering
\caption{Various evaluation settings for emotion style transfer.}
\label{tab:evaluation-settings}
\resizebox{0.95\textwidth}{!}{
\begin{tabular}{@{} l|l|l|l|cccc|c @{}}
\toprule
\textbf{Setting} & \textbf{Description} & \textbf{Src. Dataset} & \textbf{Ref. Dataset} & \textbf{Matched Spkr.} & \textbf{Matched Text} & \textbf{Seen Emo.} & \textbf{Seen Spkr.} & \textbf{\#Samples} \\
\midrule
SSST & Same Speaker Same Text & ESD & ESD & \Checkmark & \Checkmark & \Checkmark & \Checkmark & $1200$ \\
SSDT & Same Speaker Different Text & ESD & ESD & \Checkmark & \XSolidBrush   & \Checkmark & \Checkmark & $1160$ \\
DSST & Different Speaker Same Text & ESD & ESD & \XSolidBrush   & \Checkmark & \Checkmark & \Checkmark & $10800$ \\
DSDT & Different Speaker Different Text & ESD & ESD & \XSolidBrush   & \XSolidBrush   & \Checkmark & \Checkmark & $10440$ \\ \midrule
UTE  & Unseen Target Emotions & ESD & CREMA-D & \XSolidBrush & \XSolidBrush & \XSolidBrush   & \Checkmark & $1000$ \\
USS  & Unseen Source Speakers & TIMIT & ESD & \XSolidBrush   & \XSolidBrush   & \Checkmark & \XSolidBrush & $800$ \\
USUE  & Unseen Speaker Unseen Emotion & TIMIT & CREMA-D & \XSolidBrush   & \XSolidBrush   & \XSolidBrush & \XSolidBrush & $2000$ \\
\bottomrule
\end{tabular}}
\end{table*}
\subsection{Implementation}
The content encoder utilizes soft-HuBERT representations which are quantized using a k-means clustering algorithm with $K=100$ clusters~\cite{polyak2021speech}. The speaker encoder is trained for $10$ epochs with $\lambda_{adv}^{emo} = 10$ (Eq.~\ref{eq:lossspkr}) and a batch size of $32$. The emotion classifier is trained with $\lambda_{adv}^{spk} = 1$ (Eq.~\ref{eq:lossemo}) and is jointly optimized with the pitch reconstruction network and duration predictor using the total loss function in Eq.~\ref{eq:losstot}. The weighting coefficients are set as $\lambda_{e} = 1000$, $\lambda_{f} = 1$, and $\lambda_{d} = 10$, where the values were set based on  validation loss. 
This joint training process is performed for $200$ epochs with a batch size of $32$ and a learning rate of $1e-4$. During inference, the predicted durations for the tokens are constrained to lie within $\pm40\%$ of the original source token durations. 
 \textcolor{black}{This constraint is applied as a post-processing step to prevent extreme duration deviations that may result in unnatural rhythm or unstable pitch reconstruction. We observed that,  without this constraint, occasional outlier predictions can lead to excessive stretching or compression of segments. The $\pm40\%$ bound was selected empirically to balance expressive variation and speech naturalness.}
The cross-attention model in the pitch reconstruction module has $4$ attention heads and a hidden dimension of $256$, while the duration predictor consists of a $1$D-CNN network with a kernel size of $3$ and a hidden dimension of $256$.
Finally, the speech reconstruction module, BigVGAN~\cite{leebigvgan}, is trained with a batch size of $16$ and a learning rate of $1e-4$. 
All models are optimized using the AdamW optimizer~\cite{loshchilov2017decoupled}. \textcolor{black}{For the S2S-ZEST-diverse variant, the same architectural configuration and hyperparameters are used, with the training performed on the augmented dataset described in Section~\ref{est_dataset}.} The code and samples are open-sourced.\footnote{\url{https://github.com/iiscleap/A2A-ZEST}}.
\subsection{Evaluation Settings}
\label{sec:est_evaluation}
There are two broad evaluation settings,
\begin{itemize} 
    \item \textbf{Seen}: The source speech contains unseen content from the test data of ESD, but the speakers and emotions are present in the training data.
    \item \textbf{Unseen}: Either the source speaker or the reference speech emotion or both are unseen during training.
\end{itemize}
The different test evaluation settings are outlined in Table~\ref{tab:evaluation-settings}. For the four evaluation settings where both source and reference speech are drawn from ESD~\cite{zhou2021seen}, the source speech is always neutral, and the reference speech belongs to one of the four emotion categories: ``happy'', ``angry'', ``sad'', and ``surprised''. In the SSDT and DSDT settings, we randomly select $10$ neutral utterances (one per speaker) from ESD to serve as source speech.\\
For unseen evaluation settings, we use CREMA-D for unseen emotions and TIMIT for unseen source speakers. In the Unseen Target Emotion (UTE) setting, we select 50 utterances each from the “fear” and “disgust” categories in CREMA-D as reference speech, paired with 10 neutral utterances (one per speaker) from ESD, yielding 1000 evaluation utterances. In the Unseen Source Speaker (USS) setting, 100 utterances from TIMIT are used as source speech, each paired with 8 emotional utterances from ESD (2 per emotion, excluding neutral), resulting in 800 evaluation utterances. 
In the Unseen Speaker Unseen Emotion (USUE) setting, we pair the same 100 TIMIT source utterances with 10 utterances each from the “fear” and “disgust” categories in CREMA-D as reference speech, for a total of 2000 evaluation utterances.

\begin{table*}[t!]
\caption{Objective evaluation results. Here, WER - Word Error Rate, Emo. Sim. - Emotion Similarity according to emotion2vec embeddings, Spk.-Sim. - Speaker Similarity according to the Resemblyzer embedding. Word PCC refers to the word speaking rate Pearson correlation coefficient. Different test settings are described in Table~\ref{tab:evaluation-settings}. $*$ indicates that these entries are not \textbf{marked in bold} or \underline{underline} as StarGANv2-EST is not applicable to $2$ out of the $7$ test settings.}
\label{tab:obj results}
\centering
\renewcommand{\arraystretch}{1.2}
\resizebox{\textwidth}{!}{
\begin{tabular}{@{}l|l|l|ccccccc|c@{}}
\toprule
\textbf{Category} & \textbf{Method} & \textbf{Metric} & SSST & SSDT & DSST & DSDT & UTE & USS & USUE & Avg \\
\midrule
\multirow{5}{*}{\textbf{Emotion Transfer}}
& StarGANv2-EST~\cite{li2021starganv2} &  & $0.39$ & $0.37$ & $0.37$ & $0.37$ & - & $0.25$ & - & $0.35$ \\
\cmidrule{2-2}\cmidrule{4-11}
& VEVO~\cite{zhangvevo}  &\multirow{5}{*}{Emo.-Sim. ($\uparrow$)}  & $-0.01$ & $-0.03$ & $-0.01$ & $-0.04$ & $-0.01$ & $0.37$ & $-0.02$ & $0.04$ \\
\cmidrule{2-2}\cmidrule{4-11}
& ZEST~\cite{dutta2024zero} &  & $0.59$ & $0.62$ & $0.52$ & $0.51$ & $0.53$ & $0.48$ & $0.41$ & $0.52$ \\
\cmidrule{2-2}\cmidrule{4-11}
& S2S-ZEST &  & $\mathbf{0.69}$ & $\mathbf{0.71}$ & $\mathbf{0.58}$ & $\mathbf{0.56}$ & $\mathbf{0.64}$ & $\mathbf{0.59}$ & $\mathbf{0.51}$ & $\mathbf{0.61}$ \\
\cmidrule{2-2}\cmidrule{4-11}
& S2S-ZEST\textcolor{black}{-diverse} &  & \textcolor{black}{$\underline{0.67}$} & \textcolor{black}{$\underline{0.70}$} & \textcolor{black}{$\underline{0.55}$} & \textcolor{black}{$\underline{0.52}$} & \textcolor{black}{$\underline{0.59}$} & \textcolor{black}{$\underline{0.57}$} & \textcolor{black}{$\underline{0.50}$} & \textcolor{black}{$\underline{0.59}$} \\
\midrule

\multirow{5}{*}{\textbf{Rhythm Transfer}}
& StarGANv2-EST~\cite{li2021starganv2} &  & $0.54$ & $0.03$ & $0.43$ & $0.02$ & - & $0.02$ & - & $0.21$ \\
\cmidrule{2-2}\cmidrule{4-11}
& VEVO~\cite{zhangvevo} & \multirow{5}{*}{Word PCC ($\uparrow$)} & $\underline{0.65}$ & $0.10$ & $\mathbf{0.59}$ & $\mathbf{0.16}$ & $0.01$ & $0.08$ & $0.01$ & $0.23$ \\
\cmidrule{2-2}\cmidrule{4-11}
& ZEST~\cite{dutta2024zero} & & $0.53$ & $0.04$ & $0.44$ & $4\times10^{-4}$ & $5\times10^{-3}$ & $0.02$ & $2\times10^{-4}$ & $0.15$ \\
\cmidrule{2-2}\cmidrule{4-11}
& S2S-ZEST &  & $\mathbf{0.68}$ & $\mathbf{0.18}$ & $\underline{0.58}$ & $\underline{0.14}$ & $\mathbf{0.07}$ & $\mathbf{0.11}$ & $\mathbf{0.06}$ & $\mathbf{0.26}$ \\
\cmidrule{2-2}\cmidrule{4-11}
& S2S-ZEST\textcolor{black}{-diverse} &  & \textcolor{black}{$\mathbf{0.68}$} & \textcolor{black}{$\underline{0.16}$} & \textcolor{black}{$\underline{0.58}$} & \textcolor{black}{$0.12$} & \textcolor{black}{$\underline{0.05}$} & \textcolor{black}{$\underline{0.09}$} & \textcolor{black}{$\underline{0.04}$} & \textcolor{black}{$\underline{0.25}$} \\
\midrule \midrule

\multirow{5}{*}{\textbf{Content Preservation}} 
& StarGANv2-EST~\cite{li2021starganv2} &  & $5.72$ &$8.28$ & $6.61$ & $\underline{6.02}$ & - & $\mathbf{9.41}$ & - & $7.21^{*}$ \\
\cmidrule{2-2}\cmidrule{4-11}
& VEVO~\cite{zhangvevo} & \multirow{5}{*}{WER ($\downarrow$)} & $\underline{4.55}$ & $7.65$ & $\underline{4.72}$ & $7.68$ & $6.16$ & $14.36$ & $14.06$ & $\underline{8.45}$ \\
\cmidrule{2-2}\cmidrule{4-11}
& ZEST~\cite{dutta2024zero} &  & $5.11$ & $5.32$ & $5.19$ & $6.11$ & $\underline{5.92}$ & $16.57$ & $16.42$ & $8.66$ \\
\cmidrule{2-2}\cmidrule{4-11}
& S2S-ZEST &  & $5.58$ & $\underline{4.94}$ & $5.55$ & $6.48$ & $6.03$ & $13.53$ & $\underline{12.43}$ & $\underline{7.79}$ \\
\cmidrule{2-2}\cmidrule{4-11}
& S2S-ZEST\textcolor{black}{-diverse} &  & $\textcolor{black}{\mathbf{3.65}}$ & \textcolor{black}{$\mathbf{3.15}$} & \textcolor{black}{$\mathbf{3.51}$} & \textcolor{black}{$\mathbf{4.82}$} & \textcolor{black}{$\mathbf{5.02}$} & \textcolor{black}{$\underline{11.63}$} & \textcolor{black}{$\mathbf{10.84}$} & \textcolor{black}{$\mathbf{6.09}$} \\
\midrule

\multirow{5}{*}{\textbf{Speaker Preservation}}
&  StarGANv2-EST~\cite{li2021starganv2}&  & $\underline{0.76}$ & $\underline{0.76}$ & $0.67$ & $0.65$ & - & $\underline{0.65}$ & - & $0.70$ \\
\cmidrule{2-2}\cmidrule{4-11}
&  VEVO~\cite{zhangvevo}& \multirow{5}{*}{Spk.-Sim.($\uparrow$)} & $\mathbf{0.86}$ & $\mathbf{0.87}$ & $\mathbf{0.86}$ & $\mathbf{0.86}$ & $\mathbf{0.85}$ & $\mathbf{0.87}$ & $\mathbf{0.86}$ & $\mathbf{0.86}$ \\
\cmidrule{2-2}\cmidrule{4-11}
& ZEST~\cite{dutta2024zero} & & $0.74$ & $0.75$ & $0.74$ & $0.72$ & $0.73$ & $0.54$ & $0.53$ & $0.68$ \\
\cmidrule{2-2}\cmidrule{4-11}
&  S2S-ZEST &  & $0.74$ & $0.73$ & $0.74$ & $0.72$ & $0.73$ & $0.53$ & $0.53$ & $0.68$ \\
\cmidrule{2-2}\cmidrule{4-11}
& S2S-ZEST\textcolor{black}{-diverse} &  & \textcolor{black}{$0.75$} & \textcolor{black}{$0.74$} & \textcolor{black}{$\underline{0.76}$} & \textcolor{black}{$\underline{0.74}$} & \textcolor{black}{$\underline{0.76}$} & \textcolor{black}{$0.61$} & \textcolor{black}{$0.59$} & \textcolor{black}{$\underline{0.71}$} \\
\bottomrule
\end{tabular}}
\end{table*}
\subsection{Objective Evaluation Metrics} \label{sec:metrics}

\begin{itemize}
    \item \textbf{Emotion transfer w.r.t reference}: We extract embeddings from emotion2vec~\cite{ma2024emotion2vec} for both the reference and converted speech and compute their mean cosine similarity (denoted by Emo.-Sim.).
    \item \textbf{Rhythm transfer w.r.t reference}: We force align the reference and the converted speech signals with their corresponding text transcripts using a pre-trained ASR model~\cite{baevski2020wav2vec}. Following this, the word speaking rate is computed for the reference and the converted speech. The Pearson Correlation Coefficient (PCC) between the reference speaking rate and the converted speaking rate is used as measure of rhythm transfer. This is similar to the metric proposed by Barrault et al.~\cite{barrault2023seamless}.
    \item \textbf{Content preservation w.r.t source}: We transcribe the converted speech using an automatic speech recognition (ASR) system. Specifically, we employ the pre-trained Whisper-large-v3 model~\cite{radford2023robust} for performing speech recognition.  The word error rate (WER) is measured using the ground-truth transcripts of the source speech.
    
    \item \textbf{Speaker preservation w.r.t source}: We report the average cosine similarity of speaker embeddings 
     between the source and the converted speech.
     To derive these speaker embeddings, we use the Resemblyzer tool~\footnote{\url{https://github.com/resemble-ai/Resemblyzer}}. This is denoted as Spk.-Sim.    
\end{itemize}
\subsection{Comparison with baselines}
Three baseline systems are considered:
\begin{itemize}
    \item \textbf{StarGANv2-EST}: Li et al.~\cite{li2021starganv2} proposed a StarGAN-v2 network for voice conversion. An auxiliary classifier was trained to classify the source speaker alongside the discriminator, enabling speaker conversion. For our experiments, we modify this architecture to perform emotion transfer by using the auxiliary network to classify emotions instead. We refer to this modified model as StarGANv2-EST. It is trained on the same dataset as S2S-ZEST. Note that this baseline cannot handle emotions unseen during training, and is therefore not evaluated on the UTE and USUE test settings.
    \item \textbf{VEVO}: We use the pre-trained VEVO model proposed by Zhang et al.~\cite{zhangvevo}. Since this model was trained on a much larger corpus than S2S-ZEST, we do not fine-tune it for our test settings. Instead, we use VEVO’s style imitation pipeline for comparison.
    \item \textbf{ZEST}: We also compare with  prior work - ZEST~\cite{dutta2024zero}.
\end{itemize}

\subsection{Objective Results}\label{sec:obj_results}

The results for these objective tests are shown in Table \ref{tab:obj results}. 
The following are the insights drawn from these  evaluations.
\begin{itemize}

    \item \textbf{Emotion Transfer}: S2S-ZEST achieves the highest emotion similarity across both seen and unseen settings. For seen speakers and emotions, it achieves $0.69$ similarity in the SSST setting and $0.71$ in SSDT, outperforming all baselines. In the unseen target emotion (UTE) scenario, the S2S-ZEST achieves $0.64$ similarity, demonstrating effective transfer of emotions, not observed during training. For unseen source speakers (USS), the model achieves $0.59$, demonstrating robust generalization to novel speakers. In the most challenging unseen speaker and unseen emotion (USUE) setting, S2S-ZEST attains $0.51$, where both the speaker and emotion are unseen. In contrast, StarGANv2-EST and other prior-works achieve low similarity scores, indicating their limited ability for effective emotion style transfer. \textcolor{black}{The S2S-ZEST-diverse exhibits similar emotion similarity on average ($0.59$ vs.\ $0.61$).}

    \item \textbf{Rhythm transfer}: For same-text settings, S2S-ZEST achieves $0.68$ correlation in SSST and $0.58$ in DSST, showing strong rhythm transfer. In different-text settings, SSDT and DSDT, the correlation drops to $0.18$ and $0.14$, respectively, reflecting lower alignment due to text differences. A reduced correlation of $0.11$, $0.07$ and $0.06$ are observed for USS, UTE and USUE settings. However, across all settings, on average, S2S-ZEST ($0.26$) outperforms ZEST ($0.15$) and StarGANv2-EST ($0.21$), while its performance is similar to VEVO ($0.23$). \textcolor{black}{The S2S-ZEST-diverse variant maintains similar rhythm transfer performance.}

    \item \textbf{Content Preservation}: S2S-ZEST demonstrates consistent WER across test settings where the source speech is from ESD (i.e., all settings except USS and USUE). Although VEVO is trained on $60$K hours of speech, S2S-ZEST achieves lower WER in five out of the seven applicable test settings (SSDT, DSDT, UTE, USS, USUE). While StarGANv2-EST attains the lowest WER in some individual settings (DSDT and USS), it is not applicable for UTE and USUE, limiting its overall evaluation.
Overall, S2S-ZEST achieves the best average WER ($7.79$), demonstrating strong content preservation while performing emotion style transfer. \textcolor{black}{The S2S-ZEST-diverse variant substantially improves content preservation across all test settings, reducing the average WER from $7.79$ to $6.09$. Notably, improvements are also  observed in unseen-speaker conditions (USS: $13.53 \rightarrow 11.63$, USUE: $12.43 \rightarrow 10.84$), indicating enhanced linguistic stability.}

    \item \textbf{Speaker identity preservation}:Across all test settings, VEVO consistently achieves the highest speaker similarity, ranging from $0.85$ to $0.87$, demonstrating its strong ability to preserve speaker identity. \textcolor{black}{  When trained with additional speaker diversity, S2S-ZEST-diverse improves speaker similarity consistently across all settings, particularly in unseen-speaker scenarios (USS: $0.53 \rightarrow 0.61$, USUE: $0.53 \rightarrow 0.59$). These results suggest that speaker preservation in unseen conditions is influenced by training data diversity. Training with more speakers and with diverse emotional speech will form part of our future scope of work.}
    
\end{itemize}
\begin{figure}[t!]
    \centering
    \includegraphics[width=0.7\textwidth,trim={1cm 6cm 6cm 3cm},clip]{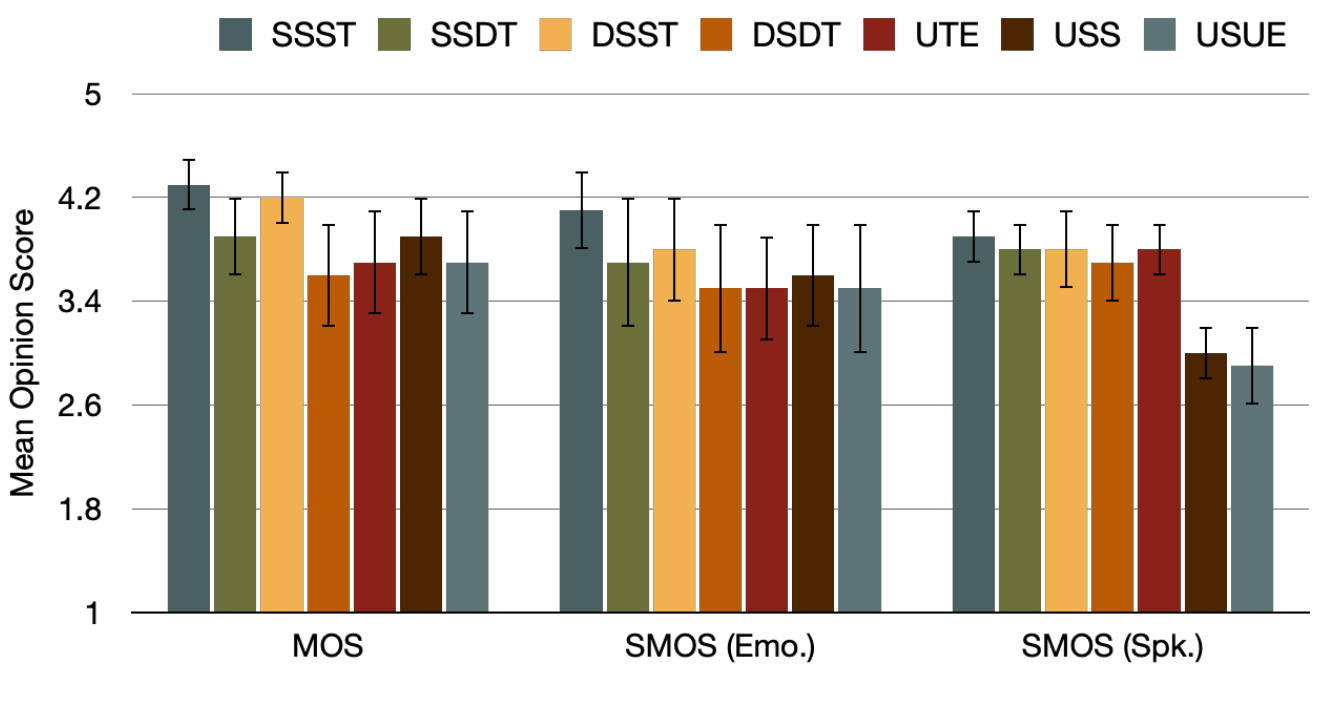}
    \caption{Subjective evaluation on the different test settings. Abbreviations used: MOS- Mean Opinion Score, SMOS - Similarity Mean Opinion Score. \textcolor{black}{The $95\%$ confidence intervals are also shown.}}
     \label{fig:subjective}
\end{figure}
\subsection{Subjective tests}
\subsubsection{Evaluation of S2S-ZEST\textcolor{black}{-diverse}}
We conduct listening tests using the Prolific\footnote{\url{https://www.prolific.co}} platform.
We recruited $30$ participants to perform the subjective evaluation. 
We chose $44$  recordings, with $8$ recordings from each of the $4$ test settings (SSST, SSDT, DSST, DSDT) and $4$ recordings each from UTE, USS, and USUE settings. The recordings were presented in a random order. The participants were also provided with training examples to clarify the objective of the test.
 
All the participants in the survey were asked to give their opinion score on the speech files (range of $1$-$5$) based on three criteria - i) Emotion similarity between the converted and the reference signal, ii) Quality of the converted speech, and iii) Speaker similarity between the converted and the source signal. 
The subjective evaluation results (in terms of mean opinion score (MOS)) are reported in Figure~\ref{fig:subjective}. The key observations from the subjective tests are as follows:
\begin{itemize}
    \item  The performance of S2S-ZEST\textcolor{black}{-diverse} is best on all the three criteria for the SSST test setting. This is expected as transfer of emotional speaking style from the reference to the source, when they share the same content and speaker, is the easiest of all the test settings. 
    \item When the source and reference share the same content but have different speakers (DSST), subjective scores indicate that emotion transfer is easier compared to the SSDT setting (same speaker, different content).
    \item The SMOS (Emo.) scores in the UTE and USUE setting are comparable to the DSDT setting, indicating that generalizing to unseen emotions is not any more challenging compared to transferring seen emotions across different speakers and content. 
    \item  \textcolor{black}{The USS and USUE settings exhibit lower speaker similarity scores compared to seen-speaker conditions, reflecting the inherent difficulty of preserving speaker identity for speakers unseen during training. While the diverse variant is a step in this direction, training the encoders and the vocoder with more speaker data will help improve the speaker generalization further.} 

\end{itemize}
\subsubsection{Comparison of S2S-ZEST\textcolor{black}{-diverse} with baselines}
\begin{figure}[t!]
    \centering
    \includegraphics[width=0.7\textwidth,trim={1cm 4cm 9cm 4cm},clip]{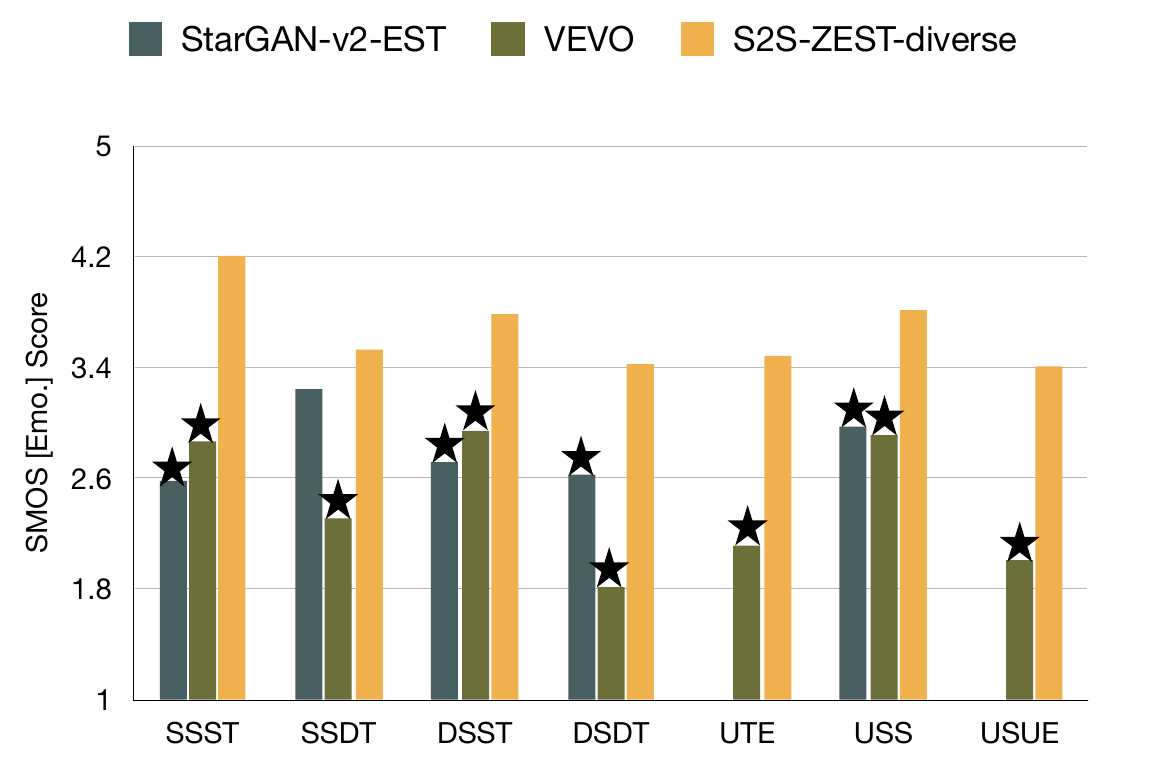}
    \caption{Subjective evaluation on the different test settings. SMOS stands for Similarity Mean Opinion Score. {\small \FiveStar}~indicates that the difference in scores between the baseline and S2S-ZEST is statistically significant (p $<$ $0.05$)}
     \label{fig:subjective_comp}
\end{figure}

We consider $4$ converted files from each of the $7$ test settings, and for each of the three methods (StarGANv2-EST does not have any outputs for the UTE and USUE test settings), and ask \textcolor{black}{$30$ participants} to rate the emotion speaking style similarity between the reference speech and the converted speech on a scale of $1$ to $5$. The results for this subjective test are shown in Figure~\ref{fig:subjective_comp}. 

The results indicate that S2S-ZEST\textcolor{black}{-diverse} outperforms both the baselines by a significant margin for all the test settings (except for StarGANv2-EST for the SSDT test setting). This showcases the superiority of S2S-ZEST\textcolor{black}{-diverse} in emotional style transfer compared to the baseline methods and follows the trends observed in the objective evaluations. 

\subsection{\textcolor{black}{Any-to-Any Emotion Style Transfer}}
\textcolor{black}{ In this experiment, the source and reference utterances share identical speaker identity and linguistic content, differing only in emotional category. This controlled setup is constructed from the ESD dataset and results in $3600$ test utterances.}

\begin{figure}[t!]
    \centering
    \includegraphics[width=0.5\textwidth,trim={1cm 7cm 11cm 5.5cm},clip]{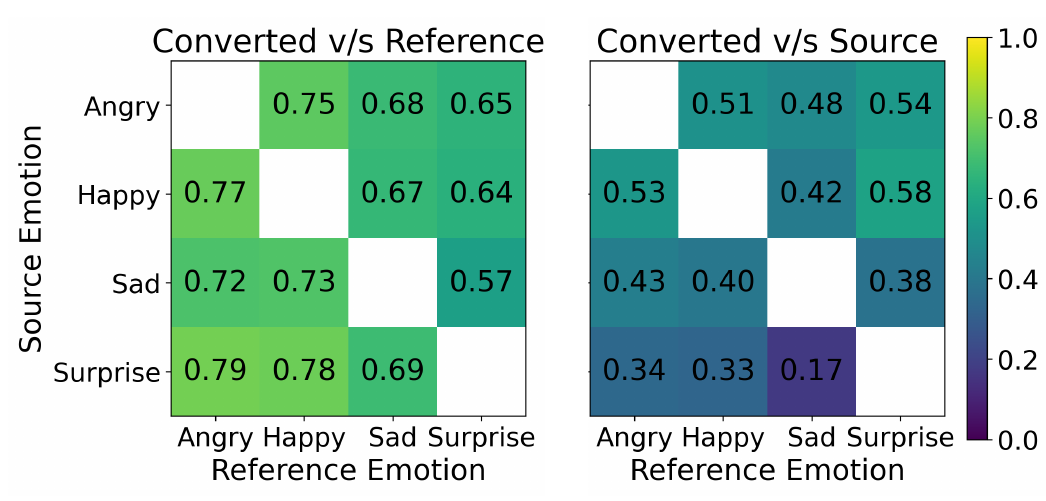}
    \caption{\textcolor{black}{Average emotion similarity between different pairs of emotions.}}
     \label{fig:any2any}
\end{figure}

\textcolor{black}{We evaluate all $4 \times 4$ source-target emotion combinations among \{angry, happy, sad, surprise\}. Emotion similarity is computed using the metric  described in Sec.~\ref{sec:metrics}. Fig.~\ref{fig:any2any} presents two emotion similarity matrices: (i) similarity between the converted speech and the reference emotion, and (ii) similarity between the converted speech and the original source emotion. Successful emotion transfer should increase similarity w.r.t. the reference while reducing similarity w.r.t. the source.} 

\textcolor{black}{Across all source-target pairs, the converted speech consistently exhibits high similarity w.r.t. the reference emotion (e.g., Surprise$\rightarrow$Angry: $0.79$, Angry$\rightarrow$Happy: $0.75$). At the same time, similarity w.r.t. the original source emotion is comparatively lower for most pairs, suggesting that the model has moved away from the source emotion.}
\textcolor{black}{For certain transitions (e.g., Sad$\rightarrow$Surprise), the reference similarity is relatively lower, reflecting the increased difficulty of converting between these two emotional states. Overall, these results demonstrate that S2S-ZEST\textcolor{black}{-diverse} supports directional emotion style transfer across expressive source conditions, beyond neutral-source scenarios.}
\begin{table}[t]
\centering
\caption{\textcolor{black}{Robustness under additive MUSAN noise in the neutral-to-emotion setting.}}
\begin{tabular}{l|c|c}
\toprule
\textcolor{black}{Condition} & \textcolor{black}{Emo. Sim. $\uparrow$} & \textcolor{black}{WER (\%) $\downarrow$} \\
\midrule
\textcolor{black}{Clean} & \textcolor{black}{$0.67$} & \textcolor{black}{$3.65$} \\
\textcolor{black}{$20$ dB} & \textcolor{black}{$0.65$} & \textcolor{black}{$4.38$} \\
\textcolor{black}{$10$ dB} & \textcolor{black}{$0.64$} & \textcolor{black}{$7.74$} \\
\bottomrule
\end{tabular}
\label{tab:noise}
\end{table}
\subsection{\textcolor{black}{Robustness to Additive Noise}}
\textcolor{black}{
To evaluate robustness under noisy conditions, we conduct experiments in the SSST test setting (Table.~\ref{tab:evaluation-settings}) with additive MUSAN noise applied to the source speech at $20$ dB and $10$ dB SNR. The reference speech remains clean. Emotion similarity and WER are computed as described in Sec.~\ref{sec:metrics}. The results are summarized in Table~\ref{tab:noise}.}

\textcolor{black}{Emotion similarity remains largely stable under moderate noise, decreasing marginally from $0.67$ (clean) to $0.65$ ($20$ dB) and $0.64$ ($10$ dB).}
\textcolor{black}{However, in terms of intelligibility, WER increases progressively from $3.65\%$ (clean) to $4.38\%$ ($20$ dB) and $7.74\%$ ($10$ dB), highlighting the   impact of additive noise on content preservation. While recognition performance degrades at lower SNR levels, emotion similarity remains relatively stable, indicating that emotion style is effectively transferred despite corruption of the source signal by noise. }
\section{Discussion}
In this section, all experiments are conducted using the base S2S-ZEST configuration trained on the ESD dataset.
\begin{figure}[t!]
    \centering
\includegraphics[width=0.7\textwidth,trim={1.5cm 4cm 9.5cm 6cm},clip]{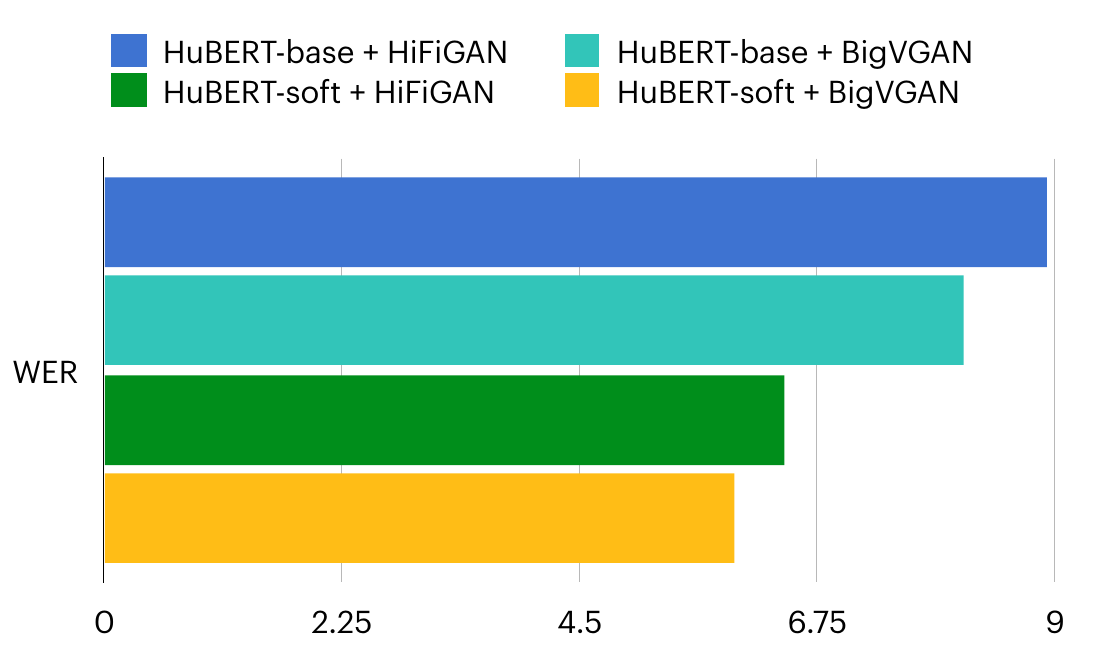}

     \caption{The WER of the resythesized speech with various combinations of the speech tokenizer and synthesizer.}
    \label{fig:token_synth}

\end{figure}
\subsection{Choice of the speech tokenizer and synthesizer}
To assess the impact of the speech tokenizer and synthesizer, we conduct an experiment where we replace the soft-HuBERT features with those from the pre-trained HuBERT-base model. We reconstruct the speech signals in the test set of ESD, with the tokens derived from HuBERT-base features. 
Additionally, we train a HiFi-GAN model as an alternative speech synthesizer with the tokens from the soft-HuBERT features. The WER results for the reconstructed speech signals with these different configurations are presented in Figure~\ref{fig:token_synth}.\\
\textbf{Key takeaways}: 1) Speech reconstructed using soft-HuBERT tokens achieves lower WER compared to HuBERT-base tokens, regardless of the synthesizer used. Prior work~\cite{van2022comparison} has shown that continuous soft-HuBERT features improve intelligibility over HuBERT-base tokens. Our results indicate that even after discretizing these features into tokens, speech reconstruction still achieves lower WER compared to HuBERT-base tokens, preserving this advantage. 2) The BigVGAN-base model consistently outperforms HiFi-GAN, irrespective of the tokenizer used, with the same number of parameters (14M). This improvement can be attributed to enhanced activation functions in BigVGAN.
\subsection{Impact of the duration predictor}\label{sec:dur_imp}
To evaluate the impact of the duration predictor, we conduct an experiment where we synthesize the converted speech using the source speech token sequence, without any duration modification.
These results are reported in Table~\ref{tab:dur_pred}.\\
\textbf{Key takeaways}: 1) Using the duration predictor leads to higher WER. However, it improves word-rate correlation, suggesting that better rhythm expression comes at the cost of reduced recognition on ASR trained with neutral speech. This aligns with observations by Maimon et al.~\cite{maimon2022speaking}. 2) The improvement of the emotion similarity across all the $4$ test settings indicates that inclusion of the duration predictor leads to effective emotion style transfer.
\begin{table}[t!]
\caption{WER, Emo.-Sim., and Pearson correlation coefficient (PCC) with respect to words for four test settings on the ESD dataset, using S2S-ZEST and S2S-ZEST without Dur. pred.}
\label{tab:dur_pred}
\centering
\renewcommand{\arraystretch}{1.2}
\setlength{\tabcolsep}{3pt} 
\resizebox{0.7\linewidth}{!}{ 
\small
\begin{tabular}{@{}c|cc|cc|cc@{}}
\toprule
\multirow{2}{*}{\shortstack{Test \\ Setting}} 
& \multicolumn{2}{c|}{WER $\downarrow$} 
& \multicolumn{2}{c|}{Emo.-Sim. (\%) $\uparrow$} 
& \multicolumn{2}{c}{Word PCC $\uparrow$} \\  
\cmidrule(l){2-7} 
& S2S-ZEST & \shortstack{$-$Dur. \\ pred.}  
& S2S-ZEST & \shortstack{$-$Dur. \\ pred.}  
& S2S-ZEST & \shortstack{$-$Dur. \\ pred.}  \\ 
\midrule
SSST & $5.58$ & $\mathbf{4.76}$ & $\mathbf{0.69}$ & $0.61$ & $\mathbf{0.68}$ & $0.51$ \\
SSDT & $4.94$ & $\mathbf{4.10}$ & $\mathbf{0.71}$ & $0.62$ & $\mathbf{0.18}$ & $0.04$ \\
DSST & $5.55$ & $\mathbf{4.85}$ & $\mathbf{0.58}$ & $0.51$ & $\mathbf{0.58}$ & $0.41$ \\
DSDT & $6.48$ & $\mathbf{5.59}$ & $\mathbf{0.56}$ & $0.52$ & $\mathbf{0.14}$ & $5\times10^{-4}$ \\ 
\bottomrule
\end{tabular}
}
\end{table}

\subsection{Importance of the adversarial losses}
\begin{figure}[t!]
    \centering
\includegraphics[width=0.8\textwidth,trim={5cm 8.5cm 2cm 5cm},clip]{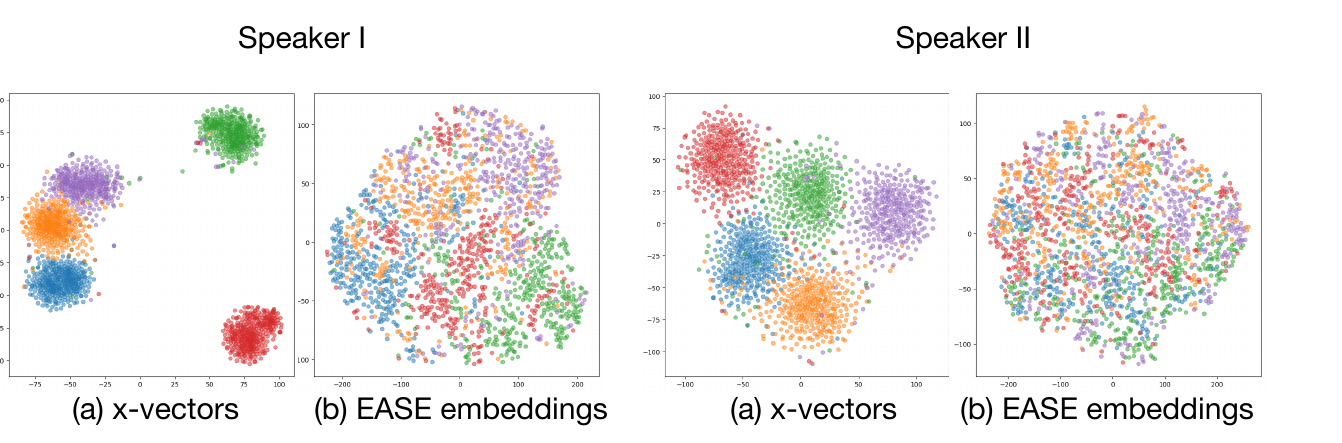}

    \caption{t-SNE plots for speaker embeddings with and without adversarial loss for two speakers I and II. In these figures, the colors correspond to $5$  emotion categories in ESD.}
    \label{fig:ease_ablation}

\end{figure}

An attentive reader might enquire about the necessity of the two adversarial losses in training S2S-ZEST, we analyze their impact using t-SNE visualizations~\cite{van2008visualizing} of x-vectors~\cite{desplanques20_interspeech} and the corresponding speaker embeddings for two speakers from the ESD dataset (see Fig.~\ref{fig:ease_ablation}). The t-SNE plots for x-vectors reveal well-defined clusters that correspond to emotion categories without the emotion adversarial training. To address this, we introduce an emotion adversarial loss (Eq.\ref{eq:lossspkr}) with $\lambda_{adv}^{emo} > 0$, forcing the speaker embeddings to be emotion-disentangled. The effectiveness of this approach is confirmed by the speaker embedding t-SNE plots in Fig~\ref{fig:ease_ablation}. \par
Additionally, we empirically assess the speaker adversarial loss in the emotion classifier (denoted as $\lambda_{adv}^{spk}$ in Eq.\ref{eq:lossemo}). The results of these ablation experiments are presented in Table~\ref{tab:adv}.\\
\textbf{Key takeaways}: 1) The removal of the emotion adversarial loss from the training of the speaker embedding leads to a significant drop in the emotion accuracy metric across all four test settings. This confirms that x-vectors inherently entangle speaker and emotion information, making disentanglement crucial. 2) The removal of the speaker adversarial loss from the emotion classifier (in Eq.~\ref{eq:lossemo}) leads to a drop in the speaker similarity for the test settings DSST and DSDT. 3) Removal of either of the adversarial losses increases WER, indicating that speaker-emotion disentanglement improves the quality of the converted speech.

\begin{table}[t!]
\caption{WER, Emo.-Sim., and Spk.-Sim. for four test settings on the ESD dataset, using S2S-ZEST, S2S-ZEST without Emo. Adv. loss ($\lambda_{adv}^{emo}=0$), and S2S-ZEST without Spk. Adv. loss ($\lambda_{adv}^{spk}=0$)}
\label{tab:adv}
\centering
\renewcommand{\arraystretch}{1.2}
\setlength{\tabcolsep}{3pt} 
\resizebox{0.7\linewidth}{!}{ 
\small
\begin{tabular}{@{}c|ccc|ccc|ccc@{}}
\toprule
\multirow{2}{*}{\shortstack{Test \\ Setting}} 
& \multicolumn{3}{c|}{WER $\downarrow$} 
& \multicolumn{3}{c|}{Emo. Acc. (\%) $\uparrow$} 
& \multicolumn{3}{c}{Spk.-Sim. $\uparrow$} \\  
\cmidrule(l){2-10} 
& S2S-ZEST & \shortstack{$-$Emo. \\ Adv.}  & \shortstack{$-$Spk. \\ Adv.} 
& S2S-ZEST & \shortstack{$-$Emo. \\ Adv.}  & \shortstack{$-$Spk. \\ Adv.} 
& S2S-ZEST & \shortstack{$-$Emo. \\ Adv.}  & \shortstack{$-$Spk. \\ Adv.}     \\ 
\midrule
SSST & $\mathbf{5.58}$ & $5.91$ & $5.64$ & $\mathbf{0.69}$ & $0.58$ & $0.67$ & $\mathbf{0.74}$ & $0.74$ & $0.74$ \\
SSDT & $\mathbf{4.94}$ & $5.02$ & $5.01$ & $\mathbf{0.71}$ & $0.47$ & $0.67$ & $0.73$ & $\mathbf{0.75}$ & $0.74$ \\
DSST & $\mathbf{5.55}$ & $5.97$ & $5.73$ & $\mathbf{0.58}$ & $0.48$ & $0.56$ & $\mathbf{0.74}$ & $0.74$ & $0.72$ \\
DSDT & $\mathbf{6.48}$ & $6.59$ & $6.51$ & $\mathbf{0.56}$ & $0.32$ & $0.51$ & $\mathbf{0.72}$ & $0.7$ & $0.71$ \\ 
\bottomrule
\end{tabular}
}
\end{table}

\begin{table}[t!]

\caption{The weighted F1-scores for IEMOCAP $4$ and $6$ class classification under different training conditions and with or without augmentation. Orig. refers to using all the training samples, while $100$ and $200$ means only $100$ or $200$ samples from each class are used for training the model. No Aug. refers to when no augmented data is used, whereas experiments using synthesized samples are mentioned as Aug.}
\centering
\label{tab:aug}
\resizebox{0.6\linewidth}{!}{
\begin{tabular}{l|ccc|ccc}
        \toprule
        Training data & \multicolumn{3}{c|}{IEMOCAP-4} & \multicolumn{3}{c}{IEMOCAP-6} \\
        \cmidrule(lr){2-4} \cmidrule(lr){5-7}
        & Orig. & $200$ & $100$ & Orig. & $200$ & $100$ \\
        \midrule
        No Aug. & $65.85$ & $59.33$ & $49.76$ & $51.67$ & $43.13$ & $36.41$ \\
        Aug. & $\mathbf{67.13}$ & $\mathbf{61.09}$ & $\mathbf{52.61}$ & $\mathbf{52.91}$ & $\mathbf{46.06}$ & $\mathbf{40.02}$ \\
        \bottomrule
    \end{tabular}}
\end{table}
\subsection{Application in Speech Emotion Recognition}\label{sec:aug}
The emotion style transfer capability of S2S-ZEST can be leveraged for data augmentation in Speech Emotion Recognition (SER). To evaluate this, we use the IEMOCAP dataset\cite{busso2008iemocap}. Details of this dataset are mentioned in Sec.~\ref{sec:setting_datasets}. For SER model training, we extract features using WavLM-base\cite{chen2022wavlm} and follow the SUPERB evaluation framework\cite{yang2021superb}.

For augmentation, we select $100$ neutral utterances and $50$ utterances per emotion class from the training split of IEMOCAP. Using these emotional utterances as reference speech signals, we generate a total of $5000$ converted utterances per emotion class with S2S-ZEST. Since the converted utterances involve unseen speakers and reference speech signals, we apply a quality filtering based on the Emo.Sim. score (Sec.~\ref{sec:metrics}), where we take the top $100$ utterances for each of the emotion classes based on this score. Thus, $300$ utterances are used to augment IEMOCAP-4 class setting, while $500$ utterances are used for the IEMOCAP-6 class setting.

 To further investigate the impact of data augmentation, we create smaller versions of the IEMOCAP dataset - $100$ and $200$ training utterances per emotion class. We report the weighted F1-scores on all the test settings for this dataset in Table~\ref{tab:aug}.
 
\textbf{Key takeaways}: 1) In the full IEMOCAP-4 dataset, augmentation improves performance by $3.69\%$ (relative). However, for IEMOCAP-4($100$) (only $100$ samples per class), the improvement is $5.67\%$, highlighting the stronger impact of augmentation in low-resource settings. A similar trend is seen for IEMOCAP-6, where improvement jumps from $2.57\%$ (full set) to $5.68\%$ (with only $100$ samples per class). 2) Even though IEMOCAP-6 has two unseen emotion classes (``excited'' and ``fear''), S2S-ZEST improves for all the three test settings, indicating the generalizable nature of S2S-ZEST.

\section{Chapter Summary}
\textit{Key Highlights:} This chapter proposes an speech-to-speech emotion style transfer by introducing S2S-ZEST, an auto-encoding framework that reconstructs speech from emotional and non-emotional factors.  While pre-trained models are used for the content and speaker, the models are trained for the emotion embedding extraction. Additionally, we develop a pitch and duration prediction pipeline conditioned on these factors. These are combined by a speech synthesis module based on the BigVGAN architecture for speech reconstruction. During style transfer, content and speaker embeddings are derived from the source, whereas emotion embeddings—derived from a reference utterance—govern the pitch and duration of the converted speech. We evaluate S2S-ZEST through objective and subjective experiments under matched and mismatched speaker/text scenarios and demonstrate its generalization to unseen speakers and target emotions. \\

\newpage\null\thispagestyle{empty}\newpage
\chapter{Summary and Future Extensions}
\label{chap:summary}

The thesis has explored the problem of modeling emotion in spoken language through the joint analysis of acoustic and semantic information, with the goal of advancing both emotion understanding and emotion synthesis. In this chapter, we first provide a consolidated overview of the contributions made across the different problem settings considered. We then discuss the limitations of the proposed approaches and outline directions for future research that build upon the insights gained in this work.

\section{Summary of Contributions}

\paragraph{Chapter~\ref{chap:pretraining}: Emotion-Aware Pre-training.}
This chapter introduced pre-training strategies for speech and text models that are better aligned with emotion understanding. The main contributions of this chapter are summarized as follows:
\begin{itemize}
    \item We proposed a novel self-supervised framework for speech emotion recognition (SER) based on dual encoders that separately model semantic and acoustic information. This model, termed CARE (\textit{Content and Acoustic Representations of Emotions}), explicitly captures complementary emotion-relevant cues in spoken language (Sec.~\ref{sec:caremodel}).
    
    
    \item We presented a supervised pre-training pipeline that leverages large language models to generate emotion-labeled text data from spoken language, facilitating emotion-aware text representation learning in the absence of manually annotated text corpora (Sec.~\ref{sec:meritsmodel}).
    
    \item Through extensive experiments, we demonstrated the benefits of acoustic-semantic modeling for improving speech emotion recognition, as well as the effectiveness of emotion-aware supervised pre-training for text-based emotion recognition (sections~\ref{sec:careresults},~\ref{sec:meritsresults}) .
\end{itemize}

\paragraph{Chapter~\ref{chap:recognition}: Multimodal Emotion Recognition from Spoken Language.}
This chapter addressed emotion recognition from spoken language in two challenging settings: Emotion Recognition in Conversations (ERC) and emotion recognition under naturalistic recording conditions. The main contributions of this chapter are summarized as follows:
\begin{itemize}
    \item For Emotion Recognition in Conversations (ERC), we proposed hierarchical and modular modeling frameworks that explicitly capture conversational structure and modality-specific dynamics. In particular:
    \begin{itemize}
        \item We introduced a hierarchical modeling paradigm in which unimodal utterance-level representations are first learned independently, followed by contextual modeling across conversational turns and multimodal fusion using cross-attention. Experimental results demonstrate that explicitly modeling conversational hierarchy leads to consistent performance improvements (Sec.~\ref{sec:hcammodel}).
        
        \item We proposed MiSTER-E, a modular mixture-of-experts framework that decouples modality-specific contextual modeling from multimodal integration through logit-level expert fusion. This design enables adaptive, per-utterance expert weighting and effectively handles modality imbalance, allowing unimodal experts to dominate when cross-modal cues are unreliable (Sec.~\ref{sec:mistere}).
        
        \item The proposed mixture-of-experts model achieved state-of-the-art performance on benchmark ERC datasets - IEMOCAP, MELD and CMU-MOSI, without relying on speaker identity information (Sec.~\ref{sec:misterd_down}).
    \end{itemize}
    
    \item For emotion recognition under naturalistic recording conditions, we presented \textit{Abhinaya}, our system developed for the Speech Emotion Recognition in Naturalistic Conditions Challenge. 
\end{itemize}

\paragraph{Chapter~\ref{chap:styletransfer}: Emotion Style Transfer in Speech.}
This chapter investigated the problem of emotion style transfer (EST) in speech, focusing on a textless and non-parallel speech-to-speech setting. The goal was to transfer emotional speaking style from a reference speech signal to a source speech signal while preserving the linguistic content and speaker identity of the source. The main contributions of this chapter are summarized as follows:
\begin{itemize}
    \item We formulated emotion style transfer as a disentangled representation learning problem, explicitly separating emotional style from linguistic content and speaker identity in speech representations. This formulation enables controlled manipulation of emotional attributes without relying on textual supervision (Sec.~\ref{sec:est_method}).
    
    \item We proposed a zero-shot emotion style transfer framework that operates in a non-parallel setting, where paired recordings of the same speaker uttering identical content in different emotions are unavailable during training. The proposed method generalizes to unseen speakers, unseen content, and unseen emotion combinations (Sec.~\ref{sec:est_evaluation}). 
    
    \item Extensive objective and subjective evaluations were conducted to assess emotional similarity, speaker preservation, content intelligibility, and naturalness of the converted speech. The results demonstrate that the proposed approach achieves effective emotion transfer while maintaining speaker identity and linguistic content (Sec.~\ref{sec:est_evaluation}).
    
\end{itemize}






\section{Limitations}

Various assumptions and modeling strategies were made without fully exploring all alternate choices. Here, we highlight some of the general limitations of the work.
\begin{itemize}
    \item The architectures proposed for the different problems are largely empirical and guided by prior works. They are proposed and analyzed based on performance metrics. The final models with the network architecture and processing is justified based on performance.
    \item Some of the works carried out in this thesis were performed when large language models (LLMs) were not ubiquitous. Such modeling paradigms may not have relevance as these large-scale models undergo continuous improvement. However, some of the approaches may still find value in low compute settings. An interesting future direction can be to investigate whether models such as CARE improve emotion understanding of speech LLMs when used as speech encoders in the LLM architecture. Further, using speech LLMs for speech-to-speech emotion style transfer is an interesting research direction to pursue.
    \item We do not provide any justification or reasoning towards the predicted emotion categories for the different models proposed in this thesis.
    \end{itemize}
    The limitations of the proposed techniques for the different problems considered in this thesis are now listed below:
    \begin{itemize}
    \item Although CARE performs better than many large models on average across the $8$ datasets considered, it still under-performs on some datasets (e.g. CMU-MOSI). Extensively training the model on even larger datasets, or improved semantic modeling might alleviate this issue.
    \item While MiSTER-E achieves strong performance across multiple benchmarks, several limitations remain,  (i) the use of large LLM/SLLM encoders introduces non-trivial computational and memory overhead, which may limit applicability in low-resource or real-time settings, despite the use of parameter-efficient fine-tuning. We have partially addressed this concern in Table~\ref{tab:our_feats}, where the proposed approach is shown to be beneficial even for non-LLM based features,  (ii) our evaluation is primarily conducted on benchmark datasets consisting of scripted or semi-scripted dialogues (e.g., TV shows and acted conversations), and performance under domain shift to spontaneous, real-world conversational settings remains an open question, (iii) emotion inference models are known to be susceptible to dataset biases related to demographic factors, language use, and annotation subjectivity and we have not performed any bias/fairness analysis in this study, and, (iv) throughout this work,  we intentionally avoid explicit speaker identity modeling, however, the utilization of speaker metadata may improve the modeling of interpersonal dynamics. Addressing some of these limitations could be pursued in future.
    \item For the naturalistic speech emotion recognition task, the ratings provided for the speech files by the different human annotators are provided as part of the dataset (Sec.~\ref{sec:abhinaya_data}). Modeling this annotator behavior may lead to better modeling of emotions from spoken language. However, in this thesis, we have focused on the majority vote among the human annotators.
    \item The base S2S-ZEST configuration is trained on the English partition of the ESD dataset (approximately $15$ hours from $10$ speakers), which limits speaker diversity during training and affects generalization to unseen voices. While the S2S-ZEST-diverse variant demonstrates that increasing training data diversity improves speaker similarity and content preservation, performance in unseen-speaker scenarios remains below large-scale systems trained on substantially larger corpora. Future work will explore training on larger and more diverse emotional speech datasets.
\end{itemize}
\begin{table*}[ht!]
\centering
\caption{The Word Error Rate (WER) (\%) for the Google ASR system on the three datasets used. Abbreviations used: Happy:\textit{Hap.}, Neutral:\textit{Neu.}, Angry:\textit{Ang.}, Excited:\textit{Exc.}, Frustrated:\textit{Fru.}, Disgust:\textit{Dis.}, Positive:\textit{Pos.} and Negative:\textit{Neg.} For IEMOCAP, the numbers refer to the combined train and validation sets.}\label{tab:asr_wer}%
\resizebox{\textwidth}{!}{%
\begin{tabular}{@{}l|ccclll|lllllll|ll@{}}
\toprule
\multirow{2}{*}{\begin{tabular}[c]{@{}l@{}}Dataset\\ Splits\end{tabular}} & \multicolumn{6}{c|}{IEMOCAP} & \multicolumn{7}{c|}{MELD} & \multicolumn{2}{c}{CMU-MOSI} \\ \cmidrule(l){2-16} 
 & \multicolumn{1}{l|}{Hap.} & \multicolumn{1}{l|}{Sad} & \multicolumn{1}{l|}{Neu.} & \multicolumn{1}{l|}{Ang.} & \multicolumn{1}{l|}{Exc.} & Fru. & \multicolumn{1}{l|}{Ang.} & \multicolumn{1}{l|}{Sad} & \multicolumn{1}{l|}{Neu.} & \multicolumn{1}{l|}{Fear} & \multicolumn{1}{l|}{Sur.} & \multicolumn{1}{l|}{Dis.} & Joy & \multicolumn{1}{l|}{Pos.} & Neg. \\ \midrule
Train & \multicolumn{1}{c|}{\multirow{2}{*}{36{}{}}} & \multicolumn{1}{c|}{\multirow{2}{*}{44{}{}}} & \multicolumn{1}{c|}{\multirow{2}{*}{31.2{}{}}} & \multicolumn{1}{l|}{\multirow{2}{*}{23.4{}{}}} & \multicolumn{1}{l|}{\multirow{2}{*}{36.5{}{}}} & \multirow{2}{*}{28{}{}} & \multicolumn{1}{l|}{50.1} & \multicolumn{1}{l|}{50.1} & \multicolumn{1}{l|}{50.2} & \multicolumn{1}{l|}{54.8} & \multicolumn{1}{l|}{59} & \multicolumn{1}{l|}{47.5} & 54.1 & \multicolumn{1}{l|}{37.3} & 37.9 \\
Val & \multicolumn{1}{c|}{} & \multicolumn{1}{c|}{} & \multicolumn{1}{c|}{} & \multicolumn{1}{l|}{} & \multicolumn{1}{l|}{} &  & \multicolumn{1}{l|}{47.4} & \multicolumn{1}{l|}{49.9} & \multicolumn{1}{l|}{46.4} & \multicolumn{1}{l|}{52.8} & \multicolumn{1}{l|}{58.8} & \multicolumn{1}{l|}{46.6} & 50.7 & \multicolumn{1}{l|}{34} & 37.9 \\
Test & \multicolumn{1}{c|}{34.5{}{}} & \multicolumn{1}{c|}{47.5{}{}} & \multicolumn{1}{c|}{30.2{}{}} & \multicolumn{1}{l|}{27.3{}{}} & \multicolumn{1}{l|}{36{}{}} & 30.5{}{} & \multicolumn{1}{l|}{50.5} & \multicolumn{1}{l|}{49.2} & \multicolumn{1}{l|}{48.1} & \multicolumn{1}{l|}{59.5} & \multicolumn{1}{l|}{61.2} & \multicolumn{1}{l|}{49.3} & 54.5 & \multicolumn{1}{l|}{38.2} & 40.4 \\ \bottomrule
\end{tabular}
}
\end{table*}
\begin{table}[t!]
\caption{\label{tab:new_asr}Weighted F1 score of our system with ASR transcripts during test time on the datasets.}
\begin{center}
\resizebox{0.6\columnwidth}{!}{
\begin{tabular}{@{}l|cc|c|c@{}}
\toprule
\multicolumn{1}{c|}{\multirow{2}{*}{\begin{tabular}[c]{@{}c@{}}Text Transcripts\\ using test time\end{tabular}}} & \multicolumn{2}{c|}{IEMOCAP} & \multirow{2}{*}{MELD} & \multirow{2}{*}{CMU-MOSI} \\ \cmidrule(lr){2-3}
\multicolumn{1}{c|}{} & \multicolumn{1}{l|}{4-way} & \multicolumn{1}{l|}{6-way} &  &  \\ \midrule
ASR & \multicolumn{1}{c|}{$84.6\%$} & $68.1\%$ & $50.2\%$ & $80.1\%$ \\
Original & \multicolumn{1}{c|}{$85.9\%$} & $70.5\%$ & $65.8\%$ & $85.8\%$\\ \bottomrule
\end{tabular}}
\end{center}
\end{table}

\section{Future Research Directions}

The work presented in this thesis opens several promising directions for future research. 
\begin{itemize}
    \item \textbf{Emotion Diarization:} In this thesis, ERC is solved when the utterances in the conversation are demarcated from each other. A more practical scenario ensues when the beginning and ending of the utterances are not marked beforehand. This problem, called Emotion Diarization, was introduced by Wang et al.~\cite{wang2023speech}, and can be considered as the next step to the problem of ERC.
    \item \textbf{Usage of ASR transcripts in ERC:} In this thesis, the provided text transcripts are utilized as the semantic component of spoken language for ERC. Instead of this, the speech signal may need to be transcribed during inference. In order to see the impact of such a change, we train the proposed HCAM model (Sec.~\ref{sec:hcammodel}) with the provided transcripts, but test it with the transcripts generated by an ASR system\footnote{\url{https://cloud.google.com/speech-to-text}\label{googlefoot}}. The word error rate (WER) of this ASR system  is reported  in Table \ref{tab:asr_wer} for each of the three datasets. As seen here, the WER on emotional conversational speech is significantly higher than those seen on other controlled datasets. The ASR performance in the case of MELD is the lowest, partly due to the high levels of background noise in the dataset. The impact of using ASR transcripts on emotion recognition is shown in Table~\ref{tab:new_asr}, where we see a significant drop for datasets like MELD. This observation leads to an interesting future research direction - attempting emotion recognition in conversations and conversational ASR jointly.
    \item \textbf{Extension of zero-shot EST to multilingual settings:} Our proposed EST system is evaluated only with speech from the English language. An interesting future direction would be to generalize this work to other languages. A more challenging aspect is to transfer the emotional style from one language to another, while training the model in a textless and non-parallel way. The challenges of training with diverse speakers across different languages in order to preserve the speaker identity will be probably higher than a monolingual setting.
    \item \textbf{Identifying mental health traits from speech:} Finally, improvement of emotion understanding from spoken language should be helpful to chatbot users who turn to these agents for mental support. The goal of pre-training these models for improved emotion understanding should be towards making them more useful to this community. This is a largely unexplored area, with the most significant roadblock being the lack of proper datasets. The collection of data, and setting up of proper benchmarks is the most important next step in this direction.
\end{itemize}

\section{Chapter Summary}
This chapter summarized the key contributions of the thesis and situated them within the broader landscape of emotion modeling in spoken language. While the proposed methods demonstrate strong performance across multiple tasks—ranging from emotion-aware pre-training to multimodal emotion recognition and emotion style transfer—they also highlight several open challenges. These include robustness to ASR system errors, computational costs and experiments limited to English. The future research directions outlined in this chapter point toward more realistic, inclusive, and generalizable emotion-aware speech systems. Collectively, the work in this thesis lays a foundation for advancing emotionally intelligent spoken language technologies while identifying clear pathways for continued progress.
\appendix

\setcounter{chapter}{0}
\renewcommand{\chaptername}{Appendix}
\renewcommand{\thechapter}{\Alph{chapter}}
\renewcommand{\thesection}{\thechapter.\arabic{section}}
\chapter{Architectural Reference Components}
\label{sec:appendix}

\section{Self-Supervised Speech Models}
\label{sec:appendix_ssl}

\subsection{PASE+}
PASE+~\cite{pascual2019learning} is among the early self-supervised models designed specifically for speech representation learning. The model is trained in a multi-task manner to predict a collection of short-term and long-term speech descriptors—such as MFCCs, filterbank energies, and prosodic features—from a corrupted version of the input waveform. By learning to reconstruct diverse speech attributes, PASE+ captures general-purpose acoustic representations. Although originally evaluated for automatic speech recognition (ASR), its self-supervised training paradigm enables its use across a wide range of downstream speech processing tasks, including speech emotion recognition.

\subsection{HuBERT}
HuBERT (Hidden-Unit BERT)~\cite{hsu2021hubert} introduces a masked prediction framework for speech representation learning. In this approach, continuous speech features are first clustered using a $k$-means algorithm to obtain discrete pseudo-labels. The model is then trained using a Transformer encoder to predict these cluster assignments for masked portions of the input sequence. By iteratively refining the clustering targets and model representations, HuBERT learns speech embeddings that have demonstrated strong performance across various speech tasks.

\subsection{WavLM}
WavLM~\cite{chen2022wavlm} extends the HuBERT framework by incorporating more diverse data augmentation strategies and a denoising training objective. In addition to masked prediction, WavLM introduces utterance mixing and noise perturbation to improve robustness and speaker-awareness in the learned representations. These modifications enable WavLM to achieve improved performance over HuBERT across multiple benchmarks, including speech recognition, speaker verification, and speech emotion recognition.

\subsection{data2vec}
data2vec~\cite{baevski2022data2vec} proposes a unified self-supervised learning framework applicable across speech, vision, and text modalities. In the speech setting, a student network is trained to predict the contextualized latent representations generated by a teacher network. The teacher receives unmasked inputs and is updated as an exponential moving average (EMA) of the student parameters. Unlike clustering-based methods such as HuBERT, data2vec directly regresses continuous hidden representations, encouraging the model to capture global contextual structure in speech. The resulting representations have been shown to generalize effectively across a wide range of downstream speech processing tasks.

\section{SONAR}~\label{sec:appendix_sonar}
SONAR~\cite{duquenne2023sonar} is a large-scale multilingual and multimodal embedding model designed to learn aligned representations across speech and text. Unlike traditional self-supervised speech models that focus solely on acoustic modeling, SONAR is trained using parallel speech–text data and contrastive objectives to produce embeddings that are semantically aligned across modalities and languages.

The model consists of modality-specific encoders for speech and text that project inputs into a shared embedding space. Training is performed using large amounts of multilingual data with translation and speech–text alignment supervision, enabling the learned representations to capture high-level semantic information. As a result, SONAR embeddings are particularly effective for cross-lingual retrieval, speech-to-text alignment, and multimodal transfer tasks.

\section{Whisper}~\label{sec:appendix_whisper}
Whisper~\cite{radford2023robust} is a large-scale encoder–decoder speech model trained in a supervised fashion for automatic speech recognition (ASR) and speech translation. The model is trained on approximately $680,000$ hours of multilingual and multitask speech data collected from the web, enabling it to generalize robustly across languages, accents, and recording conditions.

Architecturally, Whisper follows a transformer-based sequence-to-sequence framework. The encoder processes log-mel spectrogram features extracted from raw speech, while the decoder generates text tokens autoregressively. In addition to transcription, the model is trained to predict language identification and task tokens, allowing it to handle multiple speech processing tasks within a unified framework.

\section{SALMONN}~\label{sec:appendix_salmonn}
SALMONN~\cite{tang2023salmonn} is a speech large language model (SLLM) that integrates a pre-trained speech encoder (Whisper-large-v2) and an audio encoder (BEATs) with a large language model to enable speech understanding within an instruction-following framework. The architecture consists a projection layer that aligns the acoustic representations with the embedding space of a pre-trained LLM. The combined model is trained using instruction tuning so that it can process speech inputs and generate text-based responses across multiple speech understanding tasks. Unlike conventional self-supervised speech models that are optimized for reconstruction objectives, SALMONN is trained to align speech representations with high-level linguistic and reasoning capabilities of LLMs.

\bibliographystyle{unsrt}
\newpage\null\thispagestyle{empty}\newpage
\bibliography{references}
\end{document}